\documentstyle[12pt,aaspp4]{article}
\begin{document}

\newcommand{\lap}{$L_{38}^{-1/3}$}
\newcommand{\ergs}{\rm \su  erg \su s^{-1}}
\newcommand{\etal}{ {\it et al.}}
\newcommand{\porb}{ P_{orb} }
\newcommand{\Po}{$P_{orb} \su$}
\newcommand{\pdot}{$ \dot{P}_{orb} \,$}
\newcommand{\pot}{$ \dot{P}_{orb} / P_{orb} \su $}
\newcommand{\mm}{$ \dot{m}$ }
\newcommand{\mdot}{$ |\dot{m}|_{rad}$ }
\newcommand{\myr}{ \su M_{\odot} \su \rm yr^{-1}}
\newcommand{\msol}{\, M_{\odot}}
\newcommand{\ppp}{ \dot{P}_{-20} }
\newcommand{\cms}{ \rm \, cm^{-2} \, s^{-1} }
\newcommand{\pdott}{ \left( \frac{ \dot{P}/\dot{P}_o}{P_{1.6}^{3}}
\right)}

\def\p{\phantom{1}}
\def\pmu{\mbox{$^{-1}$}}
\def\ApJ{{\it Ap.\,J.\/}}
\def\ApJL{{\it Ap.\,J.\ (Letters)\/}}
\def\ApJS{{\it Astrophys.\,J.\ Supp.\/}}
\def\AJ{{\it Astron.\,J.\/}}
\def\AA{{\it Astr.\,Astrophys.\/}}
\def\AAL{{\it Astr.\,Astrophys.\ Letters\/}}
\def\AAS{{\it Astr.\,Astrophys.\ Suppl.\,Ser.\/}}
\def\MN{{\it Mon.\,Not.\,R.\,Astr.\,Soc.\/}}
\def\Na{{\it Nature \/}}
\def\SAIt{{\it Mem.\,Soc.\,Astron.\,It.\/}}
\def\kms{km^s$^{-1}$}
\def\sbu{mag^arcsec${{-2}$}}
\def\e{\mbox{e}}
\def\dex{\mbox{dex}}
\def\L{\mbox{${\cal L}$}}
\def\gte{\lower 0.5ex\hbox{${}\buildrel>\over\sim{}$}}
\def\lte{\lower 0.5ex\hbox{${}\buildrel<\over\sim{}$}}
\def\loe{\lower 0.6ex\hbox{${}\stackrel{<}{\sim}{}$}}
\def\goe{\lower 0.6ex\hbox{${}\stackrel{>}{\sim}{}$}}
\def\lessim{\lower 0.6ex\hbox{${}\stackrel{<}{\sim}{}$}}
\def\gtrsim{\lower 0.6ex\hbox{${}\stackrel{>}{\sim}{}$}}
\def\sgr{SGR~1900+14 }
\def\sgrp{SGR~1900+14}


\title{
The Giant Flare of 1998 August 27 from SGR 1900$+$14: \break
II. Radiative Mechanism and Physical Constraints on the Source
}

\author{C. Thompson\altaffilmark{1,3} and R.C. Duncan\altaffilmark{2,3}
}
\vskip 1in

\altaffiltext{1}{Canadian Institute for Theoretical Astrophysics,
60 St. George St., Toronto, ON M5S 3H8}

\altaffiltext{2}{University of Texas, Department of Astronomy \& McDonald
Observatory, Austin, TX 78712, USA}

\altaffiltext{3}{Institute for Theoretical Physics, University of California,
Santa Barbara, CA 93106, USA} 

\vskip 1in
\centerline{Accepted for publication in the Astrophysical Journal}
\centerline{11 July 2001}

\begin{abstract}
The extraordinary 1998 August 27 giant flare places
strong constraints on the physical properties of its source, SGR 1900+14.
We make detailed comparisons of the published data with the magnetar model, 
which identifies the Soft Gamma Repeaters as neutron stars endowed with
$\sim 10^{15}$ G magnetic fields.  The giant flare evolved
through three stages, whose radiative mechanisms we address in turn.
The extreme peak luminosity $L > 10^6\,L_{\rm edd}$, hard spectrum,
and rapid variability of the initial $\sim 0.5$ s spike emission all point to
an expanding pair fireball with very low baryon contamination.  
We argue that this energy must have been deposited directly through
shearing and reconnection of a magnetar-strength external magnetic field.
Low-order torsional oscillations of the star fail to transmit energy
rapidly enough to the exterior, if the surface field is much weaker.
A triggering mechanism is proposed, whereby a helical distortion
of the core magnetic field induces large-scale fracturing in the crust and
a twisting deformation of the crust and exterior magnetic field.
After the initial spike (whose $\sim 0.4$ s duration can be related to the
Alfv\'en-crossing time of the core), very hot ($T \la 1$ MeV) plasma 
rich in electron-positron pairs remains confined close to the star
on closed magnetic field lines.  The envelope of the August 27 flare can
be accurately fit, after $\sim 40$ s, by the contracting surface of such a
``trapped fireball."  The form of this fit gives evidence that
the temperature of the trapped pair plasma decreases outward from its center.
We quantify the effects of direct neutrino-pair emission on the X-ray
light curve, thereby deducing a lower bound 
$\mu_{min} \sim 1\times 10^{32}$ G-cm$^3$ to the magnetic moment of the 
confining field -- comparable to the strongest fields measured in radio 
pulsars.  The radiative flux during the intermediate $\sim 40$ s of 
the burst appears to exceed the trapped fireball fit.  The lack of 
strong rotational modulation and intermediate hardness
of this smooth tail are consistent with the emission from an extended pair 
corona, in which O-mode photons are heated by Compton scattering.  
This feature could represent residual seismic activity within
the star, and accounts for $\sim 10$ percent of the total flare fluence.   
We consider in detail the critical luminosity, 
below which a stable balance can be maintained between heating and 
radiative cooling  in a confined, magnetized pair plasma; but above 
which the confined plasma runs away to a trapped fireball in local 
thermodynamic equilibrium.  The emergence of large-amplitude pulsations
at $\sim 40$ s probably represents a transition to a pair-depleted
photosphere whose main source of opacity is electrons (and ions)
ablated from the heated neutron star surface.  The best fit temperature
of the black body component of the spectrum equilibrates at a value
which agrees well with the regulating effect of photon splitting.
The remarkable four-peaked 
substructure within each 5.16-s pulse, and the corresponding collimation
of the X-ray flux, has a simple explanation based on the strong inequality
between the scattering cross sections of the two photon polarization modes.
The width of each X-ray `jet' is directly related to the amount 
of matter advected outward by the high-cross section ordinary mode.
\end{abstract}

Subject Headings:
{gamma rays: bursts -- stars: neutron -- X-rays: stars}

\section{Introduction}

On 1998 August 27, a giant flare of hard X-rays, lasting more than 5
minutes, was detected by Konus-Wind, \it Ulysses \rm  and BeppoSAX
(Hurley et al. 1999a; Feroci et al. 1999; Mazets et al.~1999). 
It was only the second giant outburst detected from the Soft Gamma
Repeaters (SGRs), following nearly 20 years after the famous
event on 1979 March 5 (Cline et al.~1982).  The four or five known SGRs are
identified by their frequent emission of much less energetic and
shorter-duration bursts ($E\lessim 10^{41}$ ergs, $\Delta t\lessim 1$ s), 
whose peak luminosities reach nevertheless ten thousand times the 
Eddington luminosity.  Statistically, the
short bursts are similar to earthquakes and solar flares
(Cheng et al.~1996; Gogus et al.~1999; Gogus et al.~2000).  

SGR 1900+14, the source of the August 27 outburst,
was first detected in 1979 (Mazets et al.~1979a), and
became active again in 1992 (Kouveliotou et al.~1993)
and in 1998 May  (Kouveliotou et al.~1998a, Hurley et al.~1999b).
ASCA and RXTE observations after the 1998 May activity episode
revealed a periodicity of 5.16~s in the quiescent
X--ray emission  with period derivative
$\sim 1 \times 10^{-10} \, \rm s \, s^{-1}$
(Hurley et al.~1999c, Kouveliotou et al.~1999; Woods et al.~1999).
A rapid spindown $\dot P \sim 10^{-10}$ s s$^{-1}$ and 7.47 spin
period was also measured for SGR 1806-20 (Kouveliotou et al.~1998b)
These apparent rotation periods are similar to that of SGR 0526-66,
which displayed an 8~s periodicity during the giant flare of 
1979 March 5 (Mazets et al.~1979b, Barat et al.~1979).
At least two of the SGRs are associated with supernova remnants
(SNRs) and/or radio plerions (Hurley 2000).
However, the recently-verified position of SGR 1900+14
(Vasisht et al.~1994; Hurley et al.~1999d; Murakami et al.~1999; Frail,
Kulkarni \& Bloom 1999) lies just outside the edge of G43.8+0.6, a
$\loe 10^{4}$-year-old galactic SNR. \  
It is plausible that SGR 1900+14 was born in this SNR, since the 
{\it other} known giant flare source, SGR 0526-66,  has a position 
inside but near the the edge of a $\lessim 10^4$--year--old SNR 
in the Large Magellanic Cloud (Cline et al.~1982).  

The SGRs have been identified with {\it magnetars} (Duncan \& 
Thompson 1992, ``DT92"): neutron stars
in which magnetism rather than rotation or accretion is the dominant 
source of energy for radiative and particle emissions.  The hypothesized  
$\sim 10^{14}-10^{15}$ G magnetic fields of these stars 
are consistent with many observed SGR properties, including the 
very high energies and hyper-Eddington luminosities of the 
outbursts, the long spin periods, rapid spindown rates, strong
X-ray emissions in the quiescent state, and the young
ages of active stars (DT92; Paczy\'nski 1992; 
Thompson \& Duncan 1995, ``TD95";
Thompson \& Duncan 1996, ``TD96"; Thompson et al.~2000). 

A recent joint paper (Feroci et al. 2000, hereafter Paper I)
analyzed the Ulysses and BeppoSAX observations of the August 27 giant flare,
and drew some conclusions about the underlying physics of the event. 
The intense, hard spike of  gamma-rays during the
first 0.4 second was identified with a freely expanding pair fireball,
which requires a very clean energy source. 
We found that the subsequent rotation-cycle averaged flux 
diminished monotonically, or nearly so.  This strongly suggests
that most of the burst energy was released at early times, near 
the beginning of the flare, and that the energy which did not escape promptly 
during the initial hard spike ($t\lessim 0.4$ s) was 
retained by the star as a residue, to leak away on longer timescales.

In principle, such an energy residue could take several forms.  Some 
energy could be in the form of an optically-thick pair-photon
plasma contained in the magnetosphere by closed magnetic field lines (TD95), 
which we will call a ``trapped fireball."  Additional energy could 
be stored within the star itself in the form of a internal oscillations,
including a torsional Alfv\'en mode in the liquid core;
or a standing shear wave in the neutron star crust (McDermott, 
Van Horn, \& Hansen 1988; Duncan 1998).  Thus,
observations of the flare and its aftermath give clues about the way in
which the energy residue is partitioned among various components.

In this paper, we begin by discussing the general issue of SGR energy budgets 
in \S \ref{energy}, estimating the minimum magnetic field needed to
power both the bursting and quiescent emission of the SGRs.  
Then, in \S \ref{trigger} we consider the physical
states of magnetars on the threshold of a giant flare,
with a particular focus on how and where the energy that powers the
flare is stored, and how its catastrophic release may be triggered.
Our favored model involves a wound-up magnetic field that is pinned
by the neutron star crust.

With this background information developed, we then begin to consider
successive stages of the 1998 August 27 flare. 
In Paper I we showed that the envelope of the August 27 light
curve, smoothed over the 5.16-s rotation period during the interval
from 40 to 400 s after the onset of the flare,
is well-fit by a mathematical curve describing the cooling of a 
zone of very hot plasma composed of $\gamma$-rays and electron-positron
pairs.  Such a ``trapped fireball'' cools through the formation of a 
sharp temperature gradient inside its outer boundary, which
propagates inward in a ``cooling wave'' (TD95).  In particular, 
this model matches the rapid, late decline in 
the X-ray flux at $\sim 400$ s after the flare 
onset.    The observed terminal drop in the X-ray flux
is does not appear to be consistent with models in which the emissions 
are powered by cooling surface hotspots or by incremental internal 
mode damping.  In \S \ref{softail} of this paper, we relate the measured
slope of the lightcurve to the internal properties of the fireball
and study the effects of neutrino cooling on the trapped fireball.

The extreme properties of the giant flare lead to stringent bounds
on its magnetic field (DT92; Paczy\'nski 1992; TD95).
In \S \ref{momentbound} we use the condition that the trapped fireball
energy must remain confined by the magnetic field to derive lower bounds 
on the star's magnetic dipole moment.  We generalize this argument
by considering deviations from centered-dipole geometry, which imply
a lower magnetic moment for a fixed fireball energy.  We show that
the lower bound on the magnetic moment cannot be significantly weakened
without raising the temperature of the confined plasma sufficiently
that almost all its energy is lost to neutrino pair emission, rather
than X-rays.

Then, in \S \ref{peak}, we show how the extremely
high peak luminosity of the hard spike (more than a million times the
Eddington luminosity of a neutron star) points to the presence of
magnetic fields stronger than $\sim 10^{14}$ G.  Because the magnetic
field lines are tied to material of a very high density, a sudden release
of elastic stresses in the deep crust occurs only gradually compared
with the time $\sim R_{\rm NS}/c \sim 3\times 10^{-4}$ s for an
Alfv\'en-like excitation to cross the magnetosphere.  As a result,
the high measured luminosity is consistent with external shearing 
and reconnection of the external field, but only if the flux density
exceeds $\sim 10^{14}$ G.  For essentially the same reason,
an internal shear wave cannot transmit energy to the magnetosphere
at $\sim 10^6\,L_{\rm edd}$ in weaker magnetic fields.  We constrast
our results with the previous suggestion that an internal 
$p$-mode or $f$-mode excitation of a neutron star could induce 
strong shock heating of the surface layers
and electromagnetic damping in the magnetosphere (Ramaty et al. 1980;
Lindblom \& Detweiler 1983), even if the surface magnetic field were 
only $\sim 10^{12}$ G.  We explain how two basic assumptions of this 
model -- that the surface layers would be shock heated, and that 
external Alfv\'en modes could be excited with $\delta B/B \sim 1$ 
by an internal oscillation -- are very likely incorrect.   Indeed,
Blaes et al. (1989) have shown that internal shear modes have a much 
stronger coupling to the magnetosphere than do $p$- and $f$-modes.
 
Some aspects of giant flare emission spectra and spectral 
evolution are elucidated in \S \ref{spectra}.  In particular, we consider 
the hard spectrum of the intial spike in \S \ref{spikespectrum}, and argue
that the hardest $\gamma$-ray photons must be generated at more than 
$\sim 10^{10}$ cm from the star, due to the same limitations of pair
opacity that arise in models of cosmological gamma-ray bursts.
The equilibrium temperature of the pulsating soft tail fitted in
Paper I is shown in \S \ref{split} 
to agree with the calculation in TD95 of the freeze-out
of photon splitting in a strongly-magnetized scattering atmosphere.

The first $\sim 40$ seconds of the August 27 flare showed only mild
and irregular pulsations at the 5.16 spin period, with a higher flux
and harder spectrum than the remainder of the burst.  This excess
emission over the predictions of the trapped fireball curve amounts
to $20\%$ of the burst fluence (Paper I).  In \S \ref{smoothdecay}, we 
interpret this 40-s ``smooth tail'' as the signature of
a transient and extended pair corona, powered by residual creep
of the neutron star crust following the principal disruption.  
In particular, we consider the thermodynamic stability of a strongly
magnetic pair atmosphere, in which X-ray photons are subject 
simultaneously to Compton scattering and splitting.  We show that
this corona is thermally unstable if the pairs are heated at a rate
exceeding $\sim 10^{42}$ ergs s$^{-1}$ within a volume $(10~{\rm km})^3$.

Around $\sim 40$ s following the flare trigger, large-amplitude modulations 
appeared in the light curve with a remarkable four-peaked structure.
These peaks were maintained with a 5.16-s period (equal to the
spin period measured in quiescence) during most of the rest of the event.
In \S \ref{Xjets}, we examine in more detail the proposal of TD95 that
the cooling X-ray flux from a trapped fireball in a super-QED magnetic
field will be collimated along (partly) open magnetic field lines.
In this picture, the complicated pulse shape provides a template of
of higher multipoles in the neutron star's surface magnetic field.  

Finally, in \S \ref{conclusions} we summarize all of the evidence, 
based on the study of giant flares and their aftermaths, for magnetic fields  
$\ga 10\,B_{\rm QED} = 4.4\times 10^{14}$ G in SGRs, and
the main new implications of this work for the physics of magnetars.  

In what follows, we use the convenient reference field
\begin{equation}\label{bqed}
B_{\rm QED} = {m_e^2c^3\over e\hbar} = 4.4\times 10^{13}\;\;\;\;{\rm G},
\end{equation}
at which the non-relativistic Landau energy $\hbar eB/m_ec$ formally
becomes equal to the electron rest energy $m_ec^2$.

\section{Strong Magnetic Fields in the SGR Sources}\label{energy}

The electromagnetic output of an SGR source consists of both
hard X-ray outbursts and the persistent X-ray emission.
The core of the star may release an even larger energy in the
form of thermal neutrinos, over its active liftime (TD96).
We now estimate the minimal 
magnetic field needed to power the observed activity.  The close
affinity between the SGR and AXP sources (some of which have variable
persistent X-ray output even in the absence of bursts) indicates that
an SGR can dissipate magnetic energy in two distinct modes:  first,  
in an episodic manner through brittle fracturing of the crust;
and, second, through a more plastic deformation of the star (TD96).

\subsection{Energetic Requirements}

The association of several magnetar candidates (three AXPs and at least
two SGRs) with young supernova remnants, $t_{SNR} \la 10^4$ yr 
(Hurley 2000), combined with the short spindown times of these stars,
suggests an active lifetime of order of magnitude $\sim 10^4$ yr.
Two giant flares have been detected from four verified, active
SGR sources within the last $30$ years.  The implied flaring
rate is $\sim 10^{-2}$ per year per source, suggesting a total number of giant
flares $N_{\rm giant} \sim 10^2$ from each
source during its active lifetime $\sim 10^4$ yr.  
Since the measured X-ray and gamma ray output of the 1979 March 5
and 1998 August 27 flares exceed $\sim 5\times 10^{44}$ ergs and
$\sim 1\times 10^{44}$ ergs respectively, the total flare energy 
released per source is
\begin{equation}\label{emag}
E_{\rm flare} \ga 3\times 10^{46}\,\left({\dot N_{\rm giant}\over 
10^{-2}\;{\rm yr}^{-1}}\right)\,\left({t_{\rm active} \over 10^4\;
{\rm yr}}\right)\;\;\;\;\;\;{\rm ergs}
\end{equation}

The persistent X-ray luminosity
of SGR 0526-66 is $\sim 10^{36}$ ergs s$^{-1}$
(Kulkarni et al. 2000) and is approximately $10^{35}$ ergs s$^{-1}$ for 
SGR 1806-20 and SGR 1900+14 during periods of quiescence
(Murakami et al. 1994; Woods et al. 2000).   The net energy
released in persistent X-ray emission can therefore be estimated as
\begin{equation}\label{eXray}
E_X  \sim L_X\,t_{\rm active} = 
3\times 10^{46}\,\left({L_X \over 10^{35}~{\rm ergs~s^{-1}}}\right)\,
\left({t_{\rm active} \over 10^4\;{\rm yr}}\right)\;\;\;\;\;\;{\rm ergs}
\end{equation}
over the active lifetime.  Thus, the observed electromagnetic 
output of an SGR is fairly evenly divided between flares and its highly
non-thermal persistent emission.  The observation of a transient
enhancement in the persistent output of SGR 1900+14 following the
August 27 flare (Woods et al. 2000) establishes a close physical
connection between the two phenomena.  This enhanced emission has
been ascribed to persistent currents excited by non-potential deformations
of the external magnetic field, caused by the shifting positions of 
magnetic footpoints which are anchored to an evolving crust 
(Thompson et al. 2000).

It is possible that significant power is emitted  
at wavelengths intermediate between X-rays and optical in sources
such as SGR 0526-66 with soft power-law spectral components
(Kulkarni et al. 2000).  Such intermediate wavelengths are not 
observable in known galactic magnetar candidates because of the 
opacity of the intervening interstellar medium.  
In the case of the low-extinction LMC source SGR 0526-66, 
spectral measurements at $\lessim 0.5$ keV are not yet available. 
For sources with hard power-law spectra (photon index $\sim -2$ or 
less steep) extending to low energies, the correction factor to 
eqn.~(\ref{eXray}) would be of order unity only.

It is instructive to compare $E_{\rm flare}$ with the
energy detected in ordinary, short-duration SGR
bursts.  These events are distributed over a wide range of energies
$\sim 10^{37}-10^{41}$ ergs, with a power-law $d{\cal N}/dE \sim E^{-5/3}$,
so that the integrated energy is dominated by large events
(Cheng et al. 1996; Gogus et al. 2000).  Nonetheless, many fewer than
$\sim 10^3$ events with energies $\sim 10^{41}$ ergs have been detected
during the same time that two giant flares have been observed.  This
suggests that the time-averaged output in short bursts is lower than
either in the giant flares or persistent emission.

\subsection{Magnetic Field Strength}\label{magmin}

For the reasons just described,
the persistent and bursting output of an SGR source appear to
draw from the same energy reservoir, which at minimum is
$E_{\rm min} \sim 10^{47}$ ergs per source.  The present rotational energy
is much smaller, ${1\over 2}I\Omega^2 = 5\times 10^{44}(P/6~{\rm s})^{-2}$
ergs.  It has been suggested that the SGR/AXP sources were born spinning much
more rapidly, $P \la 3$ msec, and that a fraction of this initial
rotational energy was converted to a strong magnetic field through
an $\alpha-\Omega$ dynamo in the convective proto-neutron star (DT92).
This magnetic field would then provide a
reservoir of energy for later bursting activity.  An alternative form
of potential energy for bursting activity involves an elastic deformation 
of the crust, resulting from the changing figure of the rotating star 
as its rate of spin decreases (e.g Baym et al. 1969).  However,
the maximum elastic energy which can be stored by the
crustal lattice is only ${1\over 2}M_{crust}\,V_\mu^2\,\psi_{\rm cr}^2 =
1\times 10^{42}\,(V_\mu/10^8~{\rm cm~s^{-1}})^2\,(M_{crust}/10^{-2}\,M_\odot)
\,(\psi_{\rm cr}/0.003)^2$ ergs (eq. [\ref{Eelastic}]), too small to
power a single giant flare except under the most optimistic assumptions
about the breaking strain $\psi_{\rm cr}$.   (Here $V_\mu$ is the shear wave
velocity, which is nearly independent of density in the deep crust.)
Even in sources which have not emitted giant flares, the elastic energy 
would need to be replenished on a timescale short compared with the 
spindown age; but the present rotational energy is insufficient 
for that purpose.

Accretion power is not a promising energy source for the hard-spectrum, 
hyper-Eddington SGR outbursts, nor for persistent X-ray
emission from SGRs.  The observation of persistent emission from SGR 1900+14 
within 1000 s of the 27 August 1997 giant flare (Woods et al. 2000)
seems inconsistent with accretion-powered scenarios, because
even $\sim 10^{-3}$ of the radiative momentum of the flare would
excavate the accretion flow and suppress accretion for a much longer
time period (Thompson et al.~2000).  Accretion models also have
difficulty accomodating the power-law distribution of energies of 
ordinary SGR bursts, and the lognormal distribution of waiting times
between bursts (Cheng et al. 1996; Gogus et al. 1999, 2000). Recent 
HST bounds on the optical emission from SGR 0525-66 set an upper limit
of $L_{\rm opt}/L_X \sim 10^{-3}$ for reprocessed optical emission
from an accretion disk (Kaplan et al. 2001), an order of magnitude less
than is expected from an accretion disk (Perna, Hernquist, \& Narayan
2000).  Optical and infrared observations also exclude the simplest
disk models of the Anomalous X-ray Pulsars [AXPs] (Hulleman et al.~2000a;
Hulleman, van Kerkwijk \& Kulkarni 2000), which resemble SGRs in 
inactive phases.  In particular, 4U 0142+61 has a possible optical
counterpart with a much redder spectrum than is typical of LMXBs
(Hulleman et al. 2000b).  Some other arguments against accretion-powered 
models of SGRs are given in \S 7.3 of TD95.

Magnetic energy, on the other hand, can be a good ``clean" power
source for both hyper-Eddington bursts and the hard fireball component
of the giant flares (DT92; Paczy\'nski 1992; TD95).  A first estimate
of the minimum magnetic field is obtained by averaging over 
the volume of the star,
\begin{equation}\label{bmin}
B_{\rm min} = \left({6 E_{\rm min}\over \epsilon_B
R_{\rm NS}^3}\right)^{1/2} 
= 8\times 10^{14} \ \epsilon_B^{-1/2}\,
\left({E_{\rm min}\over 10^{47}~{\rm ergs}}\right)^{1/2}
\,\left({R_{\rm NS}\over 10~{\rm km}}\right)^{-3/2}\;\;\;\;{\rm G}.
\end{equation}
Here $\epsilon_B$ denotes the efficiency of conversion of magnetic
energy to X-rays.  

If the protons in the stellar interior form a Type II superconductor, 
then the magnetic field is confined to fluxoids with flux density 
comparable to the
lower critical field $B_{c1} \sim 2\times 10^{15}$ G
(e.g. Easson \& Pethick 1979).  The mean field in the core is 
related to $B_{c1}$ by the volume-filling fraction $f$ of the fluxoids,
$\langle B\rangle = B_{c1}f$, and the magnetic energy per unit volume
is $B_{c1}^2 f/8\pi = \langle B\rangle B_{c1}/8\pi$.  
When the mean field $\langle B\rangle < B_{c1}$ (that is, when the efficiency
of conversion of core magnetic energy to X-rays is less than $\sim 10$
percent) expression (\ref{bmin}) decreases by a factor 
$\sim \epsilon_B^{-1/2}(B_{\rm min}/B_{c1})$.  Otherwise, the minimum
field is not modified to first order by proton superconductivity.

If $\epsilon_B \sim 1$, expression (\ref{bmin})
is similar to the dipolar magnetic fields that are
inferred from the rapid spindown of the SGRs (Kouveliotou et al.
1998b, 1999) and AXPs (TD96).  However, 
the total magnetic energy of the star is almost certainly larger than
its minimum value, because the currents that support the field 
flow through very highly conducting material and cannot entirely
dissipate.  Very strong core magnetic fields also seem to be required 
by models of core ambipolar diffusion which include the
accelerating effects of core heating by the decaying field 
(TD96; Heyl \& Kulkarni 1998).  Only fields stronger than
$\sim 10^2 \, B_{\rm QED} \sim 5\times 10^{15}$ G are transported over
a distance $\sim 10$ km in times less than $\sim 10^4$ yr, typical of SGR
active lifetimes.  

The magnetic flux density inside an SGR/AXP source probably
lies below the upper critical field where proton superconductivity is 
fully quenched, $B_{c2} \sim 10^{17}$ G.  Nonetheless, the extended magnetic
sheaths of the fluxoids are very densely packed in fields $\ga B_{c1}$.
As a result, field transport driven by collective effects such as 
vortex-line sweeping (Ruderman 1991) is probably suppressed in 
slowly-rotating magnetars.  

Several lines of evidence suggest that SGR bursts and flares are 
powered by {\it internal} magnetic stresses acting on the deep crust and
core.  The hard initial $\gamma$-ray spikes observed
in the two giant flares had durations $\sim 0.2-0.5$ s. 
This is comparable to the time for a $\gtrsim 10^{15}$ G magnetic field to 
rearrange material in the dense stellar core (TD95), but it is much 
longer than the timescales for global rearrangements of either
the magnetosphere ($\sim R_{\rm NS}/c \sim 0.03$ ms) or the stellar crust   
($R_{\rm NS}/V_\mu \sim 10$ ms; see also~eqs.~[\ref{Vmu}] and [\ref{pshear}] 
below).  Furthermore, the cumulative fluence of short SGR bursts is observed
to rise linearly with time during periods of activity (Palmer 1999). This 
``relaxation system" behavior indicates that a reservoir of energy is 
steadily loaded and undergoes stochastic, catastrophic relaxations.  
For example, such a reservoir could encompass a fraction of the star's 
surface, steadily dragged from below by the tension of field lines 
experiencing core ambipolar diffusion (TD96).  This model is roughly
analagous to what happens in earthquakes, where relaxation-system
behavior driven by crustal plate motion is also found (see Palmer 1999; Gogus
et al.~1999, 2000, and references therein). 

The magnetic field geometry of an SGR/AXP source almost certainly involves 
higher multipoles (for which the August 27 lightcurve provides clear
evidence:  Paper I; \S \ref{Xjets} below); and quite plausibly a strong 
toroidal field component in the deep crust and core, as would result
from post-collapse $\alpha$--$\Omega$ dynamo action (DT92; Thompson
\& Duncan 1993, hereafter TD93).
This suggests that the magnetic energy of a magnetar is dominated by the 
internal field, with a probable value 
\begin{equation}\label{Emagnetar}
E_{\rm magnetic} = 3 \times 10^{48}  \ f_B
\left({B_{\rm internal}\over 100 \, B_{\rm QED}}\right)^2 \;\;\;{\rm ergs},
\end{equation}
where $f_B$ is the fraction of the star's volume filled by the
strong field. 

Comparisons of eqs.~(\ref{emag}), (\ref{eXray}) and (\ref{Emagnetar})
indicate that  {\it only a fraction $\epsilon_B \sim 0.03/f_B$
of the star's magnetic energy must be dissipated in the form of X-rays
and gamma-rays to power all observed 
SGR activity}.  Indeed, most of the core field energy is probably lost 
to thermal neutrinos or ultimately remains trapped in the core 
when the epoch of SGR activity ends.

\section{Giant Flare Mechanism}\label{trigger}

The crust of a magnetar is subjected to strong, evolving magnetic
stresses.  The star must initially relax to an equilibrium configuration
before the crust forms (but after it becomes compositionally stratified:
Lattimer \& Mazurek 1981; Reisenegger \& Goldreich 1992).  
Observations of magnetic white dwarfs prove that static, magnetized 
equilibrium states exist even in the absence of any rigidity due to
solidification. After the neutron star crust forms,
the crust, core and field evolve through a sequence of equilibrium 
states in which magnetic stresses 
are balanced by both hydrostatic forces and elastic stresses 
in the crust.  The star evolves via the very slow 
transport of the field through the core (by ambipolar diffusion) and
the crust (by Hall drift) (Goldreich \& Reisenegger 1992; TD96).
Because the crust has a finite shear strength, these equilibria
are punctuated by starquakes whenever the crust is strained past
its breaking point. 

We now review the forms in which potential energy can be stored inside
a magnetar, and how some of this stored energy can be transferred 
to the external magnetic field, to heat, and to internal torsional and 
standing shear waves.  This suggests a candidate mechanism for triggering
giant flares like the March 5 and August 27 events.  

\subsection{Stored Elastic and Magnetic Energy}\label{stored}

A Soft Gamma Repeater stores a certain amount of potential energy 
which can drive rapid rearrangements of the external magnetic field,
thereby triggering bright X-ray outbursts.
This potential energy can be divided into
an elastic component in the crust, and magnetic components
in the crust, core and magnetosphere.  The portion of the
magnetic energy which is available to do mechanical work can
easily be estimated, in a situation where the accumulation and
release of stress is caused by pinning of the internal
magnetic field by the crust.

The elastic energy corresponding to a static shear strain $\psi$ 
can be well approximated by\footnote{This generalizes the expression
given in Duncan (1998), to allow for a variable position (density) at the
base of the rigid Coulomb lattice.}
\begin{equation}\label{Eelastica}
E_{\rm elas} = {1\over 2}
\int d\Omega\int_{R_{\rm NS}-\Delta R_0}^{R_{\rm NS}} \mu \psi^2\, r^2 dr =
{1\over 2}\left({\gamma\over \gamma-0.2}\right)\,\langle \psi^2\rangle\,
M_{\rm Coulomb}\,V_{\mu\,0}^2.
\end{equation}
In this expression, the shear modulus is\footnote{The neutron star crust is 
here approximated as a body-centered cubic (bcc) Coulomb lattice
(Ogata \& Ichimaru 1990; Strohmayer et al.~1991), with ionic mass and
charge determined by a power-law fit to the equation 
of state of Negele \& Vautherin (1973).}
\begin{equation}\label{shearmod2}
\mu =  1.1 \times 10^{30} \ \rho_{14}^{0.8} \ \hbox{ergs cm}^{-3},
\end{equation}
and the shear wave speed, 
\begin{equation}\label{Vmu}
V_\mu = (\mu/\rho)^{1/2} = 1.0 \times 10^{3}\, \rho_{14}^{-0.1} \hbox{km
s}^{-1},
\end{equation}
depends weakly on the density $\rho = \rho_{14}\times 10^{14}$ g cm$^{-3}$
at densities above neutron
drip (Baym \& Pines 1971; Strohmayer et al. 1991, Fig. 3).  The integral
(\ref{Eelastica}) extends from the base of the Coulomb lattice at
a depth $\Delta R_0$ (density $\rho_0$ and shear wave speed $V_{\mu\,0}$).
It has been evaluated using the equation of hydrostatic
equilibrium $dP/dr = -\rho g$, 
and fitting a power-law relation between the pressure
and density profiles, $P(r) \propto [\rho(r)]^\gamma$.
In particular, the equation of state of Negele \& Vautherin (1973) gives
$\gamma \simeq 1.47$ and 
\begin{equation}\label{eqofst1}
P(\rho) = 3.75 \times 10^{32} \ \rho_{14}^{1.47} \ \hbox{ergs cm}^{-3}
\end{equation} 
above a density $3.6 \times 10^{12}$ gm cm$^{-3}$.  
The mass within the Coulomb lattice
\begin{equation}\label{mcoul}
M_{\rm Coulomb} =
\int_{R_{\rm NS}-\Delta R_0}^{R_{\rm NS}} \rho(r) 4\pi r^2 dr
= {4\pi R_{\rm NS}^2 P(\rho_0)\over g(R_{\rm NS})}
\end{equation}
excludes a transition layer at its base where the nuclei have strongly
aspherical shapes, forming rod-like and plane-like structures, with 
bulk elastic properties resembling those of a liquid crystal 
(Pethick \& Potekhin 1998).  
Normalizing the base density $\rho_0$ to a fraction 0.6 of the nuclear 
saturation density
$\rho_{\rm sat} = 2.7\times 10^{14}$ g cm$^{-3}$, expressions
(\ref{Eelastica}) and (\ref{mcoul}) become
\begin{equation}\label{Eelastic}
E_{\rm elas} = 1.7\times 10^{43}\,\Lambda^{-1}\,
\left({\langle\psi^2\rangle\over
(10^{-2})^2}\right)\,\left({\rho_0\over 0.6~\rho_{\rm sat}}\right)^{1.27}\,
\left({R_{\rm NS}\over 10~{\rm km}}\right)^4\,
\left({M_{\rm NS}\over 1.4\,M_\odot}\right)^{-1}\;\;\;\;\;\hbox{ergs},
\end{equation}
and 
\begin{equation}
M_{\rm Coulomb} = 0.025\,\Lambda^{-1}
\left({\rho_0\over 0.6~\rho_{\rm sat}}\right)^{1.47}\,
\left({R_{\rm NS}\over 10~{\rm km}}\right)^4\,
\left({M_{\rm NS}\over 1.4\,M_\odot}\right)^{-1}\;\;\;\;\;M_\odot.
\end{equation}
The scaling with stellar radius $R_{\rm NS}$ and mass 
$M_{\rm NS}$ assumes a fixed equation of state, and arises
because the hydrostatic pressure 
$P(\rho_0) \sim [g(R_{\rm NS})M_{\rm Coulomb}/4\pi R_{\rm NS}^2]$
at the base of the crust is fixed by the nuclear interactions
(Lorenz, Ravenhall, \& Pethick 1993).  The factor $\Lambda^{-1} = 
1-2GM_{\rm NS}/R_{\rm NS}c^2 \simeq 0.6-0.7$ is a relativistic
correction to the surface gravity, $g(R_{\rm NS}) = GM_{\rm NS}\Lambda
/R_{\rm NS}^2$.  

Comparing expression (\ref{Eelastic}) 
with the energy of the 1979 March 5th event, 
$E_{\rm March 5} \approx 5 \times 10^{44}$ ergs, it is clear that
even for the optimistic case $\psi_{\rm cr} \sim 10^{-2}$, giant 
flares cannot be 
powered by pure elastic energy stored within the crust of a neutron star.
Furthermore, it is often assumed that $\psi_{\rm cr}$
is one or more orders of magnitude less than the value 
$\sim 10^{-2}$ appropriate for a perfect bcc lattice,
since lattice imperfections inevitably weaken the solid 
(e.g. Ruderman 1991).  But the critical strain $\psi_{\rm cr}$ 
for a large-scale yield in the gravitationally-stratified 
crust of a neutron star is actually highly uncertain.  Any motion of the crust
over a scale of kilometers is severely constrained, because the hydrostatic
pressure in the deep crust exceeds the shear modulus by a factor $\sim 10^3$.  
Degeneracy pressure and buoyancy forces provide strong resistance to
bulk compressions and to vertical displacements within the crust.
Observations of giant flares probe the behavior of Coulomb solids in 
a regime of high pressure and large applied stress that has no direct 
experimental analogs.

It should be emphasized at this point the magnetic field which stresses
the crust can store much more potential energy than the Coulomb lattice
itself.  Within the crust, the force balance 
\begin{equation}\label{eeq}
{B\delta B\over 4\pi} \sim \psi \mu
\end{equation}
implies that 
\begin{equation}\label{ampfact}
{(\delta B)^2/8\pi\over {1\over 2}\psi^2\mu} 
= \left({B_\mu \over B_{\rm crust}}\right)^{2}.
\end{equation}
when $B < B_\mu$.  Here
\begin{equation}\label{Bmu}
B_\mu \equiv (4 \pi \mu)^{1/2} = 4 \times 10^{15} \rho_{14}^{0.4}
\;\;\;\;\;\;\;{\rm G}
\end{equation}
is a characteristic field above which the crust tends to respond plastically
to applied magnetic stresses.  For $ B > B_\mu$, the equilibrium condition 
(\ref{eeq}) requires that the strain angle of the Coulomb solid exceed
$\delta B/B$,
even while the field remains tied to the highly conducting solid (TD95).  
In weaker fields $B < B_\mu$, the field stores more potential energy
than the lattice.  This field energy might be tapped in short SGR bursts,
which would require that the crustal material undergoes abrupt plastic 
deformation.     However, a very weak field 
$B \ll B_\mu$ does not have the strength to move the crust, even as 
the ratio in eq.~(\ref{ampfact}) formally becomes enormous.   For example, when
\begin{equation}\label{Bfrac}
B < B_{\rm frac} = \psi_{\rm cr}^{1/2} \, B_\mu = 1 \times 10^{14} \, 
\left({\psi_{cr}\over 10^{-3}}\right)^{1/2} 
\ \rho_{14}^{0.4} \ \hbox{G}
\end{equation}
magnetic stresses within the crust are too weak to induce any yields or 
fractures.

\subsubsection{A Globally Twisted Magnetic Field}

The giant flares involve a large disturbance which probably is driven
by a rearrangement of  magnetic field in the deep crust and core (TD95).  
It is important 
therefore to consider the case in which the crust is stressed 
{\it from below} by the evolving magnetic field in the liquid interior
of the neutron star.  A related configuration has been considered by
Ruderman (1991), in a situation where mean core field is much weaker
than $10^{15}$ G, and its transport is driven by spindown.  As we
now show, a factor similar to (\ref{ampfact}) relates the available
magnetic energy {\it stored in the core} to the elastic energy built
up in the crust.  The flux density $B$ which enters this relation
turns out to be the poloidal field which threads the core-crust boundary.  

At least two lines of argument suggest that the interior magnetic
field of the SGR and AXP sources is strongly wound, with a large
toroidal component:  first, their association with neutron stars that
were formed with rapid rotation and strong differential rotation
(DT92); and, second, their non-thermal and transient persistent X-ray
emission (Thompson, Lyutikov, \& Kulkarni 2001).  We therefore focus on 
a cylindrically symmetric star with a uniform poloidal field $B_z < B_\mu$
in its interior (Fig. \ref{twist}).
In the core, the evolving field is assumed to be twisted about the
magnetic dipole axis (the axis of symmetry), generating a field component
$B_\phi$.  In static equilibrium, before the critical point is reached, 
the stress $B_zB_\phi/4\pi$ applied to the lower base of the
crust is balanced by an elastic stress $\sigma_{\varpi\phi} =
\mu \varpi(d\phi/d\varpi) = \mu\psi$ within the crust, so that
\begin{equation}\label{sbal}
{B_zB_\phi\over 4\pi} = {\Delta R_\mu\over \varpi}{d\over d\varpi}
\Bigl(\varpi \sigma_{\varpi\phi}\Bigr).
\end{equation}
Here, $\Delta R_\mu$ is the stress-averaged depth of the crust,
\begin{equation}\label{rmu}
\Delta R_\mu = {1\over \mu(\rho_0)}\int_0^{\rho_0}
\left({d P\over d\rho}\right) {\mu(\rho)\over \rho g} d\rho
= 0.27\,\Lambda^{-1}\left({\rho_0\over 0.6~\rho_{\rm sat}}\right)^{0.47}\,
\left({R_{\rm NS}\over 10~{\rm km}}\right)^2\,
\left({M\over 1.4\,M_\odot}\right)^{-1}\;\;\;\;\hbox{km}
\end{equation}
(Duncan 1998).  Expression (\ref{sbal})
remains a good approximation as long as the
spherical curvature of the neutron star crust can be neglected; that is,
at cylindrical radius $\varpi \la {1\over 2}R_{\rm NS}$.  Integrating
out from the axis of symmetry ($\varpi = 0$), and assuming the power-law
dependence 
\begin{equation}\label{psiscal}
\varpi B_\phi \propto \psi \propto \varpi^\alpha,
\end{equation}
equation (\ref{sbal}) implies
$B_zB_\phi/4\pi = (1+\alpha)(\Delta R_\mu/\varpi)\mu\psi$.  
The parameter $\alpha = 2$ corresponds to a twist angle
that is independent of $\varpi$;  whereas $\alpha = 0$ corresponds to
a current that is localized on the symmetry axis. 

Notice that the equilibrium strain $\psi(\varpi)$
increases away from the axis of symmetry for $\alpha > 0$ 
(eq. \ref{psiscal}).  This means that the crust can be expected
to break first at a radius $\varpi_{\rm frac}$ where
\begin{equation}
\psi(\varpi_{\rm frac}) \sim \psi_{\rm cr},
\end{equation}
the critical strain angle.  If the strain increases monotonically
with time, this means that fracture first occurs at $\varpi_{\rm frac}
\sim ({1\over 2}-1)R_{\rm NS}$.

The energy stored in the twisted magnetic field is
\begin{equation}
\delta E_{\rm mag} =
(2R_{\rm NS}\cdot \pi \varpi^2)\,{B_\phi^2\over 8\pi}\,
 = \pi(\alpha+1)^2\,\psi^2(\varpi)\mu\,
\left({B_\mu\over B_z}\right)^2\,(\Delta R_\mu)^2 R_{\rm NS},
\end{equation}
per logarithm of cylindrical radius $\varpi$.  This can be expressed as
\begin{equation}\label{emagb}
\delta E_{\rm mag} = 4\times 10^{44}\,
\left({\psi^2(\varpi)\over (10^{-2})^2}\right)\,
\left({B_z\over 10\,B_{\rm QED}}\right)^{-2}\,
\left({\rho_0\over 0.6\,\rho_{\rm sat}}\right)^{1.6}\,
\left({\Delta R_\mu\over 0.3~{\rm km}}\right)^2\hbox{ergs}
\end{equation}
for $\alpha = 2$.   
It is also worth comparing eq. (\ref{emagb}) with the crustal strain
energy.  The available magnetic energy in the core is the larger
by a factor
\begin{equation}\label{ecomp}
{(B_\phi^2/8\pi)\,R_{\rm NS}\cdot \pi \varpi^2
\over ({1\over 2}\psi^2\mu)\,\Delta R_\mu\cdot \pi \varpi^2} =
\left({B_\mu\over B_z}\right)^2\,
{(1+\alpha)^2R_{\rm NS}\Delta R_\mu\over \varpi^2}.
\end{equation}
This expression exceeds (\ref{ampfact}) by a geometrical factor,
which works out to $\sim 2\,(\varpi/{1\over 3}R_{\rm NS})^{-2}$
in the case of a uniformly twisted field ($\alpha =2$).
Thus, {\it in the giant flare the main role of the crust is to
serve as a gate to facilitate the storage and episodic release of interior 
magnetic energy, not as a reservoir of energy itself}. 

In this toy model, the available magnetic energy (\ref{emagb}) suffices to
power a single giant flare, if $\psi_{\rm cr} \sim 10^{-2}$.  This 
yield strain is larger than usually adopted for neutron star 
crusts (e.g. Ruderman 1991), but it may not be unreasonable given
the large hydrostatic pressure and near incompressibility of the crustal
material.  The required value of $\psi_{\rm cr}$ is reduced if 
the poloidal current flowing through the star varies in sign, 
with multiple interior zones in which 
the toroidal field has a magnitude comparable to its limiting value
(\ref{eeq}).  Then the winding energy (\ref{emagb}) is multiplied by
the number of zones $N$, and $\psi_{\rm cr}$ decreases by a factor
$N^{-1/2}$.  However, the hydromagnetic stability of such a configuration
is not certain.  

Although the energy that can be released in one giant flare is limited
by the shear strength of the crust, this simple model significantly 
underestimates the total energy that can be released over successive flares.
In addition, in this simplified geometry the twist of the core field
must be assumed to increase with time to reach a point of instability.  Both
apparent difficulties can be addressed by noting that the stratification
of the core allows the field to retain a significantly {\it larger} twist than
can be dissipated in a single flare.  

Consider an electrically conducting medium that is stratified along planes,
which run perpendicular to the direction of gravity. 
This stratification is assumed to be stable to convection, in the sense
that internal g-modes have real frequencies.  The gradient of electron
fraction in the core of a neutron star provides just such a stable
stratification (Reisenegger \& Goldreich 1992).   A cylindrical bundle of
magnetic flux threads the medium, tilted at an angle $\alpha$ with respect
to gravity  (Fig. \ref{twistb}).   We assume that the magnetic field
is weak, in the sense that the fluid motions driven by the 
${\bf J}\times{\bf B}$ force lie almost parallel to the equipotential
surfaces.  If $\alpha = 0$, then the flux bundle can unwind entirely
through such constrained motions, as long as its ends are not pinned.
By contrast, if $\alpha = \pi/2$ and the flux bundle runs perpendicular
to gravity, then any motion of the field and
entrained conducting fluid which reduces the twist must
involve the motion of fluid elements {\it across}
the equipotential surfaces.  This means (Fig. \ref{twistc}) that
a localized twist will not be able to spread out along the flux bundle --
even if the ends of the bundle are not pinned by external forces.  

For an arbitrary value of $\alpha$, it is clear that any initial twist will
be able to relax only partially.  In particular, if the field is strongly
twisted, $B_\phi \ga B_z/\tan\alpha$, then the flux bundle will relax
to a magnetostatic equilibrium without $B_\phi/B_z$ being significantly
reduced in magnitude.  To see this, consider a cylindrical bundle
of flux which is unpinned at either end.  Coordinate $z$ runs along
the axis of the flux bundle, and gravity lies in the plane $\phi = 0$.
The twisted bundle is allowed to deform from this initial condition,
subject to the constraint that the component of the velocity parallel
to gravity vanishes,
\begin{equation}
V_z\cos\alpha -V_\phi\sin\alpha\cos\phi = 0.
\end{equation}
(We neglect deviations from cylindrical symmetry, so that
in this coordinate system the radial components $B_R$ and $V_R$ both vanish.
In effect, the flux bundle and the stratified medium are assumed to be
confined with an infinitely rigid cylindrical shell.)  
The induction equation then becomes
\begin{equation}
{\partial B_\phi\over\partial t} =
{\partial\over\partial z}\Bigl(V_\phi B_z-V_z B_\phi\Bigr)
= {\partial\over\partial z}\Bigl[V_\phi\Bigl(B_z - 
B_\phi\tan\alpha\sin\phi\Bigr)\Bigr].
\end{equation}
One sees that if $B_\phi \gg B_z/\tan\alpha$, then a rotation
of surfaces of constant $z$ cannot reduce $B_\phi$ by more than
a factor of two, because the second term in the induction equation
is not symmetric about the axis of the cylinder.  

Slower transport processes such as ambipolar diffusion can change $\alpha$
over timescales much longer than the hydromagnetic time 
$R_{\rm NS}/V_{A\,z}$.  In this manner,
elastic stresses can build up if the flux bundle is tied to 
a rigid medium (e.g. the neutron star crust).

\subsection{Release of Stored Magnetic Energy}\label{exintmodes}

The two giant flares were initiated by hard $\gamma$-ray spikes 
whose $\sim 0.2-0.5$ s width is comparable to the Alfv\'en crossing
time of the star, if the internal poloidal magnetic field is 
$\sim 10^{15}$ G (TD95).  By contrast, a
relaxation of the external magnetic field has a much shorter characteristic
timescale $R_{\rm NS}/c \sim 3\times 10^{-5}$ s.
This gives evidence that the giant flares are driven
by the relaxation of {\it internal} magnetic stresses. 

Any motion of the crust of a neutron star is strongly constrained by
degeneracy and buoyancy forces, which resist bulk compressions and
vertical displacements.  A $10^{15}$ G magnetic field
contributes only $\sim 10^{-4}$ of the hydrostatic
pressure at the base of the crust.  At the same time, it is 
capable of deforming the crust along equipotential surfaces of the star.  
Over large scales (kilometers or more), the lowest energy 
deformations of the crust involve a displacement field ${\bf\xi}$ which
has a vanishing radial component and satisfies
${\bf\nabla}\cdot(\rho{\bf\xi}) = 0$, while maintaining a non-zero curl
${\bf\nabla}\times(\rho{\bf\xi}) \neq 0$.

Such a large-scale twisting motion of the crust, if it occurs suddenly,
probably involves the formation of one or more propagating fractures.
This type of deformation has been associated with the giant flares
(Thompson et al. 2000; TD95).  Since the 
magnetic potential energy of twisting exceeds the crustal elastic energy 
(eq. [\ref{ecomp}]), the crust acts as a  
gate for the release of magnetic energy.  The external magnetic
field lines are anchored to the crust, so when the fracture occurs,
the rapid turning motion 
does work on the external magnetic field.  At the same time, it
brings into contact regions of the crust where the magnetic field has
differing strength and orientation, thereby creating strong field gradients
and localized current sheets.  Within the star, the bulk rotational motion
also excites a helical Alfv\'en wave in the liquid core (where the 
magnetic field provides the dominant restoring force to a torsional motion). 
Torsional shear waves are excited in the rigid crust (McDermott et al.
1988; Duncan 1998), but are likely to carry less energy.
We consider each of these physical elements in turn.

\subsubsection{External Shear and Reconnection}\label{shrec}

Above the surface of the star, a disturbance of the
magnetosphere propagates at very nearly the speed of light, 
which is some 300 times
the shear wave speed $V_\mu$ in the deep crust.  Thus, the external
magnetic field can respond adiabatically to a {\it smooth}
deformation of the crust. However, regions of the magnetosphere
whose footpoints are strongly sheared can become subject to a 
purely magnetohydrodynamic instability (Lynden-Bell \& Boily 1994; 
Mijic \& Linker 1994).  If localized near a fault,
such an instability probably leads to reconnection and induces
magnetohydrodynamic waves outside the star.
Transverse Alfv\'en waves will have a characteristic frequency
$\omega \sim c/x$ at a distance $x$ from the fault.  The corresponding
wavenumbers are $k_\perp \ga k_\parallel \sim x^{-1}$ in directions
parallel and perpendicular to the background magnetic field.  
These waves can damp rapidly by
cascading to  a high wavenumber through non-linear interactions
(Thompson \& Blaes 1998).  When the rate of transfer of wave
energy (the `cascade luminosity') exceeds a critical value
($\sim 10^4\,L_{\rm edd}$ within a volume of $\sim (10~{\rm km})^3$), the
dissipated wave energy is locked onto the magnetic field lines
in a thermal pair-photon plasma (TD95; Section \ref{smoothdecay} below).
If excited directly in the magnetosphere, these waves are not
easily reabsorbed by the crust, since crustal shear waves of the
same frequency have a much larger wavenumber, by the ratio 
$c/V_\mu \sim 300$.  

Thus, even though the giant flares are probably driven by internal
magnetic stresses, the creation of a hot fireball outside the star
involves the rapid relaxation of external magnetic stresses.  This
suggests that the energy released promptly in X-rays and $\gamma$-rays
is roughly proportional to the {\it external} magnetic energy, after
taking into account geometrical factors.  Since a pure dipole field
with polar strength $B_{15}\times 10^{15}$ G carries an energy
$\sim 10^{47}\,B_{15}^2\,(R_{\rm NS}/10~{\rm km})^3$ ergs, the energy
dissipated is small compared to the total external field energy
if $B\sim 10^{15}$ G.

Suppose, for example, that a cap of the neutron
star crust of radius $\varpi$ were to rotate through an angle $\Delta\phi$,
while the surrounding crust remains almost stationary.  Even if this
rotation were to be aligned with the axis of a purely dipolar field,
one would expect strong shear to build up in the external field near
the boundary of the cap.  The amount of energy dissipated externally
can be estimated, by supposing that this shear relaxes in $N$ steps
after a rotation through an angle $\Delta\phi/N$.  The magnetic energy
released in each step is then $\sim \pi\varpi^2(R_{\rm NS}/3)\,
(\delta B_{\rm NS})^2/8\pi$, where $\delta B_{\rm NS}/B_{\rm NS} 
\sim \Delta \phi/N$ and we have assumed a dipole geometry with a 
radial scale length $R_{\rm NS}/3$.  The net energy which is available
for powering X-rays and $\gamma$-rays is $N$ times this,
\begin{equation}\label{burstenergy}
\delta E_{\rm mag} \sim {(\Delta\phi)^2\over 24N}\,
\left({\varpi\over R_{\rm NS}}\right)^2\,B_{\rm NS}^2 R_{\rm NS}^3
= 4\times 10^{45}\,{(\Delta\phi)^2\over N}\,
\left({\varpi\over 3~{\rm km}}\right)^2\,
\left({B_{\rm NS}\over 10^{15}~{\rm G}}\right)^2\,
\left({R_{\rm NS}\over 10~{\rm km}}\right)\;\;\;\;\;\;{\rm ergs}.
\end{equation}
Thus the fraction of the star's surface area 
that experiences slippage in a giant flare could be as small as
$(\pi \varpi^2)/(4 \pi R_{\rm NS}^2) \sim 0.02 (\varpi/ 3$ km$)^2$
under favorable assumptions. If the position of
the active region is determined by the large-scale winding of the
internal magnetic field, then subsequent flares may occur at the same
spot, and some of the surface may never flare.  
Alternatively, it is possible that independent regions
encompassing no more than $N_{\rm flare}^{-1} \sim 10^{-2}$ of the
surface area flare only once over the active lifetime of an SGR.
The peak luminosity of the initial spike is easily reproduced in this
model, as long as $\delta E_{\rm mag}$ exceeds the observed value.
This leads to a lower bound on the surface field, as discussed in
(\S \ref{peak}).  

In a more realistic case, high-order
multipoles of the field are present.  This implies the existence of some
exterior magnetic field lines which remain close to the star in a complex
geometry.  Stretching of field lines whose footpoints lie on opposite
sides of a fault provides another mechanism for fast reconnection and
flaring.

\subsubsection{Internal Torsional and Standing Shear Waves}\label{torseis}

A twisting deformation of the neutron star crust, driven by the
Maxwell stress $B_zB_\phi/4\pi$, will be accompanied by a partial unwinding
of the core magnetic field.  We consider, as before, a uniform poloidal
flux density $B_z$ threading the core.  Then the core supports a
torsional Alfv\'en mode, which propagates along the poloidal field
with a period 
\begin{equation}\label{torperiod}
P_{\rm Alfven} 
\simeq {4R_{\rm NS}\over V_{A\,z}} = 
0.4\,\left({B_z\over 10^{15}~{\rm G}}\right)^{-1}\,
\,\left({\rho\over 10^{15}~{\rm g~cm^{-3}}}\right)^{1/2}\,
\left({R_{\rm NS}\over 10~{\rm km}}\right)\;\;\;\;\;\;\hbox{s}
\end{equation}
near the symmetry axis.  Here, $V_{A\,z} = B_z/\sqrt{4\pi\rho}$ 
is the poloidal Alfv\'en speed.  This period receives a small correction
from the shear strains that build up in the crust as the field that
threads the crust/core boundary is twisted from below.

It should be noted that
the mild density gradient in the neutron star core introduces a gradient
in the wave period, with respect to the cylindrical radius $\varpi$.  This 
gradient has the effect of washing out any quasi-periodicity in the 
wave motion -- and thence in the dynamic
stress applied to the crust -- after a dozen or so wave periods.

Whether such a torsional Alfv\'en mode is strongly excited during a giant flare
depends on the rate at which the crust deforms or breaks.  Let us suppose 
that the core
field untwists through an angle $\Delta \phi < \Delta\phi_0$, where
the net twist across the core is
\begin{equation}
\Delta\phi_0(\varpi) = {B_\phi(\varpi)\over B_z}\,\left({2R_{\rm NS}\over 
\varpi}\right).
\end{equation}
This formula can be expressed as
\begin{equation}
\Delta\phi_0(\varpi) = 0.3\,\left({B_z\over 10^{15}~{\rm G}}\right)^{-2}\,
\left({M_{\rm Coulomb}\over 0.02\,M_\odot}\right)\,
\left({V_{\mu\,0}\over 10^8~{\rm cm~s^{-1}}}\right)^2\,
\left({\psi_{\rm cr}\over 10^{-2}}\right)\,
\left({\varpi\over 3~{\rm km}}\right)^{-2}\,
\left({R_{\rm NS}\over 10~{\rm km}}\right)^{-1},
\end{equation}
in radians, making use of eq. (\ref{sbal}).
If the crust moves slowly, over a timescale $\Delta t$ much larger than
the Alfv\'en crossing time $2R_{\rm NS}/V_{A\,z}$, 
then the amplitude of the resulting torsional wave is small.
We estimate $\Delta\phi_{\rm wave} \sim (\Delta\phi/\Delta t) \times 
(2R_{\rm NS}/V_{A\,z})$, or equivalently
\begin{equation}
{\Delta\phi_{\rm wave}\over\Delta\phi}
\sim {2R_{\rm NS}\over V_{A\,z}\Delta t}.
\end{equation}

On the other hand, if the motion of the crust is slowed 
only slightly by friction (compared with a free torsional 
oscillation of the core) then a significant fraction
of the released core magnetic energy goes into magnetic 
torsion modes. 
The fact that the observed durations of hard spikes in 
giant flares are comparable to a magnetar's 
torsional oscillation period, $\sim 2R_{\rm NS}/V_{A\,z} = 
{1\over 2}P_{\rm Alfven}$ for $\Delta\phi \sim \Delta\phi_0$,
is consistent with the hypothesis that the crust {\it broke}, with 
only modest frictional resistance after the event onset.
Significant torsional excitation of the 
neutron star core is thus plausible: 
$E_{\rm wave} \sim E_X \sim 10^{44}-10^{45}$ ergs. 

A similar conclusion holds for the proportion of the crustal 
strain energy (\ref{Eelastic}) which is converted to a torsional
shear wave.  The fundamental torsional mode
(with $n = 0$ radial nodes) has an amplitude which
varies weakly with depth in the crust
(McDermott et al. 1988).  Its harmonic is restricted
to $\ell \geq 2$ from angular momentum conservation, and its period is
\begin{equation}\label{pshear}
P_{n=0\,\ell} \simeq {2\pi R_{\rm NS}\over V_{\mu\,0}[\ell(\ell+1)]^{1/2}}
= 0.026\,\left({R_{\rm NS}\over 10~{\rm km}}\right)\,
\left({V_{\mu\,0}\over 10^8\,{\rm cm~s^{-1}}}\right)^{-1}\,
\left[{\ell(\ell+1)\over 6}\right]^{-1/2}\;\;\;\;\;\;\hbox{s}.
\end{equation}
(This period is shifted downward slightly in the presence of a poloidal
magnetic field, but the shift is not significant unless the component
of the field along the direction of shear is comparable to 
$B_\mu \sim 6\times 10^{15}$ G; Duncan 1998.)  Notice
that the period (\ref{pshear}) is significantly shorter than the torsional
Alfv\'en period (\ref{torperiod}).  As a result, a significant fraction
of the crustal strain energy would be converted {\it directly} to a 
torsional shear wave during a giant flare, only if this strain energy 
were released on a relatively short timescale compared with the duration of the
hard spike, $0.03$ s or less.   Otherwise, the dynamic strain excited
in the crust is dominated by the coupling to a torsional Alfv\'en
mode in the core.  In either case, the amplitudes of the
shear wave and Alfv\'en wave can be expected to equilibrate rapidly 
(over a few wave periods) according to eq. (\ref{eeq}), with the effect
that the crustal shear wave has the lower energy by the factor (\ref{ecomp}).

\subsubsection{External Torsional Deformation of the Magnetic 
Field}\label{extor}

Such an internal torsional mode, with frequency
$\omega = 2\pi/P_{\rm mode}$ (eq. [{\ref{torperiod}] or [\ref{pshear}]),
also drives a torsional excitation of the magnetic field outside
the star.  In the case of the crustal shear mode, we focus on the
fundamental mode, whose amplitude varies weakly with
depth below the surface of the neutron star (McDermott et al. 1988).  
The surface amplitude of the external magnetic twist is related to 
the surface displacement $\xi$ of the crust through
\begin{equation}
{\delta B_{\rm NS}\over B_{\rm NS}} = {\omega\xi\over c}.
\end{equation}
We now show that, in equilibrium, only a tiny fraction of the
energy of the internal mode is transferred to the external mode.
This energy transfer could take two forms:  first, an Alfv\'en-like mode 
propagating freely on very extended field lines;  and, second,
a quasi-static torsional deformation of the field lines closer to the 
star, which oscillates in sign at the same frequency as the internal
mode.  Field lines which can support a propagating Alfv\'en mode 
have a minimum length $cP_{\rm mode}$, and
extend out to a large radius $R_{\rm max} \ga {1\over 3}cP_{\rm mode}$ 
(in a dipole geometry).  This works out to
$R_{\rm max} \sim 10^{10}\,(B_z/10^{15}~{\rm G})^{-1}$ cm, using the period 
(\ref{torperiod}).  These field lines are concentrated within a surface 
polar angle $\theta^2(R_{\rm NS}) \la 3R_{\rm NS}/cP_{\rm mode}$.  

The internal mode will, in general, not be aligned with the external
dipole moment, and so the surface shear will increase the external 
field energy by a fractional amount $\sim (\xi/R_{\rm NS})^2$.  
This works out to $\delta E_{\rm mag}(R> R_{\rm NS}) \sim 
{1\over 12}B_{\rm NS}^2 R_{\rm NS}^3\,(\xi/R_{\rm NS})^2$
(here $B_{\rm NS}$ is the polar dipole field).  Comparing with 
the elastic energy $E_{\rm elas} = 
{1\over 2}\omega^2\langle\xi^2\rangle M_{\rm mode}$ stored in the 
crustal lattice, one finds
\begin{equation}\label{emagrat}
{\delta E_{\rm mag}(R>R_{\rm NS})\over E_{\rm elas}} =
{B_{\rm NS}^2 R_{\rm NS}^3\over 6 M_{\rm mode}\omega^2R_{\rm NS}^2}
= 5\times 10^{-8}\,\left({B_{\rm NS}\over 10^{14}~{\rm G}}\right)^2\,
\left({\omega R_{\rm NS}\over c}\right)^{-2}\,
\left({M_{\rm mode}\over 0.02\,M_\odot}\right)^{-1}\,
\left({R_{\rm NS}\over 10~{\rm km}}\right)^3.
\end{equation}
This ratio is small in the case of a crustal shear mode
($M_{\rm mode} \simeq M_{\rm Coulomb} = 0.02\,M_\odot$) and
even smaller in the case of a torsional Alfv\'en mode
($M_{\rm mode} \simeq M_{\rm NS}$).

The equilibrium energy in the freely propagating Alfv\'en mode
can easily be shown to be smaller than eq. (\ref{emagrat}) by
a factor $\sim R_{\rm NS}/R_{\rm max} = 3R_{\rm NS}/cP_{\rm mode}$.
Along the excited bundle of field lines, the product of the wave 
energy density $(\delta B)^2/8\pi$ and the cross-sectional area $A$ 
of the field lines is a constant, and proportional to the square
of the current $I(\theta)$ flowing at polar angles smaller than 
$\theta \sim (R_{\rm NS}/R_{\rm max})^{1/2}$.
One has $I(\theta) = (c/4\pi)\,\pi\theta^2R_{\rm NS}^2 
({\bf\nabla}\times{\bf B})_R$, where the current density 
is related directly to the surface shear through
$({\bf\nabla}\times{\bf B})_R \sim 2\pi\,(\xi/R_{\rm NS})\, 
B_{\rm NS}/R_{\rm NS}$.  The wave energy per length of field 
line is then proportional to $\theta^4$.  Multiplying by the 
length $\sim 3R_{\rm max}$ of the excited field lines gives 
a net energy proportional to $\theta^2 \sim R_{\rm NS}/R_{\rm max}$.

The damping rate of this external torsional mode is examined
in \S \ref{peak}, and compared with the observed peak luminosities
of the giant flares.

\subsubsection{Constraints from the Rise Time}

The August 27th event showed a very steep rise in intensity, 
with width $t_{\rm rise} \la
4$ ms (\cite{hurley99a,mazets99}).  If this timescale is limited by the
propagation of a fracture, then a triggering zone
moving with fracture speed $V_f \lessim V_\mu$ has size
\hbox{$l \gtrsim V_\mu \, t_{\rm rise} = 4 \, $km$ \, (t_{\rm rise}/ 4$ ms).} 
The fast rise of the 1979 March 5 event, 
$t_{\rm rise}$$\la 0.2$ ms (Cline et al.~1980; or
$t_{\rm rise} \sim 1$ ms as reported by Fenimore et al.~1996) 
may require that a fast
magnetic reconnection front was generated in the magnetosphere
at the onset of the event (Paczy\'nski 1992).
Alternatively, a fast rise could be understood 
if the large-scale, rotational motion results from self-organizing growth of 
collective motion in smaller-scale units in an ``avalanche" effect.
If these units have dimensions comparable to the depth of the crust,
$\Delta R_\mu\sim 0.3$ km, then rise times as short as 
$\sim (\Delta R_\mu/V_\mu) \sim 3\times 10^{-4}$ s are possible.

\subsection{Giant Flares in the Context of SGR Activity}\label{secsummary}

    It has been argued that the giant flares involve a large-scale
propagating fracture of the neutron star crust driven by magnetic
stresses in the core (TD93; TD95).  We have described a specific configuration
of the magnetic field, involving a strong twist in the core and crust,
which can release $\sim 10^{44}$ ergs in individual increments as long
as the yield strain in the gravitationally stratified crust is 
$\psi_{\rm cr} \sim 10^{-2}$.  
Transport effects such as ambipolar diffusion acting on a timescale
comparable to the source age of $\sim 10^4$ yr allow consecutive
departures from magnetostatic equilibrium and repeated flaring activity,
because unwinding of the core field is strongly inhibited by the
compositional statification of the neutron star interior.  
This is an efficient way to take magnetic free energy from its principal 
reservoir -- a magnetic field anchored in the deep crust and core of
a magnetar (see eq.~[\ref{Emagnetar}]),
and convert it to observable photon emissions, while evading the
tremendous neutrino losses which are inevitable for all dissipation 
occuring locally in the interior.  

The development of strong, localized shear in the exterior magnetic
field allows the rapid dissipation of a large amount of magnetic
energy {\it outside} the star, enough to power the
observed flare emissions (eq.~[\ref{burstenergy}]).  In addition,
a significant fraction of the released energy may be deposited 
in a torsional Alfv\'en wave in the core, which transfers energy 
only gradually through the crustal lattice to the exterior.
Moreover, a predominantly rotational deformation of the crust will
leave behind non-potential terms in the exterior
magnetic field which support persistent electrical currents which
dissipate in part through Compton drag off the ambient X-ray flux
(Thompson et al. 2000).  The
dramatic change in the non-thermal persistent flux and pulse shape
of SGR 1900+14 following the August 27 giant flare (Woods et al. 2000)
provides direct evidence for a strengthening of these non-potential
components during the flare.  More generally, such a corona has
been hypothesized as the source of the persistent non-thermal
emission in the SGRs and AXPs (TD96; Thompson et al. 2001), 
and for the transient non-thermal emission following the 29 
August 1998 aftershock from SGR 1900+14 (\S 5.2.3 in Ibrahim et al.~2000). 

Shorter $\sim 1-10$ s bursts were observed soon after the two
giant flares (see Mazets et al. 1979a for SGR 0525-66, and 
Ibrahim et al. 2000 for SGR 1900+14), with luminosities comparable
to the pulsating tails of the flares.  In the above model, these
shorter bursts probably involve shorter fault-line slippage than 
the flares.  The relation of 
the much more common $\sim 0.1$ s SGR bursts to the giant flares is 
perhaps more mysterious;  while these events could also involve
a propagating fracture, they may alternatively be driven by a more
localized and plastic deformation of the crust (TD95).

\section{Soft Pulsations:  the Light Curve}\label{softail}

The hard spike and the extended soft tail of the August 27 event released
comparable
energies, $E \sim (0.5-1)\times 10^{44}$ ergs, even though their durations
differed
by a factor of $10^3$.  This suggests that a large fraction of the
outburst
energy was injected during the initial $\sim 0.5$ s.  Whereas
the enormous luminosity of the spike requires a relativistic
outflow from the neutron star, the spectral stability and large
amplitude
oscillations observed during the last $\sim 300$ s of the outburst
require
instead that the radiated energy was confined close to the surface of
the neutron
star.  This energy, if released into the near magnetosphere,
forms a dense, thermal plasma of $e^\pm$ pairs and blackbody radiation
with a temperature
$T = 0.72\,(E/10^{44}~{\rm ergs})^{1/4}\,
(L/10~{\rm km})^{-3/4}$ MeV (in a volume $L^3$).
The scattering opacity of electrons and positrons was then high enough to
lock this energy onto magnetic field lines close to the neutron star,
in a ``trapped fireball''.  This fireball cools by radiative diffusion
through a thin surface layer, which contracts inward while remaining 
nearly congruent with the magnetic field (TD95).

Strong evidence for this trapped fireball model comes from the
faster-than-exponential
decline in the X-ray flux at the end of the August 27 outburst,
which is consistent with the hypothesis that the energy resevoir is 
evaporating completely in a finite time. 
Except for the first $\sim 40$ s (one tenth of the burst duration), 
the envelope of the
August 27 light curve is well fit by the following function (Paper I)
\begin{equation}\label{lxt}
L_X(t) = L_X(0)\left(1-{t\over t_{\rm evap}}\right)^\chi.
\end{equation}
If the cooling luminosity is assumed to vary as a power of the {\it remaining}
fireball energy, $L_X \propto E^a$, then $\chi = a/ (1-a)$.  
In the simplest case of a fireball with uniform energy density and
surface energy flux, the index
$\chi$ is effectively the number $D_c$ of curved directions of the fireball
surface: $\chi = D_c = 2$ for a spherical fireball; $\chi= D_c = 1$ 
for a fireball filling bundle of flux lines that is (locally) cylindrical;
and $\chi = D_c =  0$ for the case of a bundle of flux lines that 
is (locally) a thin slab (Fig. \ref{confinement}).

This function (\ref{lxt}) accurately captures the sudden final drop in flux
seen in the \it Ulysses \rm and BeppoSAX data (Paper I).  The best-fit
fireball index is $\chi = 3$ (or $a = 0.75$), and the evaporation time
$t_{\rm evap} = 375$ s (Figs. 2-3 of Paper I).  Note that the fit
is excellent following the initial $\sim 40$-s of the outburst. 
In comparison, the best
fit exponential profile $\exp(-t/t_{\rm exp})$ adequately describes the
intermediate portion of the decay (Fig. 1. of Paper I), with
a favored time constant $t_{\rm exp} = 78$~s in the \it Ulysses \rm
data; but significantly overshoots the final $\sim 40$ s of the
outburst.

During the first $\sim 40$ s,  the measured burst flux
significantly overshoots both the exponential and trapped
fireball fitting functions.  The mild spin-modulation of the light curve
during this smooth tail requires a more extended and variable photosphere.
which we argued in Paper I is pair-dominated and results from
a continuing creep of the neutron star crust in the active region
(or possibly a standing torsional wave in its interior).  The cooling of
a continuously heated, pair-dominated corona is discussed in detail in 
\S \ref{smoothdecay}.

\subsection{Fireball Index}\label{fireballindex}
\def\dim{D}

We now consider some simple physical models for a contracting fireball.
Because the best-fit fireball index $\chi$ is larger than 2 (the maximum
number of curved directions for the fireball surface in three spatial
dimensions), a homogeneous fireball cannot explain the data.  In this
section, we consider the effects of temperature and magnetic field
gradients.  

The surface of the fireball will be assumed to have
$D_c$ curved directions and $2-D_c$ flat directions, and to 
be fully symmetric with a single perpendicular coordinate $r$.
Before the cooling wave propagates into the fireball, the temperature
and magnetic field vary as powers of $r$:
\begin{equation}
T \propto r^\gamma  \ ;  \ \ \ B \propto r^\beta.
\end{equation} 
Temperature gradients perpendicular to ${\bf B}$ easily 
persist over the duration of the August 27 event, because
the photon diffusion time across a distance $\sim R_{\rm NS}$
through the fireball is much larger than the observed burst duration
(or the time for the cool boundary layer to propagate into the fireball:
TD95 \S 3.2 and \S 3.4).
 
If the surface flux from a trapped fireball varies as
\begin{equation}\label{sflux}
F \propto T^{\sigma_1}   B^{\sigma_2}.
\end{equation} 
then, because the emitting area scales as $A \propto r^{D_c}$, the 
luminosity is
\begin{equation}
L_X = L_X(0)  
\left({r\over r_o} \right)^{D_c + \sigma_1 \gamma + \sigma_2 \beta}.
\label{Lx1}
\end{equation} 
The energy density within the fireball includes contributions from 
both photons and electron-positron pairs.  In two regimes of
relevance, it has the form 
\begin{equation}
U \propto T^{\mu_1}   B^{\mu_2}.
\end{equation}
In particular,
if many Landau levels are populated and $T\gg m_e$, then one recovers
the usual expression for a relativistic pair plasma,
\hbox{$U = (11/4) a T^4$} where $a$ is the Stefan-Boltzmann constant,
and $\mu_1 =4$ and $\mu_2 = 0$.  The condition for a such 3-dimensional
pair gas is\footnote{In units with
$\hbar = 1 = c$, used frequently hereafter.}  $2.7\,T \ga (2eB)^{1/2}$ or
equivalently
\begin{equation}\label{landau}
T \ga 0.3\,\left({B\over B_{\rm QED}}\right)^{1/2}\;\;\;{\rm MeV}
\end{equation}
in a magnetic field $B \ga B_{\rm QED}$ (eq. \ref{bqed}).
This condition is easily satisfied near
marginal confinement, when the pressure of pairs and photons 
is close to the limiting value $B^2/8\pi$ (TD95).  

When the magnetic
pressure greatly exceeds the thermal pressure, and
$T$ is not much less than $m_e$, the pairs dominate the thermal
energy density: \hbox{$U \approx U_{e^+ e^-} \approx (1/12) e B T^2$} 
(in units where $\hbar = c = 1$).  The indices $\mu_i$ are then
$\mu_1 =2$ and $\mu_2 = 1$ 
[see eq.~(53) and the discussion following eq.~(56) in TD95; see also
Figure 5 in Duncan 2000 and Kudari 1997]. 

The surface X-ray luminosity is related to surface area $A$ of the 
fireball and its rate of contraction through $L_X = - A U (dr/dt)$, or
\begin{equation}
L_X = - L_X(0)  
\left({r\over r_o} \right)^{D_c + \mu_1 \gamma + \mu_2 \beta}
{dr/dt\over (dr/dt)_0}.
\label{Lx2}
\end{equation} 
Equations (\ref{Lx1}) and (\ref{Lx2}) together give a simple differential
equation for $r(t)$.  The solution implies that $L_X(t)$ has
the form of eq.~(\ref{lxt}) with
\begin{equation}\label{chieqn}
\chi = {D_c + \sigma_1 \gamma + \sigma_2 \beta \over 
1 + \gamma(\mu_1 - \sigma_1) + \beta (\mu_2 - \sigma_2)}
\end{equation}

The case of a structureless trapped fireball,  $\gamma = \beta = 0$, 
reduces to $\chi = D_c \le 2$.  Since $\chi \simeq 3$ 
empirically, we infer that the fireball must have had structure, with
indices which satisfiy
\begin{equation}\label{gambest}
(3\mu_1-4\sigma_1)\gamma + (3\mu_2-4\sigma_2)\beta = D_c - 3.
\end{equation}

Specific values of the structure indices $\gamma$ and $\beta$, for 
several plausible assumptions about the fireball geometry
and physical conditions, are given in Table 1.   

These values of $\gamma$ and $\beta$ depend on specific
choices for the parameters $\sigma_i$ in the 
surface X-ray flux (\ref{sflux}) from the trapped fireball.
\ One generally expects a positive dependence on temperature
and magnetic field because of the thermal nature of the emission
($\sigma_1 > 0$), and because Compton 
scattering is suppressed in a strong magnetic field ($\sigma_2 > 0$). 
We have considered three cases which span the possible range of cooling-wave
behaviors: one case in
which $F$ is most sensitive to $B$; one in which $F$ is sensitive 
to $T$, and one with sensitivity to both variables.  
The case $\sigma_1 = 0, \,  \sigma_2 = 4/3$ corresponds to a 
simple Eddington-limited flux from an atmosphere that is held down 
by gravity, when the magnetic suppression of 
scattering is taken into account (eqs.~[45] and [93] in TD95).   
The case $\sigma_1 = 5/2$,  
$\sigma_2 = 0$ corresponds to a magnetically-confined fireball 
in which the confined plasma fills many Landau levels
($\mu_1 = 4, \ \mu_2 = 0$), {\it and} in which the
integrated surface flux across {\bf B} is limited
by the rate at which energy is advected {\it along} the magnetic
field, so as to smooth out pressure imbalances
(eq.~[89] of TD95).   Finally, the case $\sigma_1 = 3/2, \, \sigma_2 = 1$
(eq.~[86] of TD95)
corresponds to a similar process of advection-limited cooling, but
in the opposite regime where the confined plasma is too cool to 
fill higher Landau levels
(so that the radiative opacity across {\bf B} is suppressed by
the strong magnetic field).
In these last two cases, $T$ is interpreted as the temperature just
inside the geometrically-thin radiative layer at the surface of
the fireball.

Table 1 shows that, in all physical regimes considered,
a fireball that is slightly hotter in the center than near the edges,
$\gamma < 0$, can fit the observed light curve.  Such a 
pattern of energy-loading in the magnetosphere could plausibly occur.  
An alternative,
less attractive possibility, is that the temperature inside the fireball
is uniform while the magnetic field {\it declines} slightly toward its
center, $\beta > 0$.

\subsection{Effects of Neutrino Cooling}\label{nucooling}

The trapped fireball light curve will also be perturbed by neutrino pair
emission,
$e^+ + e^- \rightarrow \nu + \bar\nu$, at a rate 
$\dot U(e^+e^-\rightarrow \nu\bar\nu) = 
1.3\times 10^{25}\,(T/{\rm MeV})^9$ ergs cm$^{-3}$ s$^{-1}$ 
(Dicus 1972; Schinder et al. 1987).  This can reduce the fireball energy
on a time scale comparable to the duration of the giant outburst (\S 5 in
TD95):
\begin{equation}\label{nucooltime}
\tau_{\nu} \equiv {{11\over 4}aT^4\over \dot
U(e^+e^-\rightarrow\nu\bar\nu)}
= 28\,\left({T\over 1~{\rm MeV}}\right)^{-5}
\;\;\;\;\;{\rm s}.
\end{equation}
This expression assumes that the temperature is high enough to excite
pairs in the upper Landau levels (eq. [\ref{landau}]).

The effects of pair neutrino cooling on a fireball light curve
are shown in Figs. \ref{neutrinoa} and \ref{neutrinob} for several 
initial values of the dimensionless ratio $(E/L_X\tau_\nu)$ and a fixed
evaporation time $t_{\rm evap}$.  In Fig. \ref{neutrinoa}, we plot
fireball light curves for various geometries ($D_c = 1,2$), assuming
that the trapped pair gas is 3-dimensional ($\mu_1 = 4$, $\mu_2 = 0$)
and that the cooling wave is limited by advection 
($\sigma_1 = 5/2$, $\sigma_2 = 0$; see eq.~[89] in TD95).  The
neutrino cooling time is parameterized by the initial value of the
dimensionless ratio $E/L_X\tau_\nu$, where $E$ is the trapped fireball 
energy.  One sees that neutrino cooling introduces additional
curvature in the (logarithmic) light curve, predominantly during the
initial decline.  This curvature is absent
for a homogeneous fireball ($\gamma = \beta = 0$), as well as for
a fireball with powerlaw temperature and magnetic field profiles
(constant $\gamma$, $\beta$).  In Fig. \ref{neutrinob}, we consider
a combination of neutrino cooling and temperature gradients, with
the value of $\gamma$ perturbed slightly from the best-fit value
(\ref{gambest}).  (Here the initial value of $(E/L_X\tau_\nu)$ is
calculated using the initial temperature at the outer boundary of
the fireball.)

To summarize:  neutrino cooling introduces curvature
in the trapped-fireball light curve, which cannot be compensated
by a power-law temperature gradient.  A large value of 
$(E/L_X\tau_\nu) \ga 10$ does not fit the observed
light curve, but one can not rule out a combination of slower
neutrino cooling and more complicated temperature profiles.

\section{Bound on the Stellar Dipole Moment}

In this section, we quantify how the long duration and enormous
peak luminosity of a giant flare set lower bounds to the 
magnetic moment of its source.  First, by including the effects of
neutrino cooling on the declining X-ray flux of a trapped pair-photon 
plasma, we are able to set a lower bound on
$B_{dipole}$ that is less dependent on assumptions about the
configuration of the field than those made in TD95.  Second,
by considering in more detail the coupling between a low-frequency
internal (e.g. torsional) mode of the star and its magnetosphere,
we are able to show that such a low-frequency mode cannot easily
lose energy at the rate observed in the hard initial spike:
even under the most optimistic assumptions about the damping of
the external magnetic shear, the surface flux density must be
very high (in excess of $10^{14}$ G).

\subsection{Confinement of the Radiating Plasma During the Soft
Pulsations}\label{momentbound}

A strong lower bound on the surface magnetic field is deduced from
confinement of a large amount of energy, $E \sim 10^{44}$ ergs,
close to the SGR source during a giant flare.  Assuming that
the field is a centered dipole, and requiring that $B^2/8\pi > 
P_\gamma + P_{e^\pm}$ at the outer bondary of the fireball, one
deduces 
\begin{equation}
B_{\rm dipole} > 2\times 10^{14}\,
\left({E_{\rm fireball}\over 10^{44}~{\rm ergs}}\right)^{1/2}
\,\left({\Delta R\over 10~{\rm km}}\right)^{-3/2}\,
\left[{1+\Delta R/R_{\rm NS}\over 2}\right]^3
\;\;\;\;\;\;{\rm G}
\end{equation}
for a relativistic plasma (TD95).  Here $\Delta R$ is the 
characteristic size of the fireball.

The question which next arises, is how this bound is modified if
the magnetic field is not a centered dipole.  Indeed, the remarkable
four-peaked pattern of the August 27 pulsating tail provides direct
evidence for higher multipoles (Paper I; Section \ref{Xjets}).
Consider first how this bound is modified as the footpoints of
the confining magnetic field are brought closer together.  The minimum
pressure of the confining magnetic field increases with
decreasing footpoint separation $L$, $B_{\rm min}^2/8\pi \sim
E/3L^3$, but the net magnetic moment of the confining field {\it decreases},
$\mu_{\rm min} \sim B_{\rm min}L^3 \propto E^{1/2}L^{3/2}$.  At the
same time, the temperature of the confined plasma (assumed to
have a 3-D distribution) increases, $T \propto L^{-3/4}$.  

The bound on $B_{\rm dipole}$ cannot, however, be reduced arbitrarily
by shrinking the fireball.  The reason is that the
neutrino cooling time of the fireball decreases rapidly with
decreasing $L$,
\begin{equation}
\tau_\nu \propto T^{-5} \propto L^{15/4}
\end{equation}
at fixed energy $E$, even while the photon cooling time 
(eq.~[\ref{lxt}]) hardly changes:
\begin{equation}
t_{\rm evap} \sim {L/V_{\rm cool}} \sim L^{-1/8}
\end{equation}
in the absence of neutrino cooling.  (The propagation speed of
the cool boundary is related to the surface
X-ray flux by $V_{\rm cool} = F_X(T)/[{11\over 4}aT^4] \propto T^{-3/2}$.)

To get a sense of how far $B_{\rm dipole}$ can be reduced and remain
consistent with magnetic confinement of the radiating plasma, let
us consider a small offset dipole, centered at the outer edge of the
star (Fig. \ref{confinement}).  
Neutrino cooling has a large effect on the fireball lightcurve
when $\tau_\nu \la t_{\rm evap}$, where
the evaporation time can be written as
\begin{equation}
t_{\rm evap} = (\chi + 1)  \left({E\over L_X}\right)_0.
\end{equation}
Indeed, in the case of rapid neutrino cooling, one can divide the
cooling process into two phases:  an initial neutrino-dominated
phase where the fireball volume hardly decreases;  and a subsequent
photon-dominated cooling phase.  Integrating the neutrino losses
up to a time $\sim t_{\rm evap}/(\chi+1)$, the fireball pressure
is reduced by a factor 
\begin{equation}
{P_{e^\pm} + P_\gamma\over (P_{e^\pm} + P_\gamma)_0} =
\left({T\over T_0}\right)^4 = 
\left[1+{5\over 4}\left({E\over L_X\tau_\nu}\right)_0\right]^{-4/5}
\end{equation}
Thus, as $E/L_X\tau_\nu \propto L^{-31/8}$ is pushed above unity,
the lower bound on the magnetic moment does not change significantly:
$\mu_{\rm min} \propto (E_0 L^3)^{1/2}$ is almost independent of $L$.
For the purposes of estimating $\mu_{\rm min}$, we conservatively take
$(E/L_X\tau_\nu)_0 = 10$ in what follows.  This corresponds to 
$\tau_\nu \ga 0.025 t_{\rm evap}$ when $\chi = 3$.

This lower bound on the neutrino cooling time yields an
{\it upper} bound on the fireball temperature through
eq.~(\ref{nucooltime}),
\begin{equation}\label{tmin}
T_{\rm max} =  1.2\,\left({t_{\rm evap}\over 400~{\rm s}}\right)^{-1/5}
\;\;\;\;{\rm MeV}.
\end{equation}
In order to contain the energy radiated in the August 27 
pulsating tail, the volume of the fireball must exceed
\begin{equation}
V_{\rm fireball} = {E\over (11/4)aT^4} =
(4.9~{\rm km})^3\,\left({E\over 10^{44}~{\rm ergs}}\right)\,
\left({T\over T_{\rm max}}\right)^{-4}\,
\left({t_{\rm evap}\over 400~s}\right)^{4/5}.
\end{equation}
The confining magnetic field at its weakest point must be stronger than
$B_{\rm min}^2/8\pi = P_\gamma + P_{e^\pm} = {11\over 12}aT^4$, which
combined
with eq. (\ref{tmin}) implies $B_{\rm min} = 
8.3\times 10^{13}(T/T_{\rm max})^2\,(t_{\rm evap}/400~{\rm s})^{-2/5}$ G.
Given that the plasma fills half the dipole magnetic field, and
only along field lines which never drop below this flux density, 
the magnetic moment can be calculated to be
\begin{equation}
\mu_{\rm min} = {105\over 32\pi}B_{\rm min}V_{\rm fireball} = 1\times
10^{31}\,
\left({E\over 10^{44}~{\rm ergs}}\right)\,\left({T\over
T_{\rm max}}\right)^{-2}\,\left({t_{\rm evap}\over 400~s}\right)^{2/5}
\;\;\;\;\;\;{\rm G-cm^3}.
\end{equation}
In this expression, $B_{\rm min}$ is the minimum flux density at
the magnetic equator.

This expression provides only a lower bound to the {\it net} dipole
moment of the neutron star, for two reasons.  
First, no change was observed in the spindown rate of SGR
1900+14 in the few months following the August 27 event (Woods et al. 1999),
which implies that the dipole moment of the confining magnetic field
was only a fraction of the total (Thompson et al. 2000).  
A conservative upper bound of $\sim 30$ percent
to the fractional change in  $\dot P$ translates into a maximum change
$\delta\mu_{\rm net}/\mu_{\rm net} \sim 0.15$ in the magnetic moment 
in the case of magnetic dipole spindown ($\dot P \propto \mu_{\rm net}^2$).
The tolerable change in the
magnetic moment could be larger, $\delta\mu_{\rm net}/\mu_{\rm net} \sim 0.3$,
if the spindown were accelerated by a persistent wind of particles and
Alfv\'en waves, assuming that the wind's power $L_W$ does not change
across the flare
 ($\dot P \propto \mu_{\rm net} L_W^{1/2}$; Thompson \& Blaes 1998;
Harding et al.~1999; Thompson et al.~2000).  On the other hand, the giant 
outburst is expected
to produce a change $\delta\mu_{\rm net}/\mu_{\rm net} \sim 
\mu_{\rm min}/\mu_{\rm net}$ in
the net magnetic moment, and so we arrive at 
\begin{equation}\label{munetb}
\mu_{\rm net} \ga
\left({\delta\mu_{\rm net}\over\mu_{\rm net}}\right)^{-1}\,\mu_{\rm min} =
2\times 10^{32}\,\left({\delta\mu_{\rm net}/\mu_{\rm net}\over 0.1}\right)^{-1}
\left({E\over 10^{44}~{\rm ergs}}\right)\,\left({T\over
T_{\rm max}}\right)^{-2}
\;\;\;\;\;\;{\rm G-cm^3}.
\end{equation}
In this case, the effective polar dipole field is
\begin{equation}\label{munetc}
B_{\rm dipole} = {2\mu_{\rm net}\over R_{\rm NS}^3}
\ga 4\times 10^{14}\,\left({E\over 10^{44}~{\rm ergs}}\right)^{1/2}
\,\left({T\over T_{\rm max}}\right)^{-2}\,
\left({R_{\rm NS}\over 10~{\rm km}}\right)^{-3}\;\;\;\;\;\;{\rm G}.
\end{equation}

A single dipole of magnitude (\ref{munetb}) barely contains
enough energy to power a single giant outburst, which suggests that either
the entire external field of SGR 1900+14 was strongly sheared during the 
flare;  or the actual field is much stronger than (\ref{munetc}); or the 
surface of the star is covered with many dipoles $\ga \mu_{\rm min}$.  
If the net energy output in $N$ giant flares is comparable to the external
magnetic energy, then we estimate
\begin{equation}\label{munet}
\mu_{\rm net} \ga N^{1/2}\,\mu_{\rm min} = 1\times 10^{32}\,\left({N\over
100}\right)^{1/2}
\left({E\over 10^{44}~{\rm ergs}}\right)\,\left({T\over
T_{\rm max}}\right)^{-2}\,\left({t_{\rm evap}\over 400~s}\right)^{2/5}
\;\;\;\;\;\;{\rm G-cm^3}.
\end{equation}
(Note that this expression describes both
a single large dipole, and also the incoherent
superposition of $N$ mini-dipoles of magnitude $\mu_{\rm min}$; 
see also~\S 12.3 in TD93.)

\subsection{Peak Luminosity and Bound on the Surface Magnetic 
Field}\label{peak}

The initial spike of the August 27 flare reached a peak 
luminosity\footnote{This applies after allowing for dead-time corrections; this
value remains below that inferred previously for the peak luminosity
of the March 5 event at the distance of the LMC (Mazets et al. 1979).}
$L \ga 10^{44}$ ergs s$^{-1}$ (Mazets et al. 1999), approaching
a million times the Eddington luminosity of a neutron star.  
Such a high rate of release of energy allows us to set
a stringent lower bound on the magnetic field in the source, which
is compatible with the value deduced from the total expected flaring
output (\S \ref{energy}).  
Note that the opacity of the emitting material is not directly relevant
to this bound, since the hard spike appears to have involved
an expanding fireball.

Motions of a neutron star crust will release magnetic energy directly
by inducing tangential discontinuities in the field, which lead to
fast reconnection and high frequency wave motion 
(Thompson 2000; \S \ref{shrec} above).  The output induced
by a horizontal motion through a distance $\sim R_{\rm NS}$ over
a timescale $\Delta t_{\rm spike}$ is
$L_{\rm spike} \sim \epsilon_X(B^2/8\pi)R_{\rm NS}^3/\Delta t_{\rm spike}$.
Here $\epsilon_X$ is the fraction of the magnetic energy which is dissipated
and converted to high energy photons, and is estimated in eq.
(\ref{burstenergy}).
The implied r.m.s magnetic field is 
\begin{equation}\label{spikeBbound}
B \ga 2\times 10^{14}\,\left({\epsilon_X\over 0.1}\right)^{-1/2}\,
\left({\Delta t_{\rm spike}L_{\rm spike}\over 10^{44}~{\rm ergs}}\right)^{1/2}
\;\;\;\;\;\;{\rm G}.
\end{equation}

Much of the dissipated energy would initially be confined by the 
(dipole) magnetic field.  Nonetheless,
the confined pair plasma generated in such a giant flare
can easily exceed the dipole pressure (which decreases
as $R^{-6}$) at a short distance above the surface of the neutron star.
This provides a mechanism for driving a relativistic, pair-loaded
fireball.
The rise time of the August 27 hard spike could, in fact, be limited by
the breakout of this confined pair plasma.  Evidence for repeated
breakouts
of pair plasma is provided by the $\sim 20$ ms modulation of the hard
spike of the March 5 event (Barat et al. 1983).

Let us also examine the alternative possibility that 
the magnetic field merely acts as a couple between an internal
mode of the neutron star and external Alfv\'en-like excitations.
Of the possible internal modes of a neutron star, a crustal shear wave 
has the strongest coupling to the magnetosphere (Blaes et al. 1989).  
The required magnetic field turns out to be similar, even after making
the most optimistic assumptions about the damping rate of the external
mode.  This is because the field lines are anchored in very dense
material, within which a shear disturbance propagates much more slowly
than it does in the magnetosphere.  (Equivalently, the equilibrium 
energy stored in the external mode is a tiny fraction of the energy 
in the internal mode; \S \ref{torseis}.)  A related mechanism involving
the coupling of an internal $p$- or $f$-mode to the magnetosphere was
examined soon after the 5 March 1979 flare, by Ramaty et al. (1980)
and Lindblom \& Detweiler (1983).  As we argue below, the damping rate
was significantly overestimated by these authors, and is in fact much
closer to the rate for vacuum dipole radiation.

Global toroidal modes in the crust can be excited directly through a 
large-scale fracture, as is expected in a giant flare, or indirectly 
through coupling to a torsional Alfv\'en mode in the core 
(\S \ref{torseis}; Duncan 1998).  
The fundamental toroidal mode (with $n = 0$ radial nodes) has a frequency
\hbox{$\omega \simeq \sqrt{\ell(\ell+1)}(V_{\mu\,0}/ R_{\rm NS})$} 
for $\ell \geq 2$ (McDermott et al. 1988).  Whereas the torsional Alfv\'en
mode has a frequency $\omega = \pi V_{A,z}/2R_{\rm NS}$ (in a uniform
poloidal magnetic field $B_z$ with corresponding Alfv\'en speed
$V_{A,z} = B_z/\sqrt{4\pi\rho}$; eq. [\ref{torperiod}]).  To begin,
let us consider an excitation confined to the crust.
Suppose that an energy $E_{\rm elas} \sim 10^{44}$ ergs was deposited
in toroidal modes (of various harmonics $\ell$) during the initial $\sim 0.5$ s
spike.  The r.m.s. mode amplitude $\langle\xi^2\rangle^{1/2}$ at the neutron
star surface is releated to the mode energy $E_{\rm elas}$ by substituting
$\psi^2 \simeq \ell(\ell+1)(\xi/R_{\rm NS})^2$ in equation (\ref{Eelastica}),
which gives $E_{\rm elas} = {1\over 2}\ell(\ell+1)\,M_{\rm Coulomb}
V_{\mu\,0}^2\,(\xi/R_{\rm NS})^2$.  We have in turn
\begin{equation}\label{eshear}
\langle\xi^2\rangle^{1/2} 
= 0.1\,\left({E_{\rm elas}\over 10^{44}~{\rm ergs}}\right)^{1/2}\,
\left[{\ell(\ell+1)\over 6}\right]^{-1/2}\,
\,\left({M_{\rm Coulomb}\over 0.02\,M_\odot}\right)^{-1/2}\,
\left({V_{\mu\,0}\over 10^8~{\rm cm~s^{-1}}}\right)
\;\;\;\;\;\;\hbox{km}.
\end{equation}

The resulting harmonic displacement of the neutron star magnetosphere
can be divided into two components (\S \ref{extor}):  a quasi-static 
deformation of field lines which close within a radius $R_{\rm max}
\sim {1\over 3}c(2\pi/\omega)$; and an Alfv\'en mode which propagates
freely along more extended field lines, with
a group velocity $d\omega/d{\bf k} = c\hat B$.

To set a conservative {\it lower} bound on the rate of dissipation
of energy in the magnetosphere, let us assume that the quasi-static
twist is damped within a substantial solid angle $\Delta\Omega$
on a short timescale $2\pi/\omega \ga t_{\rm damp} \ga R_{\rm NS}/c$.
(Notice that the light-crossing time $R_{\rm NS}/c$ is
much shorter than the mode period, by a factor $\sim V_{\mu 0}/c$.)
The amplitude of the deformed field grows at a rate $d\delta B_{\rm NS}/dt
\sim (\omega\xi/R_{\rm NS}) B_{\rm NS}$, leading to an equilibrium
amplitude $\delta B_{\rm NS}/B_{\rm NS} \sim (ct_{\rm damp}/R_{\rm NS})
(\omega\xi/c)$.  The torsional mode energy then decreases at the rate 
\begin{equation}
-{dE_{\rm elas}\over dt} \simeq (\Delta\Omega R_{\rm NS}^2)\,
{B_{\rm NS}^2\over 8\pi}c\,\left({\omega\xi\over c}\right)^2
\,\left({ct_{\rm damp}\over R_{\rm NS}}\right).
\end{equation}
This expression is maximized at a damping time $t_{\rm damp}
\sim P_{\rm mode} = 2\pi/\omega$ (above which the damping rate
scales inversely with $t_{\rm damp}$).  It can be inverted to 
yield the {\it minimum} mode energy needed to power a given rate of
dissipation, which we set equal to the spike luminosity $L_{\rm spike}$:
\begin{equation}\label{Eelaspike}
E_{\rm elas} = 1\times 10^{43}\,
\left({L_{\rm spike}\over 10^{44}~{\rm ergs~s^{-1}}}\right)\,
\left({P_{\rm mode}\over 0.03~{\rm s}}\right)^{-1}\,
\left({B_{\rm NS}\over 10^{14}~{\rm G}}\right)^{-2}\,
\left({M_{\rm Coulomb}\over 0.015~M_\odot}\right)\,
\left({\Delta\Omega\over 4\pi}\right)^{-1}\;\;\;\;\;\;{\rm ergs~s^{-1}}.
\end{equation}
One sees that even if the mode is damped over a large 
solid angle $\Delta\Omega \sim 4\pi$, a surface magnetic field of at
least $B_{\rm NS} \sim 3\times 10^{13}\,
(P_{\rm mode}/0.03~{\rm s})^{-1/2}$ G is required.  Otherwise, the
mode must carry much more energy than is radiated.  Indeed, crustal 
shear oscillations are barely capable of storing sufficient energy 
(eq.~[\ref{Eelastic}]).

It is perhaps more natural for a large energy to be deposited in a torsional
Alfv\'en mode in the liquid interior, for the reasons discussed in
\S \ref{torseis}.  The mode energy would then be even larger,
$E_{\rm Alfven} \sim {1\over 2}M_{\rm NS} (\xi \omega_{\rm Alfven})^2$.
In order to power the hard spike via Alfv\'en waves from the excited
star, a condition similar to  eq.~(\ref{Eelaspike}) must be satisfied, 
but with the mode mass $M_{\rm Coulomb}$ replaced by $M_{\rm NS}$.
The lower bound on the magnetic field is then more severe,
$B_{\rm NS} \ga 3\times 10^{14}\,(P_{\rm mode}/0.03~{\rm s})^{-1/2}$ G.

These bounds on the mode energy are conservative:  
it is less than clear that a smooth periodic deformation
of the external magnetic field will damp in a time as short as 
$P_{\rm mode}$, and only a small fraction $\sim 3R_{\rm NS}/cP_{\rm mode}$ 
of the external field lines are able to support a propagating shear 
Alfv\'en mode with a frequency comparable to that of the internal mode.
By contrast, dissipation will occur effectively outside the star 
through the build-up of strong, localized shear 
(e.g. near the site of a fracture).

Could an internal toroidal mode power the more modest output
of $L_X \sim 10^{42}$ ergs s$^{-1}$, above and beyond the trapped-fireball
light curve, which is observed during the initial $\sim 40$
s of the August 27 flare?  We ascribed this excess emission
to a continuously heated pair-dominated corona in Paper I.   
(Fatuzzo \& Melia (1993) and Melia \& Fatuzzo (1995) have 
explored the related possibility that short SGR bursts may be 
powered by the continuous dissipation of sheared Alfv\'en waves 
propagating away from a more weakly magnetized neutron star, but 
did not address the relation between the Alfv\'en wave flux and 
the mode energy.)  Setting the mode energy equal to this excess 
X-ray output of $\sim 10^{43}$ ergs, one deduces from equation 
(\ref{Eelaspike}) that the external Alfv\'en mode must be damped 
through a solid angle $\Delta\Omega \simeq 4\pi\,
(B_{\rm NS}/10^{14}~{\rm G})^{-2}\,(P_{\rm mode}/0.03~{\rm s})^{-1}\,
(M_{\rm mode}/M_{\rm NS})$ Sr.  
For the reasons just described, this angle is uncomfortably but
not impossibly large.  The excess emission during the first 
$\sim 40$ s could, alternatively, have been powered by a continued 
shearing of the external magnetic field, driven by a diminishing 
creep of the crust in the active region; or by the gradual decay 
of static magnetospheric currents.

Finally, let us compare our results with the model of Ramaty et al. 
(1980) and Lindblom \& Detweiler (1983).  It was suggested in these
papers that an internal $f$- or $p$-mode would be damped by shock-heating
of the surface layers of the neutron star, and also by a strong flux 
of Alfv\'en waves into the magnetosphere.  Repeated shocking of the
surface layers of a star can occur only if the time for a sound
wave to cross a pressure scale-height is {\it longer} at shallower
depths.  (Otherwise, an upward-propagating sound wave will be reflected
downward before it has a chance to steepen into a shock.)  This condition
is indeed satisfied in the envelope of a pulsating AGB star, but it
fails to be satisfied in the crust of a neutron star.  We start
with the equation of hydrostatic equilibrium, $dP/dr =
-\rho GM(<r)/r^2$, where $P(z) \propto [\rho(z)]^\gamma$ at depth
$z = R_{\rm NS} - r$.  (The index $\gamma \simeq
{4\over 3}$ for $\rho \sim 10^6-4\times 10^{11}$ g cm$^{-3}$ and
$\gamma \simeq {5\over 3}$ at lower densities.)  Integrating gives
$P(z)/\rho(z) \simeq (\gamma-1)\,z\,(GM_{\rm NS}/R_{\rm NS}^2)$,
along with the sound speed $c_s \simeq \sqrt{\gamma P/\rho}$ and the
pressure scale-height $\ell_P = P/(-dP/dr)$ as functions of
depth.  In a neutron star, the sound-crossing time decreases outward,
\begin{equation}
{\ell_P\over c_s} = \left({\gamma-1\over\gamma}\right)^{1/2}\,
\left({z\over R_{\rm NS}}\right)^{1/2}\,
\left({R_{\rm NS}^3\over GM_{\rm NS}}\right)^{1/2},
\end{equation}
and so the surface layers are manifestly stable to repeated shocking.

Now let us consider the electromagnetic damping of an $f$- or $p$-mode.
The damping rate into Alfv\'en waves is even lower, because
an $f$- or $p$-mode does not shear the external magnetic field lines.  
The mode could have a frequency as high as $\omega \sim 10^4$ Hz,
and so a much wider bundle of external field lines would be excited
than by a lower-frequency torsional mode.
(We assume for now that the magnetosphere is able to respond to the surface 
motions as a conducting fluid.)  To set a conservative upper bound on the
energy flux into the magnetosphere, we allow all the field lines
to carry away Alfv\'en waves of an amplitude $\delta B_{\rm NS}/
B_{\rm NS} = \xi\omega/c$ and a surface energy flux 
$c(\delta B_{\rm NS})^2/8\pi$.  The mode energy 
$E_{\rm mode} \simeq {1\over 2}M_{\rm NS}\omega^2\langle\xi^2\rangle$ 
is then damped at a rate
\begin{equation}
-{dE_{\rm mode}\over dt} \simeq 4\pi R_{\rm NS}^2 c {B_{\rm NS}^2\over 8\pi}
\left({\omega^2\langle\xi^2\rangle\over c^2}\right) 
= {E_{\rm mode}R_{\rm NS}^2 B_{\rm NS}^2\over M_{\rm NS}c}.
\end{equation}
For a neutron star mass $M_{\rm NS} = 1.4\,M_\odot$ and radius 
$R_{\rm NS} = 10$ km, this expression becomes
\begin{equation}
-{dE_{\rm mode}\over dt} \simeq 1\times 10^{42}\,
\left({E_{\rm mode}\over 10^{44}~{\rm ergs}}\right)\,
\left({B_{\rm NS}\over 10^{15}~{\rm G}}\right)^2\;\;\;\;{\rm ergs~s^{-1}}.
\end{equation}
As a check, consider the damping rate due to vacuum dipole radiation.
As before, we normalize the mode energy $E_{\rm mode} 
\simeq {1\over 2}M_{\rm NS}\xi^2\omega^2$ to the observed X-ray 
output of the flare.   One has 
\begin{equation}
-\left({dE_{\rm mode}\over dt}\right)_{\rm MDR} = {2\omega^4\over 3c^3}\,
{\cal M}^2\,\left({\xi\over R_{\rm NS}}\right)^4
\simeq {1\over 3}\,\left({E_{\rm mode}\over M_{\rm NS}}\right)\,
{B_{\rm NS}^2 R_{\rm NS}^4\omega^2\over c^3},
\end{equation}
where ${\cal M} = {1\over 2}B_{\rm NS} R_{\rm NS}^3$ is the magnetic
moment and $B_{\rm NS}$ the polar surface magnetic field.
For a mode frequency $\omega \sim 10^4$ Hz, $M_{\rm NS} = 1.4\,M_\odot$
and $R_{\rm NS} = 10$ km, this expression becomes
\begin{equation}
-\left({dE_{\rm mode}\over dt}\right)_{\rm MDR} = 
4\times 10^{40}\,\left({E_{\rm mode}\over 10^{44}~{\rm ergs}}\right)\,
\left({B_{\rm NS}\over 10^{15}~{\rm G}}\right)^2\,
\left({\omega\over 10^4~{\rm Hz}}\right)^2\;\;\;{\rm ergs~s^{-1}}.
\end{equation}
Using either calculation, the rate of dissipation fails by 
orders of magnitude to accomodate the observed output of the spike --
even if the surface field is in the magnetar range.

\section{Flare Spectra and Spectral Evolution}\label{spectra}

\subsection{Spectrum of the Hard Initial Spike: 
An Expanding Pair Fireball}\label{spikespectrum}

The first part of the August 27 event had a much harder
spectrum than the remainder of the outburst.  A 
very hard component $dN/dE \propto E^{-1/2}$ was measured
by BeppoSAX during the first 67 seconds, with 
a much softer power-law component $dN/dE \propto E^{-4.5}$
remaining during the following 128 seconds (Feroci et al.~1999; Paper I).  
The detection by Konus/Wind of a dramatic drop in the $> 250$ keV emission
following $\sim 0.7$ sec post-trigger (Mazets et al. 1999)
suggests that the very hard spectrum component
was mostly emitted during the initial spike.  

The large-amplitude fluctuations detected in $> 250$ keV photons 
at the end of the hard spike are also remarkable (Mazets et al. 1999).  
At $0.2-0.7$ sec after the event's onset, 
these fluctuations were were more profound in $> 250$ keV photons 
than in the 15--250 keV band, and occured on timescales as short
as $\sim 10^{-2}$ s.  This behavior is reminiscent of 
high-redshift, classical gamma-ray bursts (GRBs), in which the 
light curve is often much smoother at lower energies (Pendleton et al. 1997).  

The peak luminosity of the August 27 spike ($L > 4\times 10^{44}$
ergs s$^{-1}$;  Mazets et al. 1999) is intermediate, on a logarithmic
scale, between a thermonuclear flash and a high-redshift GRB.  
But we have much more reliable, direct 
information about its source, SGR 1900+14, than we do 
about the sources of cosmological GRBs.  Thus, detailed studies 
of the August 27 flare may pay dividends in understanding the more
mysterious classical GRB sources,  
even though the integrated spectrum of the August 27 flare is not
typical of a classical GRB.  

The volume of the emission region is strongly constrained by the
rapid variations in the $> 250$ keV flux.  Although
the hard power-law component is detected only up to $\sim 700$ keV
(Feroci et al. 1999), there is no evidence for a high energy cutoff.
Therefore, let us consider the consequences of a source function
that extends above an energy $\sim m_ec^2$ in the (relativistically)
expanding frame of the outflowing particles and magnetic field
that power the hard spike.  Pair creation will attenuate the flux
above a measured energy $\Gamma m_ec^2$, where $\Gamma$ is the bulk
Lorentz factor of the outflow.  Because the photon flux is dominated
by high-energy photons ($d{\cal N}/dE \propto E^{-1/2}$), one expects
copious pair creation,  and the luminosity in relativistic
pairs is comparable to the photon luminosity. Pairs then make the outflow
very thick to scattering, out to a considerable radius
\begin{equation}
R_{\tau=1} \sim \left({\Delta E\sigma_T\over 4\pi\Gamma m_ec^2}\right)^{1/2}.
\end{equation}
Here, $\Delta E \sim L\Delta t$ is the energy carried by an
{\it individual spike} of peak luminosity $L$ and width $\Delta t$.
In order for the spike not to be smeared out as the hot
shell of ejecta expands to a thickness $\sim R/2\Gamma^2$,
one requires $c \Delta t > R_{\tau = 1}/ 2 \Gamma^2$, or equivalently
\begin{equation}
\Gamma > \Gamma_{\rm min} = 
7\,\left({L\over 10^{43}~{\rm ergs~s^{-1}}}\right)^{1/5}\,
\left({\Delta t\over 0.01~{\rm s}}\right)^{-1/5}.
\end{equation}
Most of the dissipation that produces the hard non-thermal spectrum
must therefore occur at a large radius,
\begin{equation}\label{remission}
R > R_{\tau=1} \sim 3\times 10^{10}\,
\left({L\over 10^{43}\,{\rm ergs~s^{-1}}}\right)^{2/5}\,
\left({\Delta t\over 0.01~{\rm s}}\right)^{3/5}\,
\left({\Gamma\over\Gamma_{\rm min}}\right)^{-1/2}\;\;\;\;\;\;{\rm cm}.
\end{equation}

The emission region of the non-thermal spectral component is
very large, but comparable to the speed-of-light cylinder of 
SGR 1900+14 ($cP/2\pi = 2.5\times 10^{10}$ cm).  It sits far
enough out that direct synchrotron cooling off the dipole field
of the source can be neglected --- even if the field lines are
combed out into a $\sim R^{-2}$ geometry by the pressure of
the escaping particles and radiation.   By contrast, 
the ejecta from the outburst
could carry a signficant magnetic field.  For example, 
$B \sim R^{-1}$ if $\Gamma$ grows linearly with radius -- as 
is expected in an expanding fireball which contains a magnetic field
but whose pressure is predominantly thermal at the source.
Alternatively, inverse-Compton cooling by advected X-ray photons
will still be a powerful coolant at a radius (\ref{remission}).
The cooling time of a particle of random energy
$\gamma_em_ec^2$ in the bulk frame is $t_{\rm Compton} \sim
\gamma_e^{-1}(\sigma_T L/4\pi R^2\Gamma^2 m_ec^2)^{-1}$, and so the ratio of
the cooling time to the flow time 
\begin{equation}
{t_{\rm Compton}\over R/c\Gamma} \sim \gamma_e^{-1}\,\left({R\over
R_{\tau=1}}\right)\,\left({\Gamma\over\Gamma_{\rm min}}\right)^{5/2}.
\end{equation}

\subsection{The Pulsating Tail Spectrum: Temperature Regulation through 
Photon Splitting}\label{split}

We now consider the temperature of the radiation escaping from
a trapped fireball, and compare it with the best (blackbody)
temperature fit to the pulsating tail of the August 27 flare.
The temperature appears to have remained
remarkably constant during the phase of large-amplitude pulsations, 
even as the X-ray flux continued to decline (Mazets et al. 1999).
The best two-component spectral fit (black body $+$ powerlaw)
yields $kT_{bb} =  10.8$ keV during interval B (Table 2 of Paper I) 
and 12.2 keV during interval C (Table 3 of Paper I).  These values
are consistent with the {\it minimum} photospheric temperature
of a trapped fireball in super-QED magnetic fields, as calculated by TD95.

Near the photosphere, the spectral shape is determined by
two coupled processes:  Compton scattering and the creation of new
photons through splitting $\gamma \rightarrow \gamma + \gamma$ (TD95).  
The photons flowing out from the base of the trapped fireball are 
expected to dominate the specific heat and so, in the absence of an 
extended pair corona, the mean energy per photon will remain approximately
constant {\it outside} the fireball photosphere.
Within the outer layers of the optically thick fireball, 
the photons can maintain a Planckian distribution at temperatures well
below the internal fireball temperature of $\sim 1$ MeV.
The mean energy of the escaping photons depends directly on the 
splitting rate as a function of frequency.  In marked contrast with the
strong $B^6$ scaling of the splitting rate in sub-QED magnetic fields, 
the splitting rate approaches a $B$-independent value in fields much
stronger than $B_{\rm QED}$,
\begin{equation}\label{gamsp}
\Gamma_{\rm sp}(\omega,B,\theta_{kB}) =
{\alpha_{\rm em}^3\over 2160\pi^2}\,\left({m_ec^2\over\hbar}\right)\,
\left({\hbar\omega\over m_ec^2}\right)^5\,\sin^6\theta_{kB}
\end{equation}
(Adler 1971; Thompson \& Duncan 1992).  This implies immediately that 
an E-mode photon propagating a distance $R_{\rm NS} \sim 10$ km through 
a super-QED B-field will split if $\hbar\omega > 
38\,(R_{\rm NS}/10~{\rm km})^{-1/5}$ keV (TD95; Baring 1995).

Compton scattering becomes strongly anisotropic in
a background magnetic field, with a frequency-dependent
cross-section.
Near the surface of the star, the energy of the first Landau excitation
[about $(2B/B_{\rm QED})^{1/2}\,m_ec^2$] is much higher than the
temperature of the emerging X-rays.  In this situation, there is
a strong suppression of the E-mode's scattering cross-section:
$\sigma_E = (\omega m_e c /eB_0)^2\,\sigma_T$ (e.g. Herold 1979).
By contrast, the O-mode scatters with a cross-section near Thomson
(except for propagation almost parallel to the background field).
This suppression of the E-mode scattering opacity greatly increases
the radiative transport rate close to the neutron star.  One has
\begin{equation}\label{taueval}
{\tau_E(T_e)\over \sigma_T (n_{e^+}+n_{e^-}) R}
 = 5\pi^2\,\left({kT_e\over m_ec^2}\right)\,
\left({B\over B_{\rm QED}}\right)^2,
\end{equation}
when the dielectric properties of the medium are dominated 
by vacuum polarization (Silan'tev \& Iakovlev 1980; TD95).
First considered as a way to enhance emission from a 
hydrostatic atmosphere that is confined by gravity 
at the surface of an ultra-magnetized neutron star
(Paczy\'nski 1992; Ulmer 1994; Miller 1995), the suppression
of E-mode scattering has potentially more direct
applications to radiative transport {\it across} the confining
magnetic field lines of a trapped fireball (TD95).  

The two polarization modes are also distinguished by their ability
to split.  Only the E-mode can
split because -- when vacuum polarization dominates the dielectric
properties of the medium -- only the energy and momentum of E-mode
photons can be conserved by dividing into two obliquely propagating daughter
photons.  Splitting occurs predominantly via 
\begin{equation}
E \rightarrow O + O,
\end{equation}
with $E \rightarrow E + O$ having a lower rate (e.g. Berestetskii 
et al. 1980).  
When the photon distribution is close to Planckian, net transport
out of the E-mode by splitting is suppressed by the inverse process
of merging $O + O \rightarrow E$ (TD95).
A cascade of photons from high to low X-ray energies
cannot occur purely through splitting, as conjectured
by Baring (1995), even in magnetic fields as strong as $\sim 10^{16}$ G.  

Even in the region where the E-mode is able to stream freely,
the O-mode can still undergo many Compton scatterings and relax close to a
Bose-Einstein distribution.  This permits a very simple generalization
of the LTE diffusion formalism to an anisotropic, magnetized plasma at
large E-mode scattering depth, in which the photon energy and number
fluxes are expressed as linear superpositions of gradients in the
temperature $T$ and photon chemical potential $\mu$ (TD95).  
As a result, there is a critical temperature above which the distributions
of the E- and O-modes both become thermal, which works out to
\begin{equation}
kT_{\rm sp} = 11\,\left({R_{\rm NS}\over 10~{\rm km}}\right)^{-1/5}
\;\;\;\;\;\;{\rm  keV}
\end{equation}
(eq.~[133] in TD95).  This value
matches well the best fit temperatures for the August 27 event.    The mean
energy per escaping photon cannot fall below $\sim 2.7 kT_{\rm sp}$, in
the absence of new photon creation processes or strong adiabatic cooling.

Now let us consider the spectral variations within each 5.16-s rotation
period.  The emission
is harder at the peaks than in the valleys (see the hardness ratio
plots in Fig. 7a of \cite{mazets99}).  A positive correlation between
hardness and intensity is expected of thermal emission from an optically
thick plasma.   Nonetheless, the large-amplitude oscillations show overall
an anti-correlation between hardness and intensity (Paper I).
A deep trough is present in each rotation cycle after $\sim 40$ s, 
and plausibly involves occulation of the trapped
fireball by the neutron star.
The spectrum evolves from hard to soft across the trough, which
contributes
to an overall sawtooth-like pattern in the hardness ratio
(see Fig. 1 in \cite{feroci99}).  
This hard-to-soft evolution suggests a significant deviation
from mirror symmetry in the confining magnetic field with respect to the
center of the trough.

\section{The Smooth $\sim 40$ Tail: A Continuously Heated Pair Corona}
\label{smoothdecay}

The giant flares share two main features:  an extremely bright, hard,
and brief $\sim 0.5$-s transient at the onset of the burst; and an 
extended train of softer, large-amplitude pulses which repeat coherently
at the rotational frequency of the star.   The initial spike has 
the properties expected 
of an expanding, pair-dominated fireball;  and, as shown in Paper I, 
the declining amplitude of the pulsations in the August 27 flare can 
be well fit by the contracting surface of a trapped fireball.  

Nonetheless, this simple picture of a sudden release of thermal energy within
the first $\sim 1$ s, part of which escapes directly and the rest of
which remains confined close to the star, does not appear to provide a
complete description of the August 27 flare.  Within the first $\sim 40$
seconds, the X-ray flux declines smoothly and shows only a mild 
modulation at the 5.16-s spin of the source (Figs. 6 and 7 of Paper I). 
The flux {\it exceeds} the favored trapped fireball model during
this interval (Fig. 3 of Paper I), and the spectral
hardness decreases by a factor $\sim 1.5$ before flattening out after
$\sim 40$ s (Mazets et al. 1999).  This behavior
was ascribed in Paper I to an extended, pair-dominated corona that
scatters and
heats the thermal radiation emerging from the inner, trapped fireball
(Fig. \ref{hotcorona}).

This $\sim 40$ s coronal component of the August 27 flare emitted
about 20 percent of the total energy (excluding
the hard $\sim 0.5$ s initial spike; \S 6.3 of Paper I).  
At $t\sim 15$ s, the total luminosity was twice
the envelope of the trapped fireball, which according to eq. (\ref{lxt})
at that time hardly deviates from its initial value
$L_X(0) = (1 + \chi) E_{\rm fireball}/t_{\rm evap}$.  
Here $\chi \simeq 3$ is the fireball index, $E_{\rm fireball}$
the initial fireball energy, and $t_{\rm evap} \simeq 375$ s the 
time at which the fireball evaporates.  

The cooling time of this corona is very short given the high
burst luminosity.  For an optical depth to scattering
$\tau_{es} \ga 1$ (as required by the radiative model discussed 
below) the cooling time is set by radiative diffusion to
a magnitude $\sim \tau_{es} R_{\rm corona}/c
= 10^{-3}(\tau_{es}/10)(R_{\rm corona}/30~{\rm km})$ s. 
Thus, the coronal heating must be almost continuous.  It could be
powered in at least two ways:  through a persistent creep of the 
crust at the heated fracture site, which results in continued 
shearing of the external field;  or possibly by a coupling of
internal shear oscillations to the magnetosphere (\S \ref{torseis}).
The flare light curve does not provide a simple diagnostic between these
alternatives.

\subsection{Steady Compton Heating of the O-mode}

Why should a continuing release of seismic energy lead to the formation
of a distinct scattering
atmosphere, instead of merely adding energy to a trapped fireball which
forms during the first $\sim 0.5$ s of a giant flare?  We now show that
there is a {\it critical rate} for injection of energy into the trapped
fireball, above which the energy becomes trapped before it can leak
out radiatively;  but below which a steady balance between heating and
diffusive radiative cooling is possible.  This critical luminosity
depends on the nature of the photon source.  When double-Compton
emission dominates, the answer is given in Thompson (1997).  In the
present
context, photon splitting $\gamma \rightarrow \gamma + \gamma$ is the
most
effective source of new photons above a critical temperature $T \sim
11$ keV (TD95)   

We first summarize the properties of a pair corona with a temperature
$kT_e \ll m_ec^2$ in a very strong magnetic field $B \ga B_{\rm QED}$.
A realistic heating mechanism involves electrostatic acceleration
of the pairs through a turbulent cascade of interacting Alfv\'en waves:
the current density is driven
just above the value that can be supported self-consistently by the
available particles, and a displacement current is induced parallel to the
background magnetic field (Thompson \& Blaes 1998).  

The pairs cool primarily by
Compton upscattering the O-mode, due to its much larger cross section. 
Although at large scattering depth 
radiation diffuses fastest via the E-mode, the orthogonal O-mode will be
regenerated\footnote{Indeed, the inability of high
energy O-mode photons to split, combined with their rapid Compton heating 
by pairs, provides a mechanism for generating the non-thermal X-ray 
spectra of SGRs in their quiescent (non-bursting) state.}
by scattering (with cross section 
$\tau_{EO} \simeq \tau_{E}$; M\'esz\'aros 1992)
and by splitting (TD95; Miller 1995).  Moreover, in some circumstances
the optical depth may be too small to allow effective mode transfer
$O \rightarrow E$ by scattering.  (Conversion of the $O$-mode to the
$E$-mode does not occur by splitting, except at enormous optical depths
where the plasma dominates the dielectric properties of the medium.)

In the following discussion, we will assume that the electrostatically
heated pairs are sub-relativistic and pinned in the lowest Landau
level.  Indeed, at the very high compactness 
($\ell \equiv L_X\sigma_T/4\pi m_ec^3
R_{\rm NS} \sim 10^6$) characteristic of the August 27 smooth tail, the pairs
will equilibrate at a temperature close to that of the O-mode photons.
Since the effective temperature of the
photons is $kT_O \sim 10-20$ keV, the spectral distribution of
Comptonized photons will have an exponential cutoff $F_\nu \propto
\exp(-h\nu/k T_e)$ at high energies $h\nu \ga kT_e$. 

The question which we address here is:  what is the range of temperature
for which the pair plasma is stable to an upward excursion in the density
$n_{e^+} + n_{e^-} \simeq 2n_{e^+}$ of electrons and positrons?
An accompanying increase in the pair temperature $T_e$ forces a
runaway of the pair density (due to the higher flux of Comptonized
photons
above the pair-creation threshold $h\nu \sim m_ec^2$).  This runaway
is inevitable if the pair plasma is in local thermodynamic equilibrium 
(TD95).

If, instead, $T_e$ decreases when $n_{e^+}$ increases, then a stable
balance
between electrostatic heating and diffusive photon cooling is possible. 
Then the distribution functions of E-mode and O-mode are both approximately
Wien.  Such a steady balance requires a continuous source of O-mode photons,
which can be provided either internally by splitting the E-mode, or
externally by scattering incident E-radiation to the O-mode.\footnote{Double
Compton scattering of the O-mode --- $O + e^\pm \rightarrow O + O + e^\pm$ ---
is an effective source of new photons only at larger scattering depths
than are needed to convert the E-mode.}  Splitting can be expected to
dominate where the magnetic field is stronger than $B_{\rm QED}$;  but
an external source of E-mode photons is necessary in a more extended
corona around a bursting magnetar.  We consider each case in turn.

\subsection{Equilibrium Temperature of the Corona}\label{cortem}

We first consider the equilibrium temperature of the corona,
given a steady balance between the rate $L_{\rm corona}$ at which the 
pairs are electrostatically heated, and radiative transport out of the
magnetosphere.  The result depends on the mechanism by which radiative
transport occurs (through the E-mode or the O-mode).
We idealize the corona as a sphere of radius $R$.

If the source of photons lies below the corona -- i.e. in the
form of a cooling trapped fireball -- then photons diffuse
into the corona predominantly via the E-mode, before being converted 
to the O-mode and heated.  One deduces that $\tau_{EO} \ga 1$ (and that
energy is also transported {\it out} of the corona through the E-mode).
In addition, the scattering depth to the O-mode is very large,
\begin{equation}
{\tau_O\over\tau_{EO}} \sim {1\over 5\pi^2}\,
\left({B\over B_{\rm QED}}\right)^2\,
\left({kT_E\over m_ec^2}\right)^{-2}.
\end{equation}
(Here $T_E$ is the temperature of the {\it external} photon source, 
and we have made use of eq. \ref{taueval}.)  Fresh O-mode
photons are rapidly Compton heated up to a Wien distribution with
temperature $T_O \simeq T_e$ and mean energy per photon $3kT_e$.  
Furthermore, the E-mode and O-mode photons are in equilibrium
at large $\tau_{EO}$, and one has $T_E \simeq T_e$, $n_E \simeq n_O$.

The balance between heating and diffusive radiative cooling becomes
\begin{equation}\label{lcas2}
{L_{\rm corona}\over 4\pi R^2c } = 
{3kT_E n_E\over 3 (2n_{e^+}\sigma_E(T_E) R)}
\end{equation}
As just described, such a steady balance requires that photon creation
be too slow to establish local thermodynamic equilibrium.  The
densities of photons and pairs are then related by
\begin{equation}
{n_{e^+}\over n_{e^+}(LTE)} \simeq {n_O\over n_O(LTE)}.
\end{equation}
The corresponding LTE densities are 
\begin{equation}\label{nplte}
2n_{e^+}(LTE) =  {(m_ec)^3\over\hbar^3 (2\pi^3)^{1/2}}\,
\left({B\over B_{\rm QED}}\right)\,\left({kT_e\over m_ec^2}\right)^{1/2}
\,\exp\left(-{m_ec^2\over kT_e}\right),
\end{equation}
and
\begin{equation} 
2n_O(LTE) = n_O(LTE)+n_E(LTE) = 0.244\,\left({kT_e\over\hbar c}\right)^3
\end{equation}
at temperature $T_e$.   The equilibrium
temperature is then determined from eqs. (\ref{taueval}) and (\ref{lcas2}),
\begin{equation}\label{lcasa}
{L_{\rm corona}\over 10^{34}\;{\rm ergs~s^{-1}}}
= \left({kT_e\over m_ec^2}\right)^{3/2}\,
\exp\left({m_ec^2\over kT_e}\right)\,\left({B\over B_{\rm QED}}\right)\,
\left({R\over 10~{\rm km}}\right).
\end{equation}
The temperature corresponding to the brightest, short SGR outbursts 
($\sim 10^{42}$ ergs s$^{-1}$) in super-QED fields ($B/B_{\rm QED} \sim 
10$) is $kT_e \sim 25$ keV, which increases only to $\sim 50$ keV at
$L_{\rm corona} \sim 10^{38}$ ergs s$^{-1}$ (Fig. \ref{tpair}).

It is also possible that the photon source is {\it internal} to the
corona (e.g. splitting).  The corona can then maintain a much lower scattering
depth while cooling diffusively through the O-mode -- without any need for
converting photons back to the E-mode.  In this case, the balance between
heating and cooling becomes 
\begin{equation}\label{lcasb}
{L_{\rm corona}\over 4\pi R^2c} \sim {3kT_O n_O\over 3 (2n_{e^+}\sigma_T R)},
\end{equation}
since $\sigma_O \sim \sigma_T$ after averaging over angles.  
The equilibrium temperature is now determined by
\begin{equation}\label{lcasc}
{L_{\rm corona}\over 10^{34}\;{\rm ergs~s^{-1}}} 
= 44\,\left({kT_e\over m_ec^2}\right)^{7/2}\,
\exp\left({m_ec^2\over kT_e}\right)\,\left({B\over B_{\rm QED}}\right)^{-1}\,
\left({R\over 10~{\rm km}}\right).
\end{equation}
Notice the different dependence on both $T_e$ and $B/B_{\rm QED}$. 
In this case, the temperature corresponding to $L_{\rm corona} \sim 10^{42}$
ergs s$^{-1}$ in a magnetic field $\sim 10\,B_{\rm QED}$ is slightly lower,
$kT_e \sim 20$ keV, increasing to $kT_e \sim 30$ keV at $L_{\rm corona}
\sim 10^{38}$ ergs s$^{-1}$ (Fig. \ref{tpair}).

\subsection{Critical Luminosity:  Photons Created by Splitting}

We now look for the maximum energy input to the corona that allows
a steady balance between heating and diffusive cooling.  The key
criterion is that the equilibrium pair density $n_{e^+}$ be smaller
than the LTE value (\ref{nplte}).  

Photon splitting will occur within the bulk of the corona.
We can assume that photon transport out of the corona is through
the O-mode,  because splitting converts the E-mode to the O-mode 
(but not the reverse).  As a result, transport cannot occur self-consistently 
via the E-mode.  A high temperature $kT_O \ga 20$ keV is needed to
generate sufficient pairs to scatter the O-mode back to the E-mode
($\tau_{OE} \ga 1$);  but because this temperature is well above the
critical value ($\sim 11$ keV) where splitting is rapid, 
the transport of E-mode photons will be limited
by splitting, $L_{\rm corona}/4\pi R^2c \sim 3kT_E n_Ec/\Gamma_{\rm sp} R$.  
In this situation, it would not be possible to maintain a steady balance
between the creation and the outward diffusion of photons, because 
the rate of creation of photons $\dot N_\gamma
\sim 4\pi n_E R^3 \Gamma_{\rm sp}$ would exceed $L_{\rm corona}/3kT_E$ 
by a factor $\sim (\Gamma_{\rm sp} R/c)^2 \gg 1$. 

Instead, the radiative flux out of the corona is carried by
the O-mode. A steady balance between photon creation and loss
requires that each diffusing O-mode photon have a substantial probability 
($\sim {1\over 2}$) of converting to the E-mode and splitting.\footnote{Once
again, we must assume that the photon density is well below the
black body value value at temperature $T_O$ to obtain a steady
balance between heating and cooling.  Thus, the inverse
process of photon merging can be neglected.}  Thus
$\tau_{EO}\tau_O \sim 1$, which corresponds to
\begin{equation}\label{tauoval}
5\pi^2\,\left({kT_e\over m_ec^2}\right)^2\,
\left({B\over B_{\rm QED}}\right)^{-2}\,\tau_O^2 \sim 1.
\end{equation}
But 
\begin{equation}\label{tauovalb}
\tau_O \sim (2n_{e^+})\sigma_T R = 1.5\left({n_{e^+}\over 
n_{e^+}(LTE)}\right)\,\,\left({L_{\rm corona}\over 10^{42}~
{\rm ergs~s^{-1}}}\right)^{-1}\,\left({kT_e\over 20~{\rm keV}}\right)^4\,
\left({R\over 10~{\rm km}}\right)^2
\end{equation}
(making use of eqs. [\ref{nplte}] and [\ref{lcasc}]).  The bound 
$n_{e^+} < n_{e^+}(LTE)$ then implies, together with eq. (\ref{tauoval}),
\begin{equation}\label{lmin}
L_{\rm corona} < L_{\rm corona}({\rm max}) = 
4\times 10^{41}\,\left({B\over B_{\rm QED}}\right)^{-1}
\,\left({\rm kT_e\over 20~{\rm keV}}\right)^5\,
\left({R\over 10~{\rm km}}\right)^{-2}\;\;\;\;\;\;{\rm ergs~s^{-1}}.
\end{equation}
A stable balance between heating and diffusive cooling is possible
{\it below} the luminosity $L_{\rm corona}({\rm max})$.  
In such an equilibrium, 
the electron temperature is strongly buffered and remains
close to 20 keV as $L_{\rm corona}$ varies by a few orders of
magnitude around $\sim 10^{42}$ ergs s$^{-1}$ (eq. [\ref{lcasc}]).  
The right side of eq. (\ref{lmin}) scales as $\sim R$ in a dipolar
magnetic field.

This equilibrium between heating and cooling is steady, as it is
straightforward to check.  A rise in $n_{e^+}$ implies an increase
in $\tau_O$, which must be compensated by a decrease in $\tau_{EO}$
to maintain the relation $\tau_O\tau_{EO} \sim 1$ (eq. \ref{tauoval}).
By contrast, when the fireball is in local
thermodynamic equilibrium, an increase in $n_{e^+}$ is directly tied to
an increase in $T$.  In such a situation, continuing energy input will
force a runaway increase in the energy density to a trapped fireball
(TD95).

\subsection{Critical Luminosity:  External Source of Photons}

Now consider the case where the corona is sufficiently extended that
the magnetic field is weaker than $B_{\rm QED}$, and photon splitting 
can be neglected over most of its volume.   Diffusive cooling
now occurs through the E-mode, and a steady balance between heating
and cooling fixes the electron temperature through expression (\ref{lcas2}).
Given a source $\dot N_E^{\rm ex}$ of fresh E-mode photons 
(of temperature $T_E^{\rm ex}$) escaping from the
trapped fireball, the optical depth through the corona is fixed.
In the regime of interest, $\tau_{EO} > 1$, and the soft fireball photons
are heated by only a fraction $\sim \tau_{EO}^{-1}$ of the coronal input
$L_{\rm corona}$:
the equilibrium pair temperature is fixed by energy conservation.
\begin{equation}\label{teeq}
\dot N_E^{\rm ex} 
\times 3(T_E - T_E^{\rm ex}) = {L_{\rm corona}\over \tau_{EO}}.
\end{equation}
(We neglect the difference between the mean energy
per photon in the blackbody and Wien distributions.
This expression is approximately
valid even for low $\tau_{EO}$, where only a fraction $\sim\tau_{EO}$
of the fireball luminosity is captured by the corona.)
Combining this relation with $L_{\rm corona} = (3kT_E)\dot N_E^{\rm ex}$
gives
\begin{equation}
\tau_{EO} \simeq \left(1-{T_E^{\rm ex}\over T_E}\right)^{-1}
= O(1).
\end{equation}

At the high luminosities of SGR outbursts, the pairs in the outer corona 
equilibrate to a temperature only slightly different from that of the
heated O-mode photons.  During repeated scatterings, the mean energy
per photon exponentiates as $e^y$, where the Compton parameter is 
\begin{equation}
y \sim 4\tau_O^2{k(T_e-T_O)\over m_ec^2}.
\end{equation}
In equilibrium,
\begin{equation}
y = \ln\left(1+{L_{\rm corona}\over L_{\rm fireball}}\right),
\end{equation}
which implies
\begin{equation}
{T_e\over T_O} - 1 \sim {y\over 4\tau_O^2}\left({m_ec^2\over
kT_O}\right) \ll 1.
\end{equation}

We can now determine the critical luminosity below which 
a steady balance between heating and cooling is possible.
One has, from equations (\ref{lcas2}) and (\ref{nplte}),
\begin{equation}
\tau_{EO} \sim (2n_{e^+})\sigma_{E}(T_E) R
= 1.5\left({n_{e^+}\over 
n_{e^+}(LTE)}\right)\,\,\left({L_{\rm corona}\over 10^{42}~
{\rm ergs~s^{-1}}}\right)^{-1}\,\left({kT_e\over 20~{\rm keV}}\right)^4\,
\left({R\over 10~{\rm km}}\right)^2.
\end{equation}
This expression is very similar to (\ref{tauovalb}), with the
distinction that diffusive transport is through the E-mode instead
of the O-mode.  As a result,  $T_e$ is slightly higher than in the 
case where fresh photons are created by splitting (but is still 
strongly buffered through eq. [\ref{lcasa}]).   The critical luminosity 
is determined by setting $n_{e^+} < n_{e^+}(LTE)$:
\begin{equation}\label{lminb}
L_{\rm corona} < L_{\rm corona}({\rm max}) = 
1.5\times 10^{42}\,\left(1-{T_E^{\rm ex}\over T_E}\right)\,
\left({\rm kT_e\over 20~{\rm keV}}\right)^4\,
\left({R\over 10~{\rm km}}\right)^{-2}\;\;\;\;\;\;{\rm ergs~s^{-1}}.
\end{equation}

As before, this equilibrium is steady as long as the densities of O-mode and
(heated)
E-mode photons lie below the blackbody value at temperature $T_e \simeq
T_O$.
That is because a rise in $n_{e^+}$ increases $\tau_{EO}$, and hence
{\it reduces}
the equilibrium (Wien) temperature to which the escaping fireball
photons will
be heated (eq. [\ref{teeq}]). 

After the coronal heating rate $L_{\rm corona}$ drops below the 
luminosity of the cooling fireball,
this pair atmosphere evaporates and the scattering photosphere contracts
to the outer
boundary of the trapped fireball, where the opacity is dominated by
ion-electron
plasma (TD95).  This provides an explanation
for the flattening of the light curve simultaneously with
the appearance of large-amplitude oscillations.  We outline
the effects of the strong magnetic field on this Compton corona
in the next section.

\subsection{Very High Current Densities}

In the preceding, we have assumed that the pairs are heated 
electrostatically by a fluctuating current, and
derived the critical heating rate above which the pair-photon
plasma runs away to local thermodynamic equilibrium.  There is,
similarly, a critical current density $J$ above which photons will 
remain trapped in the current-carrying region.  The energy density 
of the photons exponentiates, thereby creating a pair plasma in local 
thermodynamic equilibrium.

The optical depth to the O-mode through the current-carrying charges is
\begin{equation}
\tau_O \sim (n_{e^-}+n_{e^+})\sigma_T R \sim {J\over eV}\sigma_T R,
\end{equation}
where $R$ is the size of the current-carrying region.
In a pure pair plasma, with equal numbers of charges moving in
opposite directions along the magnetic field lines, the mean frequency
shift per scatter is second order in the drift speed $V$ of the
current carriers, $\Delta\nu/\nu \sim (V/c)^2$.
In equilibrium, the temperature of the photons and (1-dimensional) pairs is
\begin{equation}
kT_O \simeq \left({V\over c}\right)^2.
\end{equation}
When $\tau_O > 1$, each O-mode photon undergoes $\sim \tau_O^2$
scatterings, and the Compton parameter 
\begin{equation}
y \sim \tau_O^2\,\left({V\over c}\right)^2 \sim
\left({J\sigma_T R\over ec}\right)^2
\end{equation}
depends only on the current density $J$ and the size $R$ of the `corona'.  
However, $T_O$ is strongly buffered by pair creation and annihilation
(\S \ref{cortem}).  Thus, above a critical current density $J$,
the scattering depth becomes too large to allow the photons to
diffuse out before their {\it number} and energy density multiplies.
For example, in the case where new photons are created by splitting,
the critical current density corresponds to an optical depth 
$\tau_O$ equal to (\ref{tauoval}).

\subsection{Alternative Models for the Smooth Tail}

Let us consider two alternative explanations for the smooth tail:  a
steady
increase in the ion-electron loading at the fireball surface;  and
neutrino
cooling.  Each of these alternatives fails in some manner.

The portion of the neutron star surface that is exposed
to the hot fireball will drive a super-Eddington wind as the fireball
begins
to contract (TD95).  This could force a steady increase in the density
of ions and
electrons, supported above the surface by the photon pressure.  At
high enough temperatures the radiative flux from the surface of the
fireball is fixed
self-consistently by the opacity of this suspended matter (TD95):
${1\over 2}\sigma_{SB}T_{eff}^4 \simeq
F_{\rm edd}\,[\sigma(T_{eff},B)/\sigma_T]^{-1}\,
\varepsilon_{ion}\,(\ell/R_{\rm NS})^{-1}$.  Here, $F_{\rm edd}$ is the
Eddington
flux
calculated with the Thomson cross-section; $\sigma(T_{eff},B)/\sigma_T
= 5\pi^2[kT/(\hbar eB/m_ec)]^2$ is the scattering opacity in the
strong
background
magnetic field; $\varepsilon_{ion}$ is the ion density compared with the
maximum
that can be supported against gravity; and $\ell$ is the thickness of
the ion-electron layer at the photosphere.  As the fireball contracts
and
$\ell$ increases (due to the increasing exposure of heated crust)
the radiative flux decreases as $\ell^{-2/3}$ and the effective
temperature decreases
as $T_{eff} \propto \ell^{-1/6}$.  Although this variant offers a nice
explanation for
the decrease in radiative flux and temperature during the smooth $\sim
40$ s tail,
it has difficulty accomodating the {\it contraction} of the photosphere
that is
required by the transition to large-amplitude pulsations.

Neutrino cooling does not
change the lightcurve of a homogeneous fireball in the manner needed
to explain the $\sim 40$-s smooth tail (Figs. \ref{neutrinoa},
\ref{neutrinob}).
The fireball could, of course, have two
components, one of which has a much higher temperature $T_{high}$ and
emits
neutrinos much more rapidly
than the cooler component.  Then $T_{high} \propto t^{-1/5}$ and
the radiative flux from the contracting surface of the fireball
decreases as\footnote{This applies when the higher Landau levels are populated;
eq. (89) of TD95.}  $F \propto T_{high}^{5/2} \propto t^{-1/2}$. 
However, it is difficult to understand why the energy radiated
from the higher temperature component should be much smaller (by a
factor $\sim 20$),
even while it covers a much larger volume (so as to explain the absence
of large-amplitude pulsations).

\section{The Four-Peaked Repetitive Pattern:  Collimated X-ray Jets}
\label{Xjets}

A dramatic four-peaked pattern emerged in the August 27 light curve
at $\sim 40$ s following the burst trigger.   This pattern repeated
with the 5.16-s rotation period of the neutron star.  At times, the flux
varied by more than an order of magnitude from peak to trough.
However, there is
no significant break in the pulse-averaged light curve during the
appearence of the four-peaked pattern  (Fig. 4 of Paper I),
which indicates that the beamed flux is redistributed over the pulse
period.

The phase stability of the X-ray jets suggests that they are tied
to surface features on the neutron star.  Indeed, it has previously
been argued that the radiative flux out of a trapped fireball could
become significantly collimated along the magnetic field lines
that open out to a few neutron star radii (TD95).  Figure \ref{xrayjet} 
outlines the basic geometry.
A $10^{14}-10^{15}$ G magnetic field is certainly
strong enough:  its pressure exceeds the radiative
momentum flux by some 9-10 orders of magnitude.   In this picture,
the approximate $\sim 1$-s periodicity apparent in the four-peaked 
pattern is
a chance byproduct of the location of the outburst, and the multipolar
structure of the neutron star's surface magnetic field (Paper I; Fig. 
\ref{xrayjetb}).  The difference in the number of sub-pulses observed
within each rotation during the August 27 and March 5 giant flares,
can then be ascribed to a difference in the number of xray jets.  
The most plausible geometry for each jet is a fan beam, which is
swept past the line of sight once or twice by the rotation of the star.  
Figure \ref{xrayjetb} illustrates one possible geometry for each event, 
where each fan beam is observed twice as two separate sub-pulses.

The large disparity between the scattering cross sections of the E-mode
and O-mode is the underlying reason for collimation.  In addition,
the scattering opacity of the E-mode rises rapidly with radius,
$\sigma_E \propto B^{-2} \propto R^6$ in a dipole geometry.
As a result, radiative transport across the magnetic field lines is
concentrated
close to the neutron star surface (TD95).  The trapped fireball heats
a thin outer skin of the neutron star crust, and as the fireball
contracts the cooling flux of X-rays drives matter off its
surface (Ibrahim et al. 2000).  This ablated material is easily
suspended by Compton scattering\footnote{By non-resonant scatterineg
off the O-mode, or by resonant scattering at the ion cyclotron fundamental
(which lies in the X-ray range in $\sim 10^{14}-10^{15}$ G magnetic
fields; Ibrahim et al. 2001).} in the magnetosphere, where it remains confined.

The rapid growth of the E-mode opacity then provides a mechanism for 
self-collimation:  the E-radiation can escape only by pushing
the suspended matter to the side (Fig. \ref{xrayjet}).  
The ion-electron photosphere of the
fireball is congruent with a set of magnetic field lines, as the result
of pressure gradient forces along the field. (The cooling time exceeds
the sound-crossing time of the fireball by a factor of a million.)
This means that the collimation
occurs primarily along magnetic field lines that open out beyond the
electron-ion photosphere.

The width of each X-ray jet depends on the amount of matter
advected
with the photons, and can be estimated as follows.  
A significant fraction of the E-mode flux near
the E-mode photosphere is converted to the O-mode by photon
splitting  (TD95). Mode changing
also occurs via non-resonant Compton scattering near the E-mode photosphere
(\cite{miller95}; TD95), as well as by resonant Compton scattering
near the ion cyclotron line (Thompson 2000).
The O-mode photons have scattering cross-sections near Thomson; i.e., the
scattering of O-mode photons is {\it not} significantly suppressed by the
strong
magnetic field.  Thus, the energy flux injected into the O-mode
near the neutron star surface is tremendously super-Eddington,
both due to the large luminosity $L_O \sim {1\over 2}L_X$ and
the small beaming angle $\Delta\Omega_{\rm jet}$:
\begin{equation}
F_O = {L_O\over R_{\rm NS}^2\Delta\Omega_{\rm jet}} 
\sim 3\times 10^3\,\left({L_X\over 10^{42}~{\rm ergs~s^{-1}}}\right)
\,\left({\Delta\Omega_{\rm jet}\over 4\pi}\right)^{-1}.
\end{equation}
The O-mode flows hydrodynamically
along the magnetic field even in the presence of a tiny amount of
matter, which can generate a large scattering depth along the magnetic field:
\begin{equation}\label{taupar}
\tau_\parallel(R_{\rm NS}) \sim {\dot M\sigma_T\over R_{\rm NS}
\Delta\Omega_{\rm jet}m_p c}
= \left({\dot Mc^2\over L_O}\right)\,\left({L_O\over L_{\rm edd}}\right)
\,\left({GM_{\rm NS}\over R_{\rm NS}c^2}\right)\,\left({\Delta\Omega_{\rm jet}
\over 4\pi}\right)^{-1}
\end{equation}

The resulting radiatively-driven outflow is probably relativistic 
(\S 6.4 of TD95). 
Indeed, it quickly becomes relativistic if the photons dominate
the specific heat and their expansion is adiabiatic.
The conserved enthalpy per photon can be expressed in terms
of the (bulk frame) photon density $n_\gamma$ and the 
bulk Lorentz factor $\gamma$ as
$\gamma (\hbar c)n_\gamma^{1/3} \sim kT(R_{\rm NS})$; and the
rate at which photons are carried through the jet is
\begin{equation}
\dot N_\gamma = (\gamma n_\gamma)\,\Delta\Omega_{\rm jet}\,R^2c \sim 
\left[{T(R_{\rm NS})\over \hbar c}\right]^3\,
R_{\rm NS}^2\,\Delta\Omega_{\rm jet}(R_{\rm NS}) c.
\end{equation}
Combining these two expressions gives
\begin{equation}
\gamma \sim \left({\Delta\Omega_{\rm jet} R^2\over \Delta\Omega_{\rm jet}
(R_{\rm NS}) R_{\rm NS}^2}\right)^{1/2},
\end{equation}
which is $\gamma \sim R/R_{\rm NS}$ in spherical geometry but
\begin{equation}
\gamma \sim \left({R\over R_{\rm NS}}\right)^{3/2}
\end{equation}
in an outflow that is channeled along a dipolar magnetic field\footnote{In
this geometry, the cross-sectional area of a flux bundle increases 
as $\sim R^3$.}.

Photons travelling along field lines, with 
the angle between wavevector ${\bf k}$ and ${\bf B}$ satisfying 
\hbox{$\theta_{\bf k B}<  (\hbar\omega/m_e c^2)^{1/2} (B/B_{\rm QED})^{-1/2}$,}
experience strong suppression of photon scattering even in the O mode
(e.g. M\'esz\'aros 1992).  Inside this wavevector cone, the O and 
E photon eigenstates are nearly circularly polarized, with opposing
helicities, rather than linearly polarized as 
at more oblique propagation angles. 
However, this cone is very narrow near the surface of the star: 
\hbox{$\theta_{\bf k B} < 0.05 \, B_{15}^{-1/2} \, (\hbar\omega/30 \, 
\hbox{keV})^{1/2}$ radians.} In the zone where the jet outflow 
is accelerated, just above the stellar surface, 
the O-mode specific intensity is nearly isotropic, and
almost all photons propagate outside the cone.

As the matter accelerates and expands, the optical depth through it 
drops off.  Let us estimate the radius 
$R_\tau$ where the matter and photons decouple, under the assumption
that $\theta_{\bf kB} \ll 1$ at this radius in the bulk frame.  
This decoupling radius 
must be smaller than the maximum radius of the confining magnetic
field lines ($\theta_{\rm jet}^{-2}\,R_{\rm NS}$ in a dipole geometry;
Fig. \ref{xrayjet}):
\begin{equation}\label{rtauin}
{R_\tau\over R_{\rm NS}} < \theta_{\rm jet}^{-2} = {\pi\over
\Delta\Omega_{\rm jet}}.
\end{equation}
To reach a radial path of polar angle $\theta$ and escape, the advected 
photons must diffuse through an angle $\sim {1\over 2}\theta$ from the
direction of the bulk streaming 
(which is tangent to the dipolar magnetic field).  Thus
the photons must diffuse a transverse distance $\Delta R_\perp
\sim {1\over 2}\theta R$,
which corresponds to an optical depth
\begin{equation}\label{tauperp}
\tau_\perp \sim {n_e(R)\over\gamma}\sigma_T\Delta R_\perp.
\end{equation}
The diffusion time $\tau_\perp R_\perp/c$ must be shorter than 
the radial flow time $R/\gamma c$, and we deduce
\begin{equation}
n_e(R)\left({1\over 2}\theta\right)^2\sigma_T R < 1.
\end{equation}
Making use of eqs. (\ref{taupar}) and (\ref{tauperp}), this becomes
\begin{equation}
{R_\tau\over R_{\rm NS}} \sim {\Delta\Omega_{\rm jet}(R_{\rm NS})\tau_\parallel
(R_{\rm NS})\over 4\pi}.
\end{equation}
Combining this result with eq. (\ref{rtauin}), we find that the
narrowness of the jet at its base {\it increases} increases with the 
optical depth:
\begin{equation}
{\Delta\Omega_{\rm jet}(R_{\rm NS})\over 4\pi} < 
{1\over 2\tau_\parallel^{1/2}(R_{\rm NS})}
\end{equation}.

Further collimation can occur when the escaping X-rays cross
the surface of the electron cyclotron resonance, which sits
at a distance $R/R_{\rm NS} = 6.7\,(\hbar\omega/40~{\rm keV})^{-1/3}\,
(B_{\rm NS}/10^{15}~{\rm G})^{1/3}$ from a neutron star
with polar field $B_{\rm NS}$.  During an SGR outburst, the 
matter suspended and confined at this radius can easily generate
a large {\it Thomson} optical depth.  This material
is constrained to move along the magnetic field (the magnetic
pressure greatly exceeds the radiative energy flux even at such
a distance), and so its angular distribution will reflect the
multipolar structure of the magnetic field near the stellar surface.
However, in the persistent emission, which has a flux well below Eddington,
the X-ray pulse profile will be strongly modified by resonant cyclotron 
scattering off magnetospheric currents (at either the electron or ion
resonances: Thompson et al. 2001).

\section{Conclusions}\label{conclusions}

The August 27 giant flare provides a Rosetta stone for
SGR 1900+14, just as the March 5 giant outburst did for SGR
0526-66.  These two remarkable events share very similar
peak luminosities, energies, and morphologies, which
suggests in turn that these two SGR sources are fundamentally alike.  

In the context of the magnetar model, we have discussed 
the mechanism by which such a giant flare may be triggered, and
the physical processes operating in each of the three principal
phases of the August 27 flare:  the initial $\sim 0.4$ sec hard spike;  
the intermediate smooth $\sim 40$ sec tail, and the final phase of 
large-amplitude pulsations that displays a striking four-peaked pattern.
In the process, we have refined physical arguments that point to 
an external (dipole) magnetic field stronger than $\sim 10^{14}$ G
in SGR 1900+14, and an even stronger $\sim 10^{15}$ G internal 
(toroidal) field which is the basic energy source for repeated flare 
activity.

\subsection{Physics of giant flares}

1.  In the flare mechanism proposed here, 
most of the potential energy that powers a giant flare is 
stored before the event in the magnetic field of the deep crust and
liquid stellar interior (TD95). This slowly-evolving field strains 
the crust from below.  The brittle crust acts as a gate for the 
catastrophic release of energy.  The elastic energy released in 
the crust is {\it smaller} than the available magnetic energy by a factor 
$\sim B^2/4\pi\mu$, where $B$ is the field in the stellar interior
and $\mu$ the crustal shear modulus, and by appropriate geometrical
factors (\S \ref{stored}).

2.  When the internal magnetic field is strongly wound,
its stored energy can be rapidly communicated to the stellar  
exterior via a propagating fracture, involving a rotational
deformation of a patch of the crust (\S \ref{exintmodes}).
Such a motion of the neutron star crust can create tangential 
discontinuities in the magnetic field, and thereby induce dissipation in
three distinct zones.  First, it induces strong magnetic shear
in parts of the magnetosphere, which can rapidly damp through reconnection and 
conversion to high frequency Alfv\'en waves; second, a torsional oscillation
of the magnetized core is excited (along with an accompanying toroidal 
deformations of the crust which involve less energy); and third, static
current sheets are excited deep in the crust.

3.  The damping of an internal shear mode by a flux of 
Alfv\'en waves into the magnetosphere has been further quantified.
In the case of a large-scale mode with harmonic 
$\ell \la R_{\rm NS}/\Delta R_\mu$ (eq. \ref{rmu}), the equilibrium
energy stored in a trapped magnetospheric Alfv\'en mode is greatly suppressed
with respect to the exciting internal mode (eq. [\ref{emagrat}]).
This implies strong lower bounds on the poloidal magnetic field 
($B \ga 10^{15}$ G) and internal mode energy $\delta E_{\rm elas} \ga
10^{44}$ ergs) needed to power the extreme peak luminosity of the initial
hard spike (\S \ref{peak}).  We also make a critical comparison of our 
results with the vibrating neutron star model of Ramaty et al. (1980), 
and review why an internal $f$- or $p$-mode couples even more weakly
to the magnetosphere than an internal shear mode.

4.  The contracting photosphere of a very hot ($T \sim 1$ MeV) confined
fireball provides an excellent fit to the envelope of the 1998 August 27 
giant flare after $\sim 40$s, and accounts for the rapid
final drop in flux (Paper I).  We show in \S \ref{softail}
that the fireball probably formed hotter 
in its center than at its edge, because the observed flux diminished
faster than it would for a homogeneous, spherical fireball.

5. The light curve appears not to be perturbed by neutrino cooling 
(Figs. \ref{neutrinoa}, \ref{neutrinob}), which implies an upper bound 
to the fireball temperature of $T \la T_{\rm max} = 0.8$ MeV.  This 
in turn provides a conservative lower bound to the volume and magnetic 
moment of the confining magnetic field.  The August 27 outburst
alone implies a strict lower bound  $\mu > 2.4\times 10^{31}\,
(E/10^{44}~{\rm ergs})\,(T/T_{\rm max})^{-2}$ G-cm$^3$. 
However, two considerations suggest that the net dipole moment
is several times larger:  the inferred number $\sim 10^2$ of
giant outbursts over the history of the source; and the absence
of a measureable change at the $\sim 10$ percent level
in the long-term spindown of SGR 1900+14 (Woods et al. 1999; Thompson
et al. 2000).

6.  The smooth $\sim 40$-sec tail of the 27 August flare
is somewhat harder than the ensuing large-amplitude
pulsations.  During this intermediate phase, the modest flux variations 
do not repeat coherently with rotational phase;  and the X-ray 
flux is significantly higher than predicted by the trapped fireball model 
which fits observations after $\sim 40$ sec.  In Paper I, we ascribed
this excess hard flux to Compton heating by an extended pair photosphere,
driven by a persistent seismic excitation of the neutron star.
In this paper, we have quantified the
behavior of a pair corona in a magnetic field stronger than 
$B_{\rm QED}$.  Cooling occurs primarily
by Compton heating of the O-mode.   We defined a critical
coronal luminosity (eqs. \ref{lmin} and \ref{lminb}) below which
a steady balance between electrostatic heating and diffusive
radiative cooling is possible.  This luminosity is $O(10^{42}\,$
ergs s$^{-1}$), which is comparable
to the observed output from SGR 1900+14 during the smooth tail. 
A magnetar flare could inject energy
into the magnetosphere at a much higher rate than 
this critical level. Indeed, the flux 
of the prompt, hard spike exceeds this minimum luminosity,
giving direct evidence that
energy was liberated fast enough to form a trapped fireball
at the onset of the August 27 flare.

7. The best fit black body temperature is stable during the
period of large-amplitude pulsations (Mazets et al. 1999; Paper I)
and agrees well with the value $kT_{\rm sp} \simeq 11$ keV at which
photon splitting freezes out (eq.~[180] in TD95).

8. We propose that the large-amplitude pulsations are due to a collimated
flux of X-rays from the base of the trapped fireball, moving along
extended magnetic field lines (see also TD95; Feroci et al. 2001).  
Collimation of the E-mode is
provided by the rapid increase in its scattering opacity with distance
from the stellar surface, $\sigma_E(B) \propto B^{-2} \propto R^6$.  
The O-mode flows hydrodynamically even
in the presence of a small flux of advected ions (and neutralizing electrons),
$\dot Mc^2/L_O \la (L_O/L_{\rm edd})^{-1} (GM_{\rm NS}/c^2)^{-1}$.
Energy can be released in both polarization modes at comparable rates.  The
width of the X-ray `jet' can be related directly to the
flux of advected matter.  The collimation becomes finer as
the matter flux increases, because the radiation-hydrodynamical
flow must extend to larger radius before the photons and matter can
decouple.

9. We have quantified some physical mechanisms that could generate
the persistent, hard spectral component of the August 27 flare. 
In particular, the burst spectrum is harder during dips, which points
to acceleration of non-thermal particles or direct Comptonization
by large-amplitude Alfv\'en waves in an extended corona.

\subsection{Evidence for Magnetars: Field Strength Estimates}\label{evidence} 

One of the principal goals of this paper is to set more model-independent
(lower) bounds on the magnetic field in the flaring SGR sources.  
Several observational properties of these sources directly require magnetic
fields stronger than $\sim 10^{14}$ G (the dipole component) to 
$\sim 10^{15}$ G (the internal toroidal field, and higher multipoles).

1.  The estimated output of $\sim 10^{47}$ ergs in giant flares
over the active history of a flaring source corresponds to a
r.m.s. magnetic field stronger than $\sim 10^{15}$ G (\S \ref{magmin}).

2. The observed lightcurve of the August 27 flare can be well fit
by the cooling of a `trapped fireball' (Feroci et al. 2001).
The confinement of a substantial fraction of the outburst 
energy [$\sim (1-3)\times 10^{44}$ ergs] in a pair-photon plasma
close to the source, which cools gradually over $\sim 300$ 
seconds, implies a lower bound $\sim 10^{14}$ G to the external dipole 
field (TD95).  This argument is refined in \S \ref{momentbound}, 
taking into account
the fit of the August 27 flare light curve to the contracting surface
of a trapped fireball (Paper I), and the restrictions on 
the size of the fireball from bulk neutrino-pair cooling.  
In addition, the intermediate $\sim 40$ smooth tail in the August 27
flare provides evidence that the seismic output of the source can
exceed $\sim 10^4\,L_{\rm edd}$ for a limited time.

3. The extreme peak luminosity
of the initial spike ($3-10\times 10^6\,L_{\rm edd}$: Fenimore et al.
1996; Mazets et al. 1999) can be powered through a sudden readjustment of
a magnetic field stronger than $\sim 10^{14}$ G \  (\S \ref{peak}),
especially if this readjustment involved a fracture of the neutron
star crust and the formation of regions of strong magnetic shear.
Such a violent event would also deposit energy into internal shear 
modes (including a torsional Alfv\'en wave in the liquid core and
standing shear waves in the rigid crust).  We have considered separately
the damping of these internal oscillations via a coupling to the
magnetophere.  Even under the most optimistic assumptions
about the rate at which an external torsional mode is converted to
radiation, the observed peak luminosity of the giant 
flares can barely be supplied if
the surface field is $\sim 10^{14}$ G, and cannot be supplied if
it is much weaker.  The luminosity of the pulsating tail, 
$L/L_{\rm edd} \la 10^4$, is consistent with the suppression of 
the electron scattering opacity of the E-mode radiation near the 
strongly magnetic surface of a trapped fireball.  In the context of 
the SGR sources, this effect was first discussed by Paczy\'nski (1992), 
under the assumption that the radiation is released from the 
cooling surface of a magnetar.   However, the physical effect is
in fact much cleaner if radiative transport occurs {\it across} the confining
magnetic field lines of a trapped fireball.

It is also worth re-evaluating other arguments for strong magnetic
field, in the light of recent observations:

4.  The long 8-s spin period of SGR 0526-66, combined with
its association with the young LMC supernova remnant N49 (e.g. 
Cline 1982), suggests that this source is rapidly spinning down.
Spindown from a much shorter period, in the age $t_{0526-66}$ of N49,
corresponds to a polar dipole field of 
$\sim 6\times 10^{14}\,(t_{0526-66}/10^4~{\rm yr})^{-1/2}$ G if the
classical magnetic dipole formula applies (DT92).
Although rotational modulation of the persistent emission of SGR 0526-66
has not yet been detected with compelling statistical confidence
(Kulkarni et al. 2000), two other sources SGR 1806-20 and SGR 1900+14 
are observed to spin down rapidly (Kouveliotou et al. 1998b, 1999).

The observed $\dot{P}$ variability of these stars (Woods et al. 1999;
2000) may be surprising at first sight, if they are identified as
isolated, non-accreting neutron stars.  However, as we have argued in
detail in this paper, the bright X-ray outbursts provide direct evidence
for sudden deformations of the external magnetic field in the SGR
sources.  The rate of spindown of an isolated neutron star is controlled
by the electrical current flowing across its speed-of-light cylinder.
Variations in the external magnetic field will, therefore, cause
a {\it modulation} of the spindown torque -- either through a
large-scale twisting of the external magnetic field (Thompson et al. 2001); 
or through a magnetically-powered wind (Thompson \& Blaes 1988; 
Thompson et al. 2000).  The wind model requires dipole fields 
$\gtrsim 1\times 10^{14}$ G for the SGRs with measured spindown, 
unless the energy in a wind exceeds the {\it observed} X-ray output 
by more than an order of magnitude.  

Note that several radiopulsars with spindown fields 
$B_{dipole} > B_{\rm QED}$ have
recently been found.  The present record value of $B_{dipole}$ inferred
from $P$ and $\dot{P}$ in a radiopulsar is $1.1 \times 10^{14}$ G (polar
field) for PSR 1814-1744 (Camilo et al.~2000).  This means that
{\it the minimum magnetic moment which we have inferred from the 
flare physics is near the upper end of the range measured in radio
pulsars.}  The dipole field of PSR 1814-1744 is a factor $10-20$ weaker
than that inferred from the spindown of SGRs 1806-20 and 1900+14 
if the magnetic dipole formula applies; but can plausibly be reduced
by a factor of $3-10$ in the presence of persistent currents
(Kouveliotou et al. 1998b, 1999; Harding, Contopoulos, \& Kazanas 1999; 
Thompson et al. 2000, 2001).

On the other hand, PSR 1814-1744
sits close in the $P-\dot P$ plane to the magnetar candidate AXP 1E2259+586.
Nontheless, the spindown age of 1E 2259+586 exceeds by a factor $\sim 10-30$
the age of the surrounding supernova remnant CTB 109 (e.g. Kaspi, Chakrabarty,
\& Steinberger 1999).  Given the 
rapid spindown observed in the other SGR and AXP sources, this suggests
that the spindown torque of 1E 2259+586 has decayed by a factor $\sim 10$
from its historic average.  Note that the interior, toroidal
magnetic fields of radiopulsars could be much weaker than those of 
SGRs and AXPs (TD93), which would provide an explanation for why
PSR 1814-1744 is a much weaker X-ray source than the AXPs (Pivovaroff,
Kaspi, \& Gotthelf 2000).  

5. The absence of a significant perturbation to the long-term spindown
rate of SGR 1900+14 in the few months following the August 27 event, 
implies only a small change in the external dipole
field coinciding with the release of $\sim 10^{44}$ ergs (Woods et al.
1999; Thompson et al. 2000).

6. The detection of extended afterglow from the heated surface of
a Soft Gamma Repeater is expected following exposure to a trapped fireball.
The amount of fireball energy absorbed by the crust increases 
linearly with the surface B-field at large magnetic flux densities
(TD95).  This effect has been observed in some short SGR bursts (Strohmayer
\& Ibrahim 1998), and has been related to an extended, faint
oscillatory tail that was observed in an August 29 outburst from
SGR 1900+14 (two days after the August 27 giant flare;  Ibrahim 
et al. 2000).

\acknowledgments
We thank Marco Feroci and Kevin Hurley for a stimulating collaboration
which inspired this work, and Lars Bildsten, Max Lyutikov, 
George Pavlov, and Mal Ruderman for conversations.  We also thank the 
Institute for Theoretical Physics at the University of California at 
Santa Barbara (NSF grant PHY99--0749) for its hospitality and support 
during the completion of this paper. CT acknowledges the support of 
NASA grant NAG5-3100, NSERC grant RGPIN 238487-01, and the Alfred P. 
Sloan Foundation.  RD acknowleges support from Texas Advanced Research 
Project grant no.~ARP-028 and NASA grant NAG5-8381.

\clearpage
\begin{deluxetable}{clll}
\tablecaption{Fireball Indices \label{tbl-1}}
\tablewidth{330pt}
\startdata
Dimensions &Energy Density  & \ \ \ Flux & Fireball Structure \cr 
&$U \propto T^{\mu_1} \, B^{\mu_2}$  & $F \propto T^{\sigma_1} \, 
B^{\sigma_2}$ & $T\propto r^\gamma$ \ \ $B\propto r^\beta$ \cr
$D_c$ & \ \ $\mu_1$ \ \ \ $\mu_2$ &\ \ $\sigma_1$ \ \ \ $\sigma_2$ 
&\ \ \ $\gamma$ \ \ \ \ \ \ \ $\beta$ \cr
\hline
1  &\ \ 4 \ \ \ \ 0 &\ 5/2 \ \ \ \ \ 0 & \ \ \  -1   \ \ \ \ \ \ * \cr  
1  &\ \ 4 \ \ \ \ 0 &\ \ 0 \ \ \ \ 4/3 & \ \ -1/6  \ \ \ \ \ 0 \cr  
1  &\ \ 4 \ \ \ \ 0 &\ \ 0 \ \ \ \ 4/3 & \ \ \  0   \ \ \ \ \ \ 3/8 \cr  
1  &\ \ 2 \ \ \ \ 1 &\ 3/2 \ \ \ \ \ 1  & \ \ \ *  \ \ \ \ \ \ \ 2 \cr  
1  &\ \ 2 \ \ \ \ 1 &\ \ 0 \ \ \ \ 4/3 & \ \ -1/3  \ \ \ \ \ 0 \cr  
1  &\ \ 2 \ \ \ \ 1 &\ \ 0 \ \ \ \ 4/3 & \ \ \ 0  \ \ \ \ \ \ 6/7 \cr  
2  &\ \ 4 \ \ \ \ 0 &\ 5/2 \ \ \ \ \ 0 & \ \ \  -1/2   \ \ \ \ \ * \cr  
2  &\ \ 4 \ \ \ \ 0 &\ \ 0 \ \ \ \ 4/3 & \ \ -1/12  \ \ \ \ 0 \cr  
2  &\ \ 4 \ \ \ \ 0 &\ \ 0 \ \ \ \ 4/3 & \ \ \  0   \ \ \ \ \ 3/16 \cr  
2  &\ \ 2 \ \ \ \ 1 &\ 3/2 \ \ \ \ \ 1  & \ \ \ *  \ \ \ \ \ \ \ 1 \cr  
2  &\ \ 2 \ \ \ \ 1 &\ \ 0 \ \ \ \ 4/3 & \ \ -1/6  \ \ \ \ \ 0 \cr  
2  &\ \ 2 \ \ \ \ 1 &\ \ 0 \ \ \ \ 4/3 & \ \ \ 0  \ \ \ \ \ 3/7 \cr  

\tablecomments{ The last two columns give the values of fireball structure
indices needed to explain the observed time-dependence of the August
27 event ($\chi = 3$). The symbol *  means that the observed 
time-evolution is insensitive to this index, i.e., any value would work.} 
\enddata
\end{deluxetable}

\clearpage
\begin{figure}
\figurenum{1}
\plotone{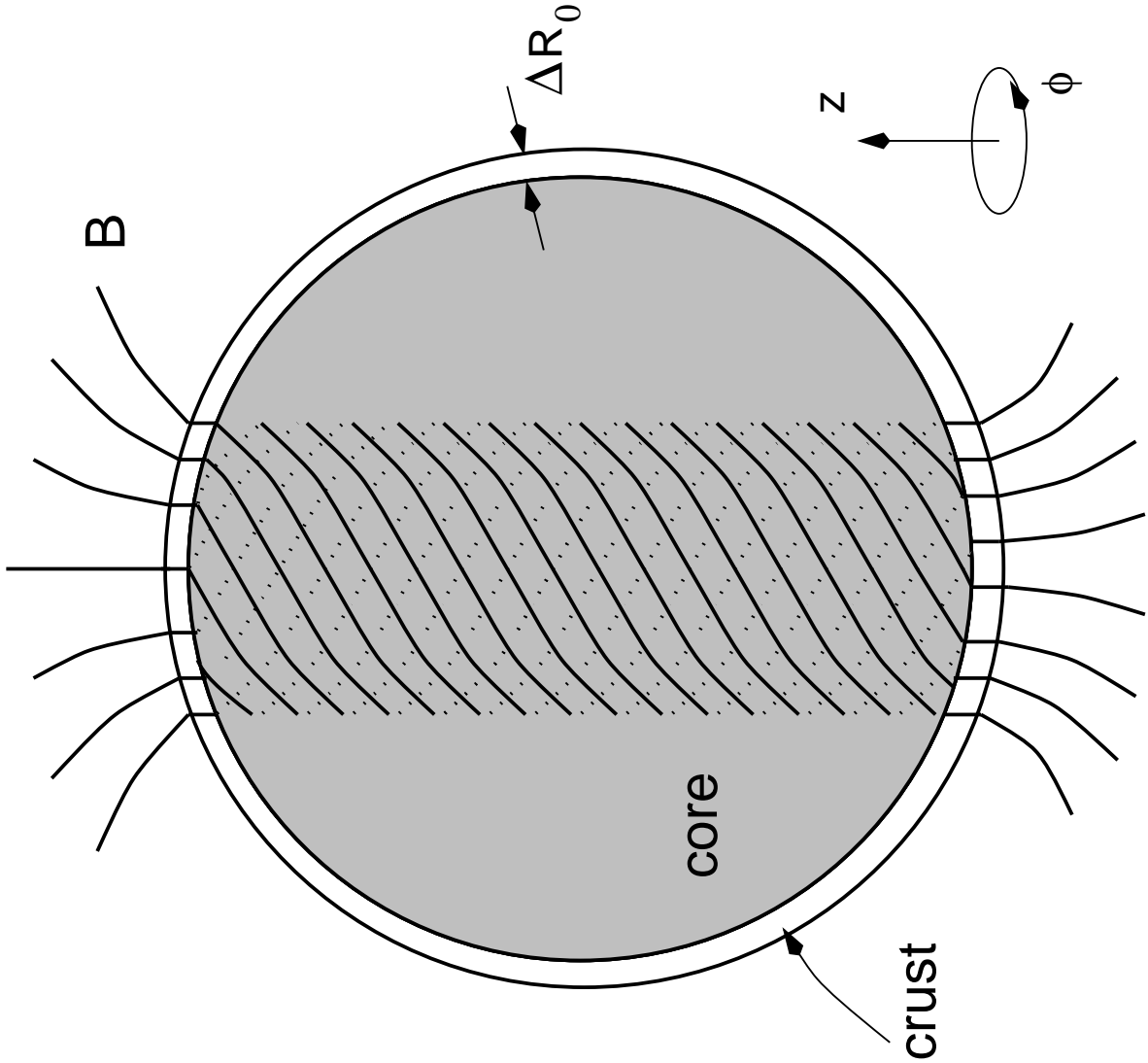}
\caption{
An almost spherical star is threaded by a uniform poloidal magnetic field.
This field threads both the liquid core and the solid, outer crust of the
star.  In the core, the field is twisted, creating a toroidal component.}
\label{twist}
\end{figure}

\clearpage
\begin{figure}
\figurenum{2a}
\plotone{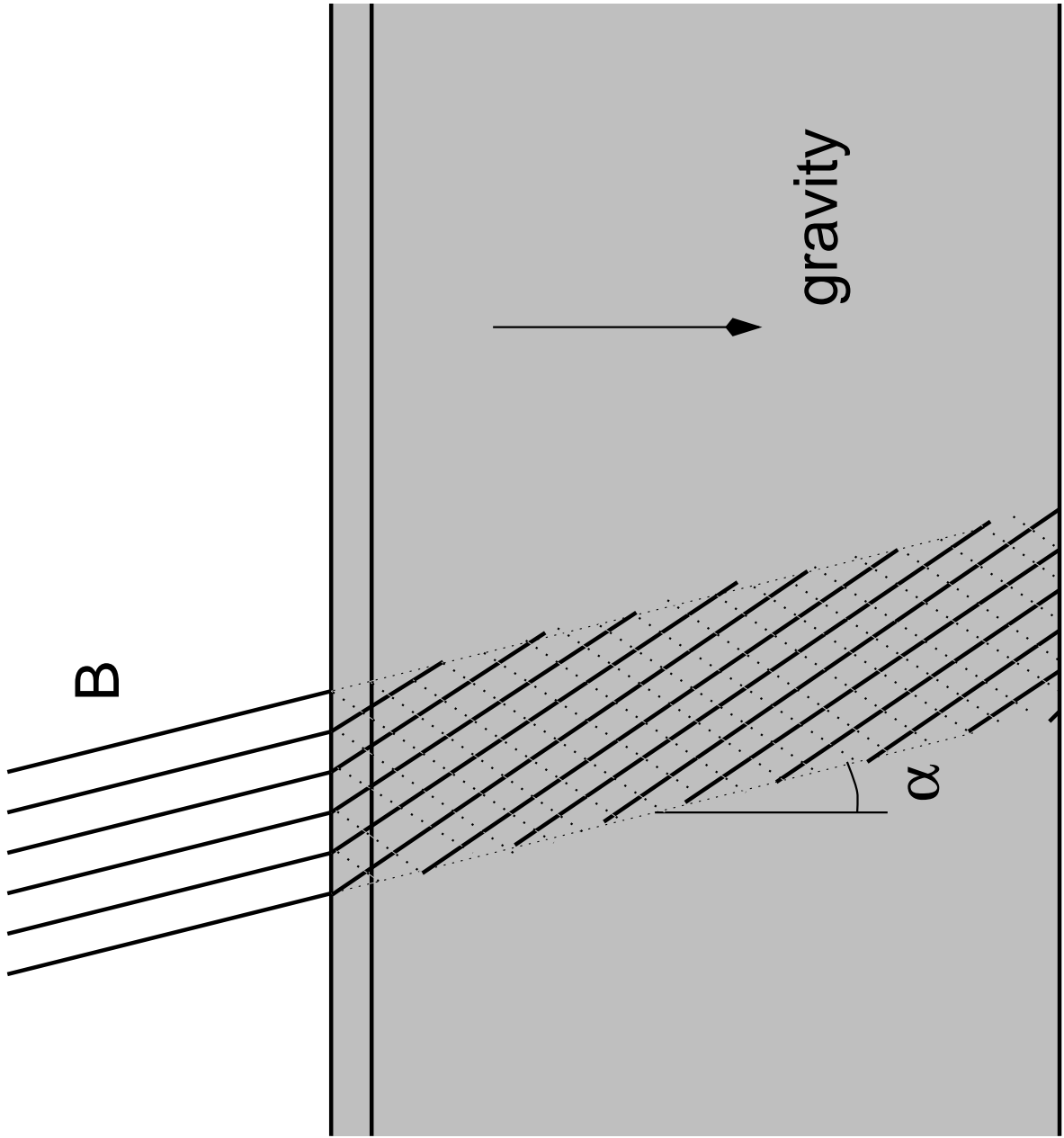}
\caption{
A twisted bundle of magnetic flux threads a conducting medium, which
is stratified along planes which run perpendicular to the direction of gravity.
The flux bundle is tilted by angle $\alpha$ with respect to gravity;
it may be anchored from above and below.}
\label{twistb}
\end{figure}

\clearpage
\begin{figure}
\figurenum{2b}
\plotone{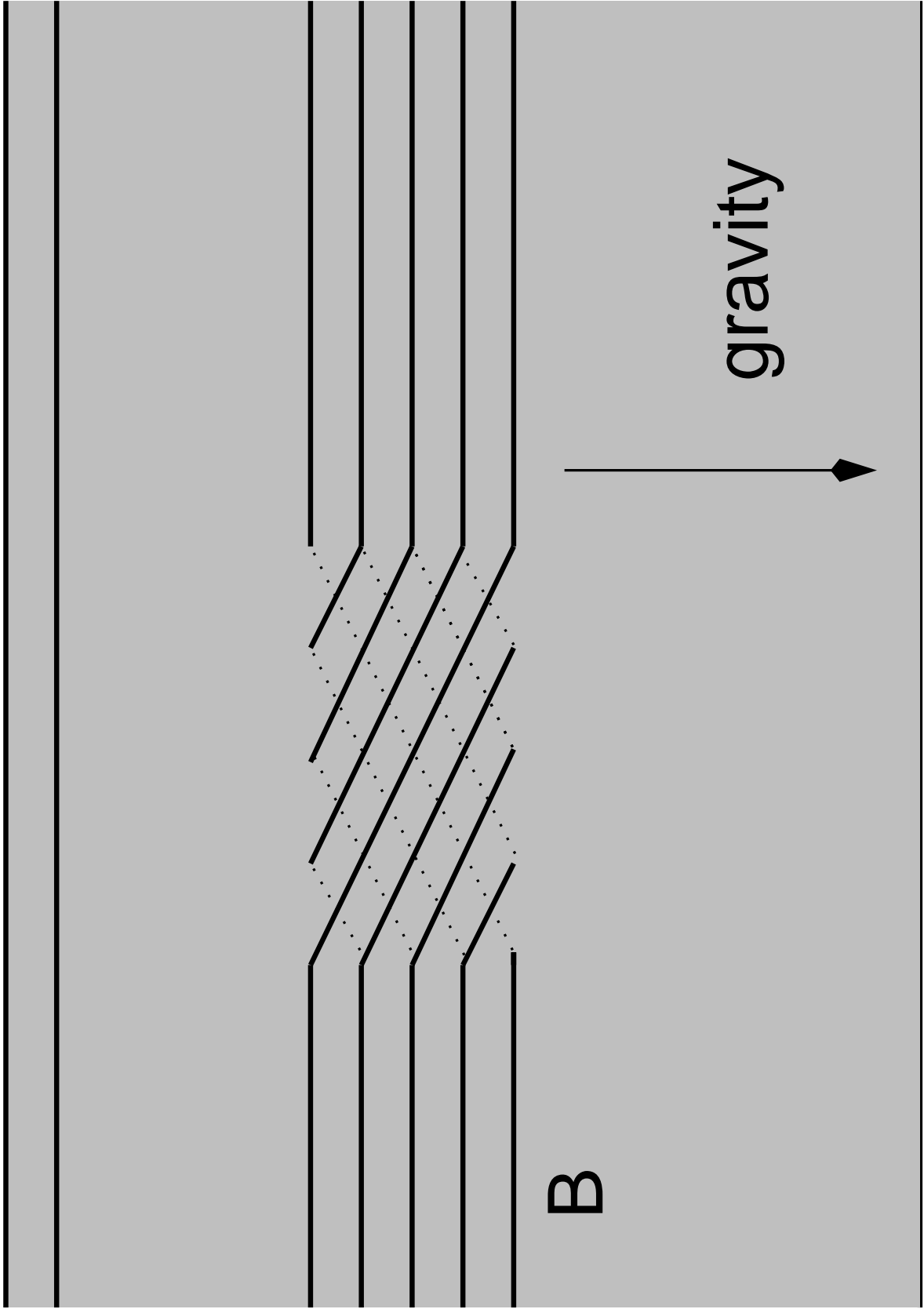}
\caption{
When the flux bundle runs perpendicular to the direction of stratification,
a localized twist is prevented from spreading out along the field
through purely hydromagnetic motions.}
\label{twistc}
\end{figure}

\clearpage
\begin{figure}
\figurenum{3}
\plotone{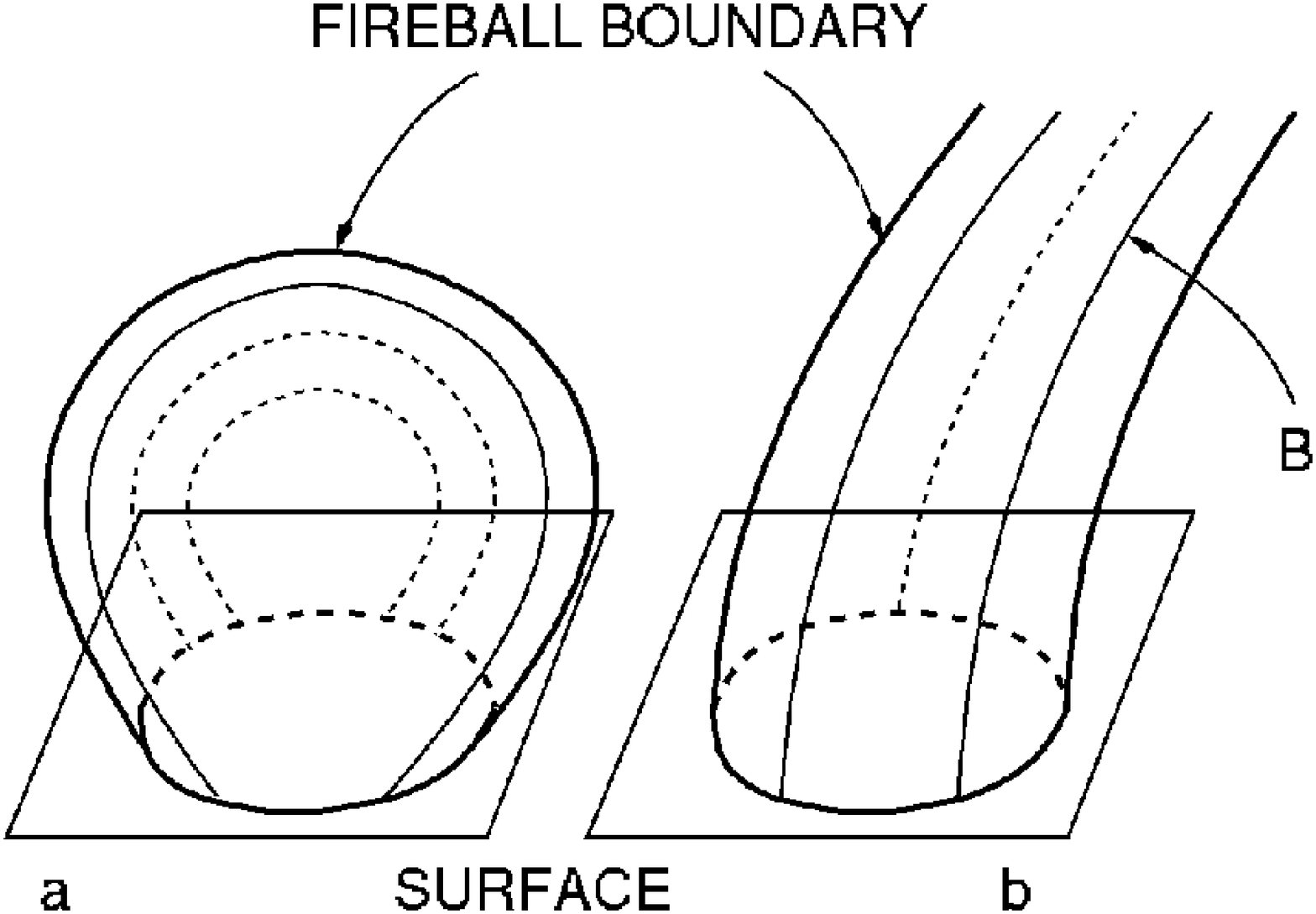}
\caption{
A trapped fireball releases energy as its cool outer surface contracts.
We illustrate two possible geometries (approximately
spherical and cylindrical) for the fireball.  
The measured fluence of the extended tail of a giant flare, combined with
the absence of a measureable perturbation to the light curve from
pair neutrino cooling, sets a lower bound to the dipole moment of the
confining magnetic field.}  
\label{confinement}
\end{figure}

\clearpage
\begin{figure}
\figurenum{4a}
\plotone{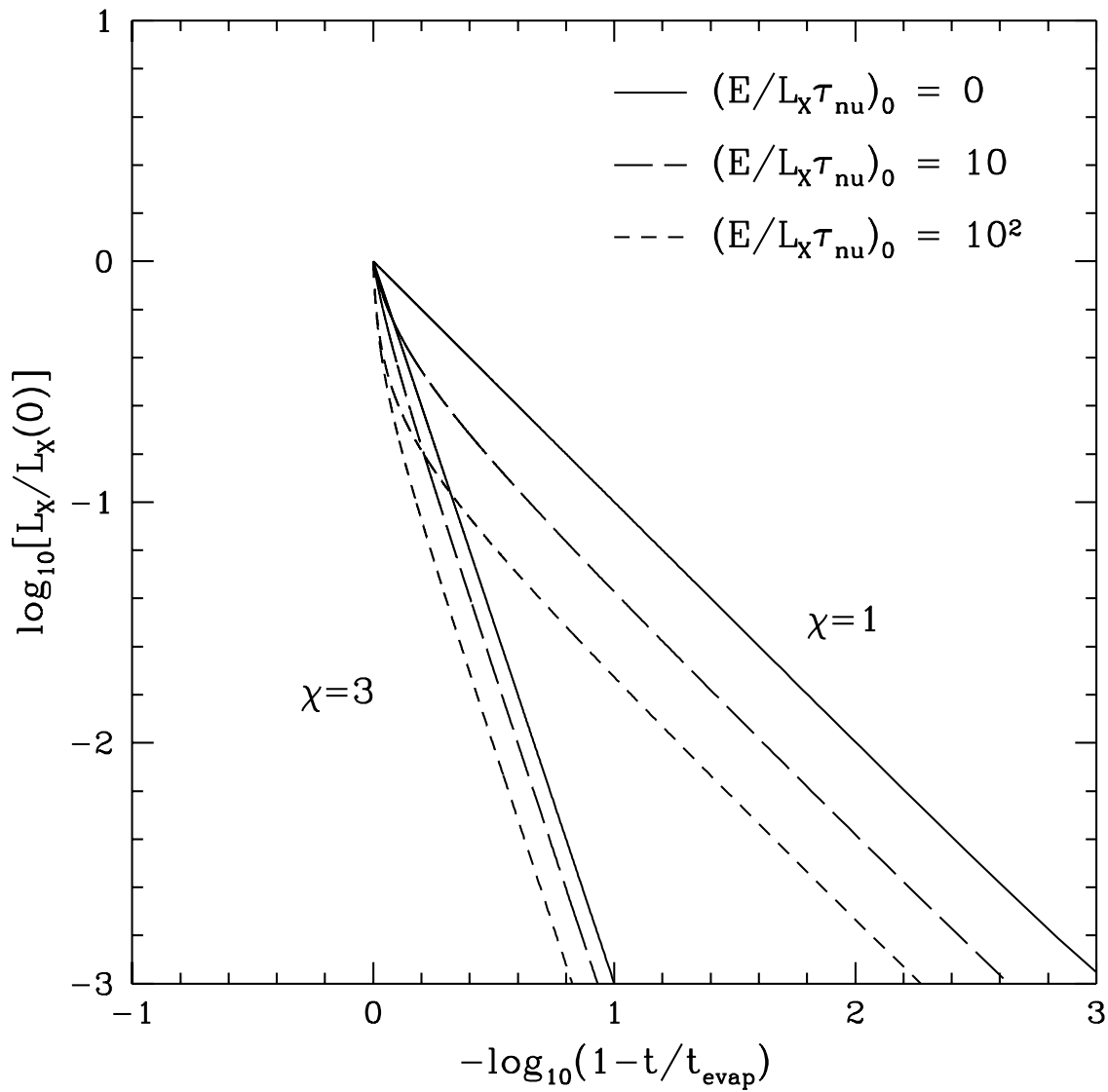}
\caption{
Surface X-ray luminosity of a homogeneous, trapped fireball, whose
radiative
area scales with the fireball volume as $A \sim V^{0.75}$ ($\chi = 3$;
lower curves) and $A \sim V^{0.5}$ ($\chi = 1$; upper curves).  
Within each set of curves, the solid curve includes no neutrino cooling.
The long-dashed and short-dashed curves correspond
to uniform pair neutrino cooling, with an initial cooling rate 1 and
10 times the initial surface X-ray luminosity.
}
\label{neutrinoa}
\end{figure}

\clearpage
\begin{figure}
\figurenum{4b}
\plotone{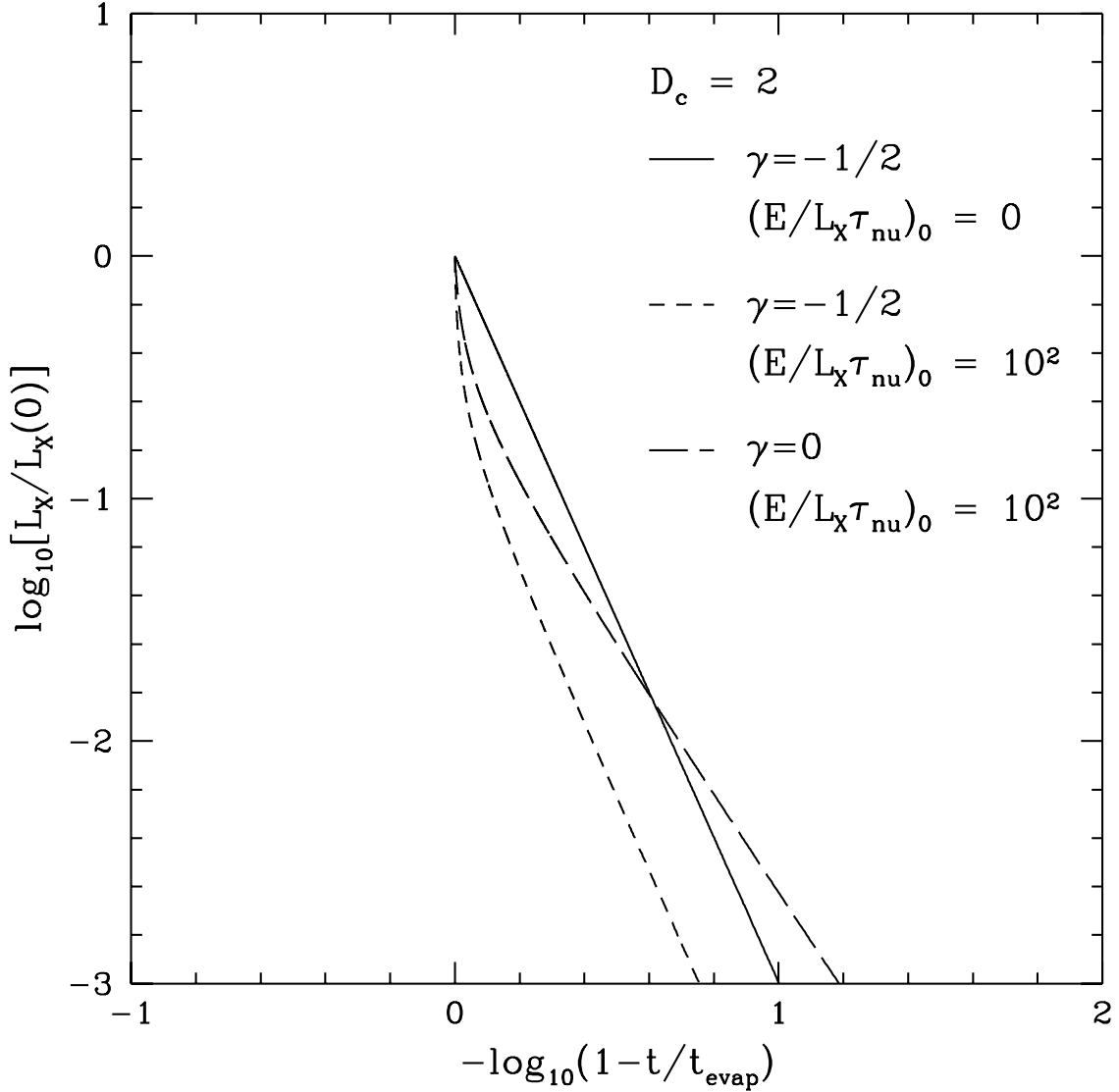}
\caption{
Surface X-ray luminosity of a spherical trapped fireball ($D_c = 2$), showing
the effects of neutrino cooling on the shape of the lightcurve.
The bold curve ($\chi = 3$) closely approximates the observed lightcurve
of the August 27 flare.  In a homogeneous magnetic field, this fireball
index corresponds to a mild temperature gradient $\gamma = -{1\over 2}$.
The short-dashed curve describes the effect of pair-neutrino cooling 
on the X-ray lightcurve, with the neutrino luminosity initially
100 times the cooling X-ray luminosity.   The same relative neutrino
cooling rate, but a different temperature structure ($\gamma = 0$), brings
closer agreement with the observed light curve, but there remains
some difference in shape (long-dashed curve).
}
\label{neutrinob}
\end{figure}

\clearpage
\begin{figure}
\figurenum{5}
\plotone{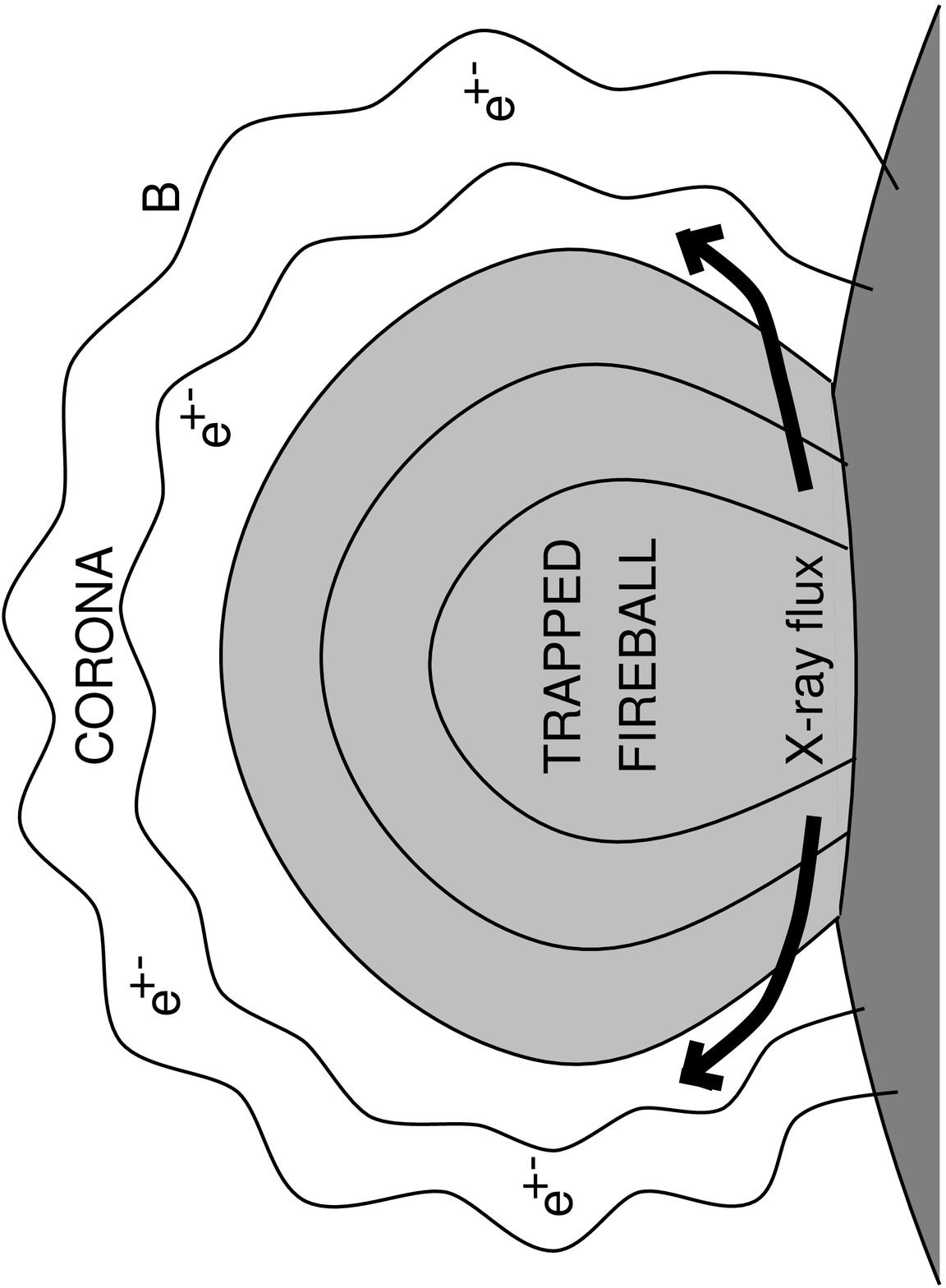}
\label{hotcorona}
\end{figure}

\clearpage
\begin{figure}
\figurenum{5}
\caption{
A persistent seismic excitation of the neutron star crust will
in turn excite sheared (current-carrying) Alfv\'en waves on 
extended magnetic field lines in the star's magnetosphere.  
Non-linear couplings between
these trapped Alfv\'en modes generate a turbulent cascade that
creates a hot corona.  At the high luminosity (and inferred compactness)
of the August 27 smooth tail, this corona consists of electron-positron
pairs with only a slightly higher temperature than the X-ray photons
(which are assumed to have a Wien distribution).  The dominant source
of fresh Compton seeds for this Corona could either be provided
by an interior trapped fireball (as depicted); or, alternatively, by
photon splitting.}
\end{figure}

\clearpage
\begin{figure}
\figurenum{6}
\plotone{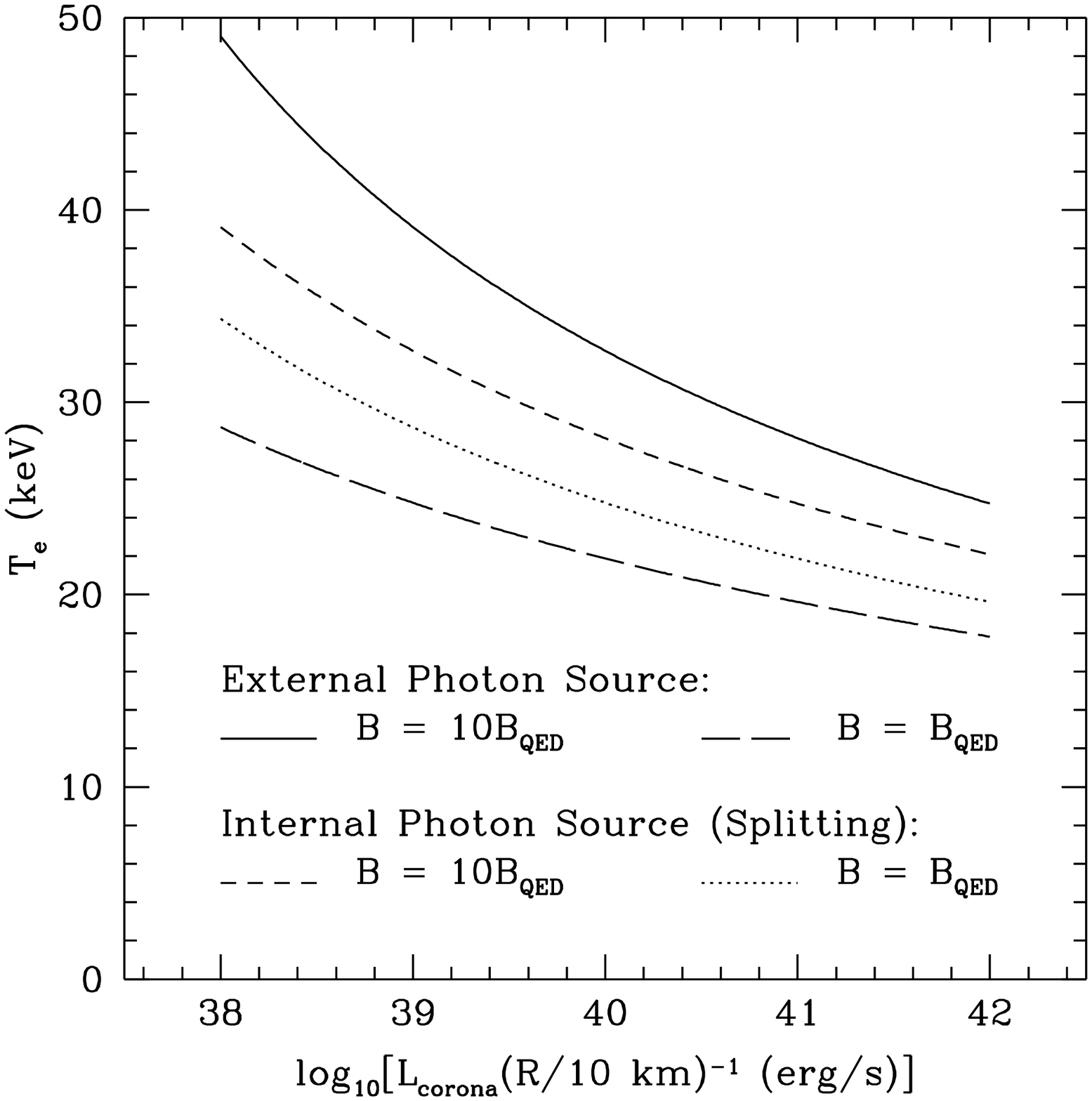}
\label{tpair}
\caption{
Equilibrium temperature of a electron-positron corona (of size $R$), under
two different assumptions about the mechanisms of photon creation
and radiative cooling.  The top and bottom curves correspond to an external
source of (E-mode) photons (eq. [\ref{lcasa}]).  The second
set of curves corresponds to an internal photon source, which
is balanced by outward diffusion through the O-mode (eq. [\ref{lcasc}]).
The maximum coronal luminosity, above which no stable balance between
heating and cooling is possible, lies close to $\sim 10^{42}$ erg s$^{-1}$
(eqs. [\ref{lmin}] and [\ref{lminb}]).}
\end{figure}

\clearpage
\begin{figure}
\figurenum{7}
\plotone{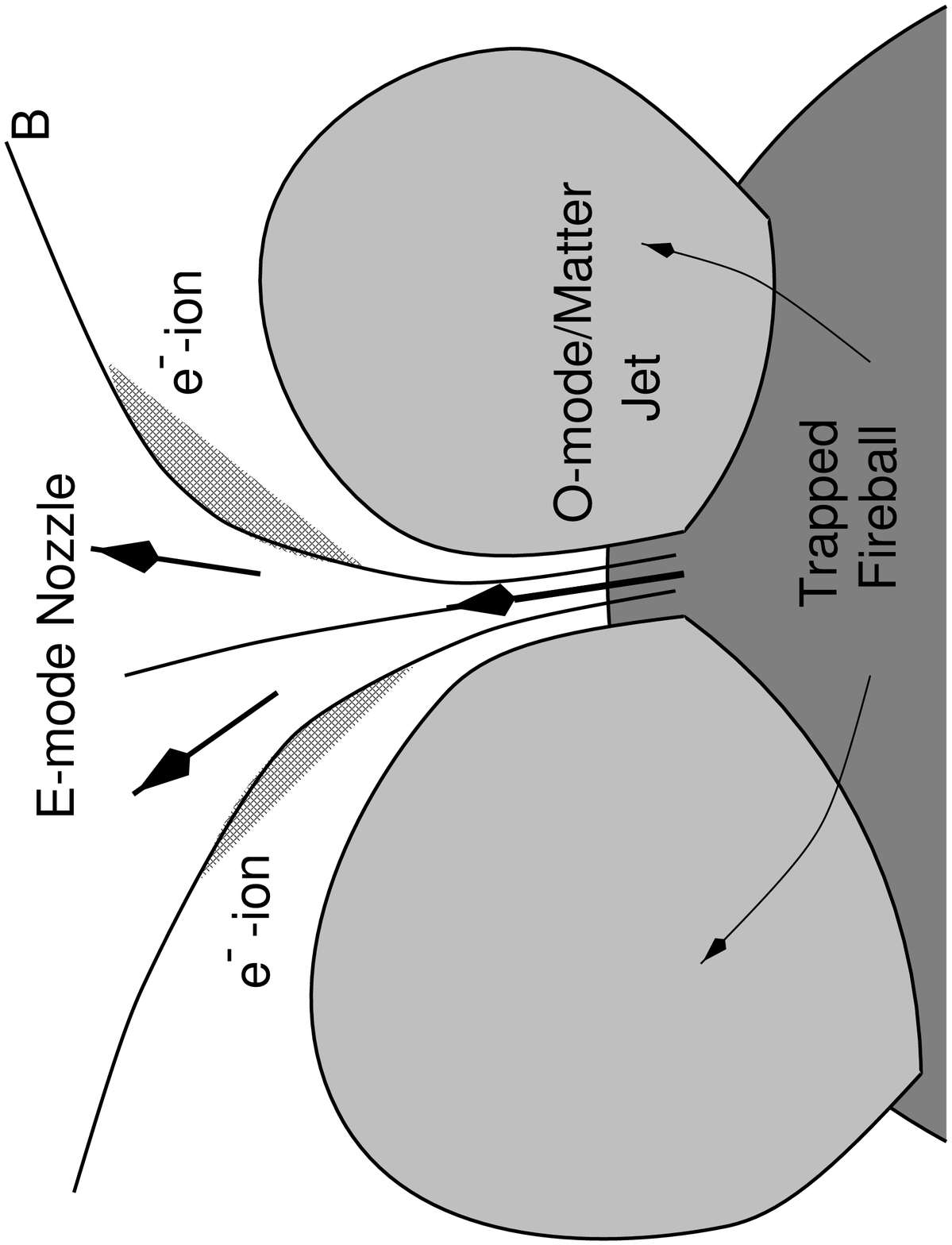}
\caption{
\baselineskip=15pt
A trapped fireball releases energy as its cool outer surface contracts.
The radiative flux across the fireball surface is concentrated close to
the surface of the neutron star, where the E-mode scattering opacity is most
strongly suppressed by the intense magnetic field.  A collimated flux of both
the O-mode and E-mode radiation can be generated along more extended
magnetic field lines.   The O-mode couples tightly to the electrons
near the stellar surface, even if a small fraction of the hyper-Eddington 
radiative flux is carried by matter;  and electrons suspended in the outer
magnetosphere provide collimation for the E-mode, out where its scattering
opacity has risen to be comparable to Thomson.}
\label{xrayjet}
\end{figure}

\clearpage
\begin{figure}
\figurenum{8}
\plotone{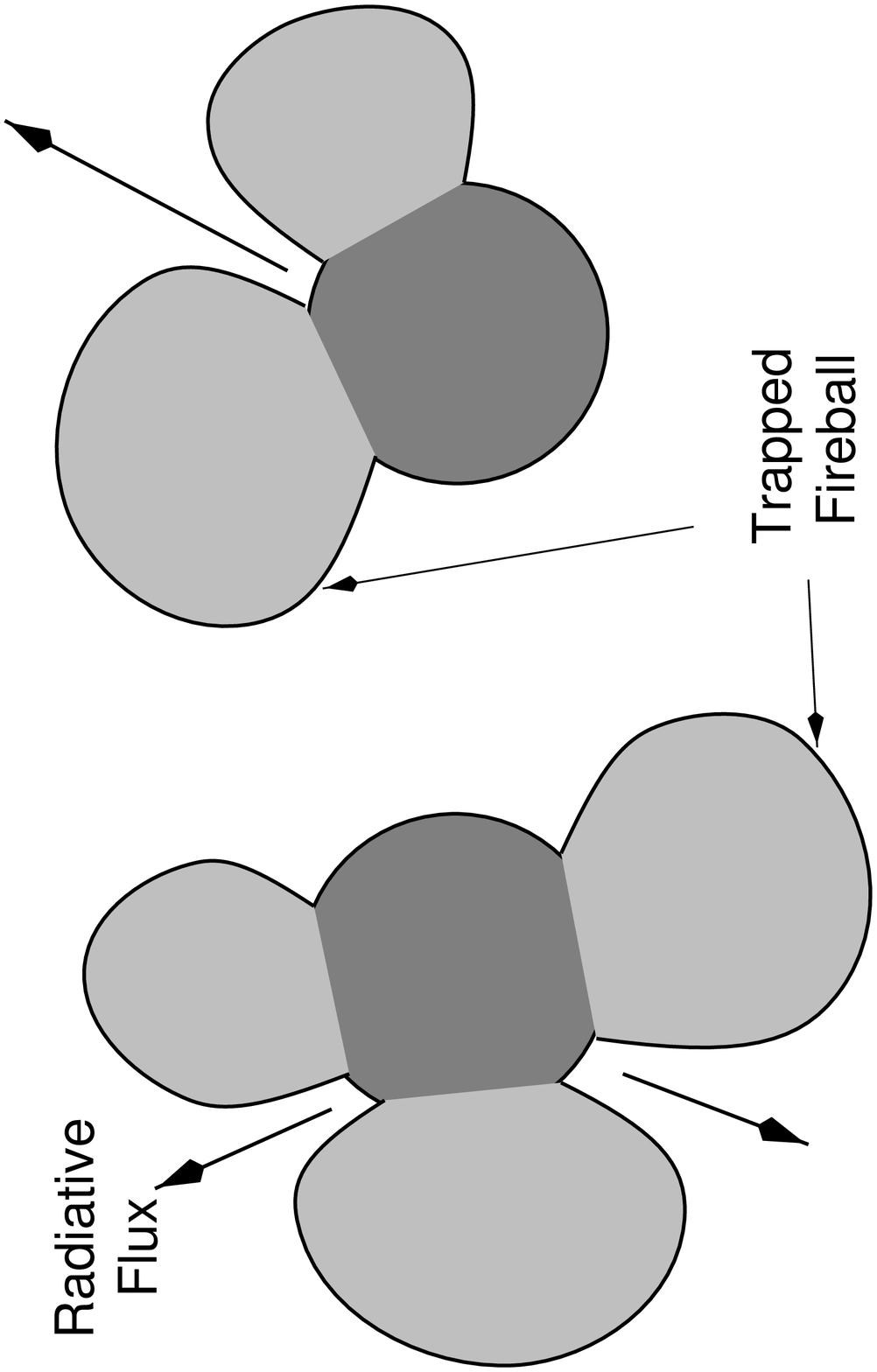}
\label{xrayjetb}
\end{figure}

\clearpage
\begin{figure}
\figurenum{7}
\caption{
A geometry is suggested both for the August 27 giant flare from 
SGR 1900+14 (left; four sub-pulses) as well as the original March 5 flare
from SGR 0525-66 (right; two sub-pulses).  Each large arrow denotes
a fan beam of X-rays and outflowing, relativistic matter.  The angular
velocity of each source is assumed to be oriented so that each fan
beam is observed twice per rotation, yielding two sub-pulses per beam.
In this model, the burst light curve is sensitive to the configuration of 
the particular subset of the closed magnetic field lines which confine 
the hot $e^\pm$-photon plasma.  Thus, the presence of higher 
multipoles in SGR 0526-66 is not excluded by the observation of
two sub-pulses within the tail of the March 5 flare.}
\end{figure}

\end{document}